\documentclass[usegraphicx,useAMS,usenatbib, a4paper]{mn2e}
\usepackage{times}
\usepackage{amssymb}
\usepackage{amsmath}
\usepackage{graphicx}
\usepackage{euscript}
\usepackage{rotating}
\usepackage{epsfig}
\usepackage{url}

\def\ghz{{\rm\thinspace GHz}}

\def\ergpspcmsq{{\rm\thinspace erg~s^{-1}~cm^{-2}}}
\def\ergps{{\rm\thinspace erg~s^{-1}}}

\def\kpc{{\rm\thinspace kpc}}

\def\kev{{\rm\thinspace keV}}
\def\ev{{\rm\thinspace eV}}

\def\Ms{{\rm\thinspace Ms}}
\def\pcm2{{\rm\thinspace cm^{-2}}}
\def\pkevpcmsqps{{\rm\thinspace keV^{-1}~cm^{2}~s^{-1}}}
\def\gx339{GX~339-4}
\def\pca{{\it PCA}}
\def\hexte{{\it HEXTE}}
\def\rxte{{\it RXTE}}

\newcommand{\eg}{e.g.\thinspace}

\begin{document}

\title{Studying the X-ray hysteresis in GX~339-4: the disc and iron line
  over one decade}
\author[Dunn, Fender, K\"ording, Cabanac \& Belloni]
{\parbox[]{6.in} {R.~J.~H.~Dunn$^1$\thanks{E-mail:
      r.j.dunn@phys.soton.ac.uk}, R.~P.~Fender$^1$,
    E.~G.~K\"ording$^1$, C.~Cabanac$^1$ and T.~Belloni$^2$\\
    \footnotesize
    $^1$School of Physics and Astronomy, University of Southampton, SO17
    1BJ, UK,\\
    $^2$INAF-Osservatorio Astronomico di Brera, Via E. Bianchi 46, I-23807 Merate (LC), Italy\\
  }}

\maketitle

\begin{abstract}
We report on a comprehensive and consistent investigation into the
X-ray emission from \gx339.  All public
observations in the 11 year \rxte\ archive were analysed.  Three
different types of model - single powerlaw, broken powerlaw and a disc +
powerlaw - were fitted to investigate the evolution of the disc, along with a fixed gaussian component at
$6.4\kev$ to investigate any iron line in the spectrum.  We show that the relative variation in flux and X-ray
colour between the two best sampled outbursts are very similar. The decay of the disc temperature during the outburst is clearly seen
in the soft state.  The expected decay is $S_{\rm Disc} \propto
T^4$; we measure $T^{4.75\pm0.23}$.  This implies that
the inner disc radius is approximately constant in the soft state.  We
also show a significant anti-correlation between the iron line
significant width and the X-ray flux in the soft state while in the
hard state the $EW$ is independent of the flux.  This results in
hysteresis in the relation between X-ray flux and both line flux and $EW$.  To compare the X-ray
binary outburst to the behaviour seen in AGN, we construct a Disc
Fraction Luminosity Diagram for \gx339, the first for an X-ray binary.  The shape qualitatively matches that produced for AGN.  Linking this with
the radio emission from \gx339\ the change in radio spectrum between
the disc and power-law dominated states is clearly visible.

\end{abstract}


\section{Introduction}

The galactic X-ray binary \gx339\ (=V821 Ara) was discovered by
the {\it OSO-7} satellite \citep{Markert73}.  The optical counterpart
to the companion star of the compact object has not been detected.
Therefore the system has been classed as a low-mass
X-ray binary (XRB) from upper limits on the luminosity of the companion
\citep{Shahbaz01}.  \citet{Hynes04} using the non-detection
and orbital parameters, conclude that the companion is probably a sub-giant, of
spectral type G or later.  Studies of the X-ray spectral and temporal characteristics conclude that it
is a black hole X-ray binary \citep{Zdziarski98,Sunyaev00}; the
characteristics of the Fourier power spectra are more similar to other
black hole X-ray binaries, rather than those of neutron stars.  For other studies of the X-ray emission from
this source see
e.g. \citet{Ueda94,Zdziarski98,Wilms99,Kong00,Wardzinski02,Zdziarski04,
Belloni05}.

Black hole X-ray binaries exhibit two main
spectral states, the low-hard and the high-soft
(e.g. \citealp{vanderKlis95, McClintock06}).  In a variety of
accretion models, the only parameter which
determines the state of an X-ray binary is the mass accretion rate
(e.g. \citealp{Esin97}), however a number of phenomena indicate that 
other parameters are also important.  The major one is the hysteresis
of the state transition -- the transition from low-hard to high-soft
occurs at higher fluxes than the return to the low-hard from the
high-soft state (e.g. \citealp{Miyamoto95}).  In addition, the luminosity of the source when the first state transition
occurs varies from outburst to outburst \citep{Belloni06}.

As described in detail in \citet{Fender04}, an average X-ray binary spends a large
fraction of its time in the low-hard state.  In this state radio
emission indicates that the black hole is producing a steady
radio-synchrotron emitting jet and the X-ray spectrum can be modelled
by a hard powerlaw (photon index, $\Gamma\sim1.5$).  As
the outburst progresses the luminosity of the binary increases and the X-ray
spectrum softens.  Eventually the jet disrupts and the X-ray spectrum
is dominated by the soft disc emission.  This is known the high-soft state.  During the outburst,
matter is accreted onto the black hole.  The mass accretion rate reduces and disc is expected to cool, and therefore
fade, as the outburst progresses.   At the end of the outburst the
spectrum hardens as the system returns to the low-hard state and the
jet reforms.

\rxte\ was launched in December 1995 and has been observing \gx339\
since June 1996.  The archive of data built up over the
intervening period allows an extensive investigation into the
behaviour of \gx339\ over a comparatively long baseline using the same
instrument.  The long
term variability allows comparisons to be made between the four clear outbursts
seen in the \rxte\ data.  The large number of individual observations
also allow detailed studies of single outbursts of \gx339.

In the hard state, \gx339\ is a weak but steady radio source, with flux
densities  $\lesssim 15$~mJy at centimetre wavelengths and a flat
spectrum (for a review see \citealp{Fender99}).  In the low-hard state
there is a non-linear positive correlation between the radio flux and both
the hard and soft X-ray
fluxes \citep{Corbel00}.  This is likely to result from
a coupling between the corona and the jet.  Radio observations during
the transition to the soft state have
also seen ejection events which have been interpreted as large-scale
relativistic radio jets \citep{Fender97, Gallo04}.

\gx339\ regularly varies by 4-5 orders of magnitude in X-ray flux.
This relatively high number of outbursts, from a very low quiescent
state to a high state, makes \gx339\ one of the most interesting
galactic X-ray binary sources.   Combining radio with X-ray
observations of the source during an outburst led to the development
of a physical explanation for the evolution of the source as the
outburst progresses \citep{Homan01, Fender04}.  

Black holes themselves are described by only two parameters,
their mass and spin\footnote{Charge is a further parameter of a black
  hole, but as no macroscopic object in the Universe has as yet found
  to be significantly charged, we ignore this parameter in this discussion.}.  The
observational properties of the X-ray binary arise from the interplay of the
parameters of the black hole (mass and spin) and those of the accretion
flow (including the mass accretion rate through the disc).  The black hole
parameters are unlikely to change much on short timescales, and so it
is the mass accretion rate and geometry of the accretion flow which is responsible for the large changes
in the emission spectrum.  

Recent developments have been made in creating 
unification schemes between X-ray binaries and Active Galactic Nuclei
(AGN).  The radio-X-ray
flux relation for \gx339\ was generalised to a large number of black
hole X-ray binaries by \citet{Gallo03}.  A scaling across a wide range of
mass scales from X-ray binaries to AGN was subsequently presented by
\citet{Merloni03} and \citet{Falcke04} and became known as the ``fundamental plane''.  Another plane using the timing
properties of black holes has recently been constructed by
\citet{McHardy06} and \citet{Koerding07}.

The observed variability timescales of AGN are many orders of magnitude longer than those
in X-ray binaries.  To study the evolution of an AGN outburst a
population study is required.  In order to be able to compare the
outbursts of black hole binaries with AGN \citet{Koerding06} create a more
general diagram to show the spectral states in AGN by comparing the
disc flux and the power-law flux.  We complement this by creating an
equivalent diagram for this X-ray binary (see Section \ref{sec:DFLD}.

We present our data reduction scheme in Sections \ref{sec:data} and
\ref{sec:model}.  Our investigations in to the iron line and the Disc
are discussed in Sections \ref{sec:FeLine} and \ref{sec:DiscT}, with the Radio
correlations presented in Section \ref{sec:Radio}.

\section{Data Reduction}\label{sec:data}

In order to fully investigate the existence and properties of the
iron line and disc in \gx339, all observations in the \rxte\ archive were
downloaded and analysed\footnote{The cut-off date for inclusion in
  this analysis was the $14^{\rm th}$
  February 2008.}.  This gave $2.59\Ms$ of \pca\ exposure over an 11 year baseline for this study,
containing the two well-sampled outbursts, as well as two other
outbursts with fewer public \rxte\ observations.  

All the data
were reprocessed so all observations had the same version of the data
reduction scripts applied.  The \hexte\ data were analysed so that the
high energy spectral slope could be well constrained.  This helps in fitting the
lower energy spectrum.  We use the data reduction tools from
HEASOFT\footnote{\url{http://heasarc.gsfc.nasa.gov/lheasoft/}} version
6.3.2.  The data reduction and model fitting were automated so that
each observation was treated in exactly the same way.  

\subsection{PCA Reduction}\label{sec:data:pca_red}

The \pca\ data were reduced according to the \rxte\
Cookbook\footnote{\url{http://rxte.gsfc.nasa.gov/docs/xte/recipes/cook_book.html}},
and only a quick summary is given here. 

We used only data from PCU-2 as it is always switched on, and so can be used
over the entire archive of data.  It is also the best calibrated of the
PCUs on \rxte.  Background spectra were obtained using {\scshape
  pcabackest} from new filter files created using {\scshape xtefilt},
from which updated GTI files were also created.  The spectra were
extracted and the customary systematic error of 1 per cent was added
to all spectra using {\scshape grppha}.  

In order that the model fitting in {\scshape xspec} was reliable and
relatively quick, we only fitted spectra which had more that 1000
background subtracted \pca\ counts.  In total there were 913 spectra extracted
from PCA data. Cutting those with fewer than 1000 counts from
further analysis, removed 191 observations (the bottom 21 per cent), leaving 722.  The excluded observations occur
throughout the light curves of the outburst, with a concentration in
the low flux periods.  The low-luminosity observations with low
counts fall into the ``stalks'' of the Hardness Intensity Diagrams.
As our analysis concentrates on the disc parameters, excluding these
observations is unlikely to bias our conclusions.

\subsection{HEXTE Reduction}\label{sec:data:hexte_red}

The \hexte\ data were also reduced according to the \rxte\ Cookbook.
Where possible, we
used both Cluster A and Cluster B data.  Background
spectra were obtained using {\scshape hxtback}.  Spectra were extracted
using the routine appropriate to the data-type (Event or
Archive). Dead-time was then calculated using {\scshape hxtdead}.  All
spectra for a given ObsID were then summed using {\scshape sumpha}, and
the appropriate responses and ancillary files were added in as header
key words using {\scshape grppha}.  In total there were 701 spectra extracted
from Cluster A data and 905 for Cluster B.  In order to accurately
determine the slope of the high energy 
power-law we require \hexte\ data to be present when fitting a model.
There have to be at least 2000 background subtracted counts in one of
the \hexte\ clusters, with the other having a positive number of
counts\footnote{The background subtraction on some observations
  resulted in a negative number of \hexte\ counts in one of the
  clusters.}.  As our analysis concentrates on the properties of the disc and any
emission line which might be present, we also require \pca\ data to be
present in the spectra which are fitted.

\section{Model Fitting}\label{sec:model}

The spectra were fitted in {\scshape xspec} (v12.3.1ao).  As our spectral
analysis concentrates on the disc properties of \gx339, we wish to
analyse the spectra to the lowest possible energies.  The
energy boundaries corresponding to channel numbers have changed
during the lifetime of the \rxte\ mission.  The calibration of
channel numbers $\leq6$ is uncertain, and we therefore
ignore all \pca\ channels $\leq6$ ($\sim 3\kev$), which allows a consistent lower
bound to the spectra, extending them to the lowest possible
energies and maintaining calibration.  All PCA data
greater than $25\kev$ were also ignored. The \hexte\ data were fitted between $25$
and $250\kev$.  To investigate the presence or absence of an iron line
and also to see whether any disc emission was present, three types of
models were fitted: Powerlaw ({\scshape power}), Broken Powerlaw
({\scshape bknpower}) and Powerlaw +
Disc ({\scshape power+discbb}).  For each of these, a version
including a Gaussian line fixed  at $6.4\kev$ was also fitted, giving
a total of six models which were fitted to each spectrum.  Galactic
absorption was modelled using the {\scshape wabs} photoelectric
absorption code, with a fixed $N_{\rm H}=0.4\times 10^{22} \pcm2$
(e.g. \citealp{Miller04a}).  Cabanac et al. (in prep.) discuss
possible variations of the $N_{\rm H}$ during an outburst of \gx339.  There is
insufficient spectral coverage at lower energies to allow the $N_{\rm
  H}$ to remain free during the spectral fitting.  To
obtain fluxes outside of the \rxte\ observing band, for the disc for example, dummy responses
were created within {\scshape xspec}.

To account for the galactic ridge emission, we add in the equivalent
of a correction file to each spectrum.  This has the greatest effect
on the low flux observations.  To model the galactic ridge emission,
we used observations of \gx339\ in Obs ID P91105 between MJD 53550 and
53650 (25 observations, 19.1~ks exposure).  We assume that the
emission in these observations is dominated by the galactic ridge,
with almost no contribution from \gx339.  We fit this data using an
absorbed powerlaw with a gaussian emission line.  The best-fit
values ($\Gamma=2.127$, $\mathcal{N}_{\Gamma}=1.153\times
10^{-4}\pkevpcmsqps$, line energy$=6.491\kev$, $\sigma=1.729 \times 10^{-3}\kev$ and
$\mathcal{N}_{\rm line}=1.07\times 10^{-6}\pkevpcmsqps$), are added into
each model fit as a fixed set of
components to take account of the galactic ridge emission.

The high energy channels from \hexte\ suffer from a high background
leading to a low number of net counts in the low exposure time
observations.  This results in the high energy flux not being well determined.  Using
{\scshape grppha}, we investigated the effect of binning.  The
channels were binned so that there was a minimum number of counts within each
bin.  To have clearly defined fluxes at all \hexte\ energies we found
that the level of binning had to be very high, and the lower energies
were then severely over-binned.  However, the change in
the value of the parameters, their uncertainties and the reduced $\chi^2$ changed very
little, on the order of 5 per cent or less.  We therefore do not
perform any binning during the analysis of the data\footnote{In the
  combined spectra in Fig. \ref{fig:spectra} we bin the \hexte\ points to improve
  the clarity of the spectra only.}.

We encountered some difficulty in fitting discs reliably.  The
response of \rxte\ is only reliable down to $\sim 3\kev$, whereas the
disc we are trying to measure has a temperature of around $1\kev$.  We
set the minimum
disc temperature to $k_{\rm B}T=0.1\kev$ in {\scshape xspec} to
prevent discs from being fit at very low temperatures.  It is likely
that in cases where this occurred, the disc was being fit to take into
account of any curvature in the powerlaw slope rather than to a true
disc component.  Before selecting the best fitting
model we penalised disc models which
had $k_{\rm B}T<0.4\kev$.  In doing this we note that we are likely to
have excluded a few disc fits which are going to be reliable.

\begin{figure}
\centering
\includegraphics[width=1.0\columnwidth]{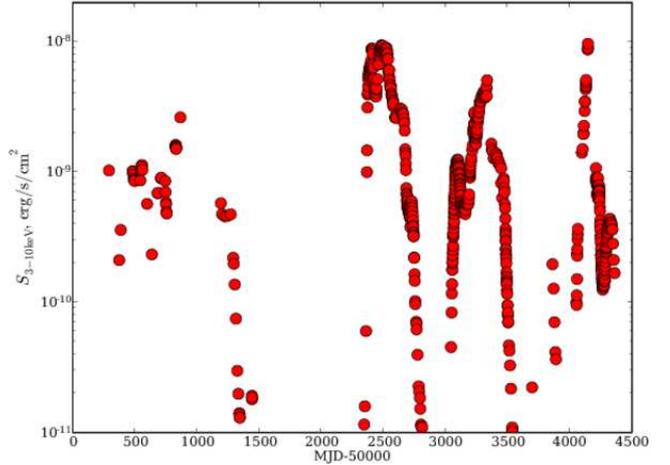}
\caption{\label{fig:lc} The \rxte-\pca\ lightcurve of \gx339 over 11
  years.   The flux is the absorbed flux between $3-10\kev$.}
\end{figure}

We initially select the model with the lowest reduced $\chi^2$.
However, if this best fitting model is a broken powerlaw or a disc+powerlaw model, we
test whether using this more complex continuum model is a significant
enough an improvement over a simple unbroken powerlaw, by performing an $F$-test.
If the $F$-statistic probability $\mathcal{P}<0.001$ then we select the more
complex continuum model.  We investigated the number of discs deemed
``significant'' using different probability cut-offs.  The number of
discs was for all intents and purposes constant regardless of whether
the probability cut-off was $0.05$ (234 disc detections) or $0.00001$\footnote{The
latter probability cut-off was chosen to be
an extreme value.} (214 disc detections).  We can therefore be certain that
almost all disc values used in the rest of the analysis are from
significant discs, what ever cut-off we use

We note that we may be missing non-dominant disc components in,
for example, the hard state.  Even though we fit all observations with
a powerlaw + disc model, the lack of sensitivity of the
\rxte\ \pca\ below $\sim 3\kev$ makes detecting non-dominant discs
difficult.  Therefore even if no significant disc is detected in our
analysis, there may be discs present at a very low level in the hard
and intermediate states -- absence of evidence does not mean evidence
of absence.  See Section \ref{sec:majorOBs} for some higher
signal-to-noise spectra and further discussion of disc components when
\gx339\ is not in the soft state.

To determine whether the gaussian component at $6.4\kev$ is likely to
be an accurate representation of a true iron line we followed the scheme outlined below.
The line was
deemed to be not well constrained when the width, $\sigma$, was larger than $1\kev$.
In these cases, although the best fitting model includes a line
component, only the base continuum model was used for further
analysis.  Subsequently, an $F$-test was used to determine
if the addition of the line was statistically significant ($\mathcal{P}<0.001$), and the
appropriate model then chosen for further analysis. 

We note that there have been recent doubts about the applicability and
accuracy of using an $F$-test to determine the presence of a line and
more complex continuum models.  \citet{Protassov02} state that
the $F$-test does not adhere to the $F$ distribution, even
asymptotically.  As we are attempting to determine the presence of
disc and line components, this is of importance to the work presented
here.  However, we note that the alternatives to an $F$-test are not
easily implemented within the current fitting proceedure.  Our
solution is therefore as follows.  A usual cut-off for a spectral
feature to be significant is $\mathcal{P}<0.001$ (99.9 per cent
confidence).  We investigated the number of lines deemed
``significant'' using different $F$-probability cut-offs. There is a
larger variation in the numbers of significant lines 
detected than in the equivalent investigation for the discs.  Using a
cut-off of $0.05$ there are 533 observations with
lines deemed significant, but using $0.00001$ there are only 322.  It
is therefore not clear which $F$ probability to use as the cut-off for
a significant line detection.  Therefore we include the estimation of
the line significance using the normalisation of the gaussian component.  

This method uses the normalisation of the line component from
{\scshape xspec} and its uncertainty.  We use the uncertainty as an
estimate of the normalisation sigma, and hence treat line components
as significant if their normalisation differs from zero by at least three sigma.  Both
of these two statistical tests have to be satisfied for the line to be
used in any further analysis.  Using an $F$-test probability of
$\mathcal{P}<0.001$ and significance of the line normalisation, results
in 400 line detections.  We do note that some may be erroneous
``detections'' but the majority will be true features in the spectrum,
see also Section \ref{sec:majorOBs} and Fig. \ref{fig:spectra} for
higher signal-to-noise spectra.

Any observation
with a $3-10 \kev$ flux from the best fitting model of less than $1\times 10^{-11} \ergps$ was
also discarded from further analysis, as were ones where the flux was
not well determined (the error on the $3-10 \kev$ flux was larger than
the flux itself).  This resulted in a final list
of 628 observations, corresponding to $2.261\Ms$, with well fitted
spectra and high enough fluxes and counts.
We extract a range of parameters and fluxes for different bands and
model components which are presented and analysed further the
following sections.  

The full \pca\ lightcurve of \gx339 is shown in
Fig. \ref{fig:lc}.  There are four outbursts covered in this period:
Outburst 1 from MJD-50000 = 0-1500, Outburst 2 from 2340-2800, Outburst 3 from
3040-3510 and Outburst 4 from 4000-.  The MJD shown is for the
midpoint of each individual
observation.  The best sampled
outbursts are the middle two, and these are the ones on which some of
the remaining analysis concentrates on as these have the most
observations.

\begin{figure}
\centering
\includegraphics[width=1.0\columnwidth]{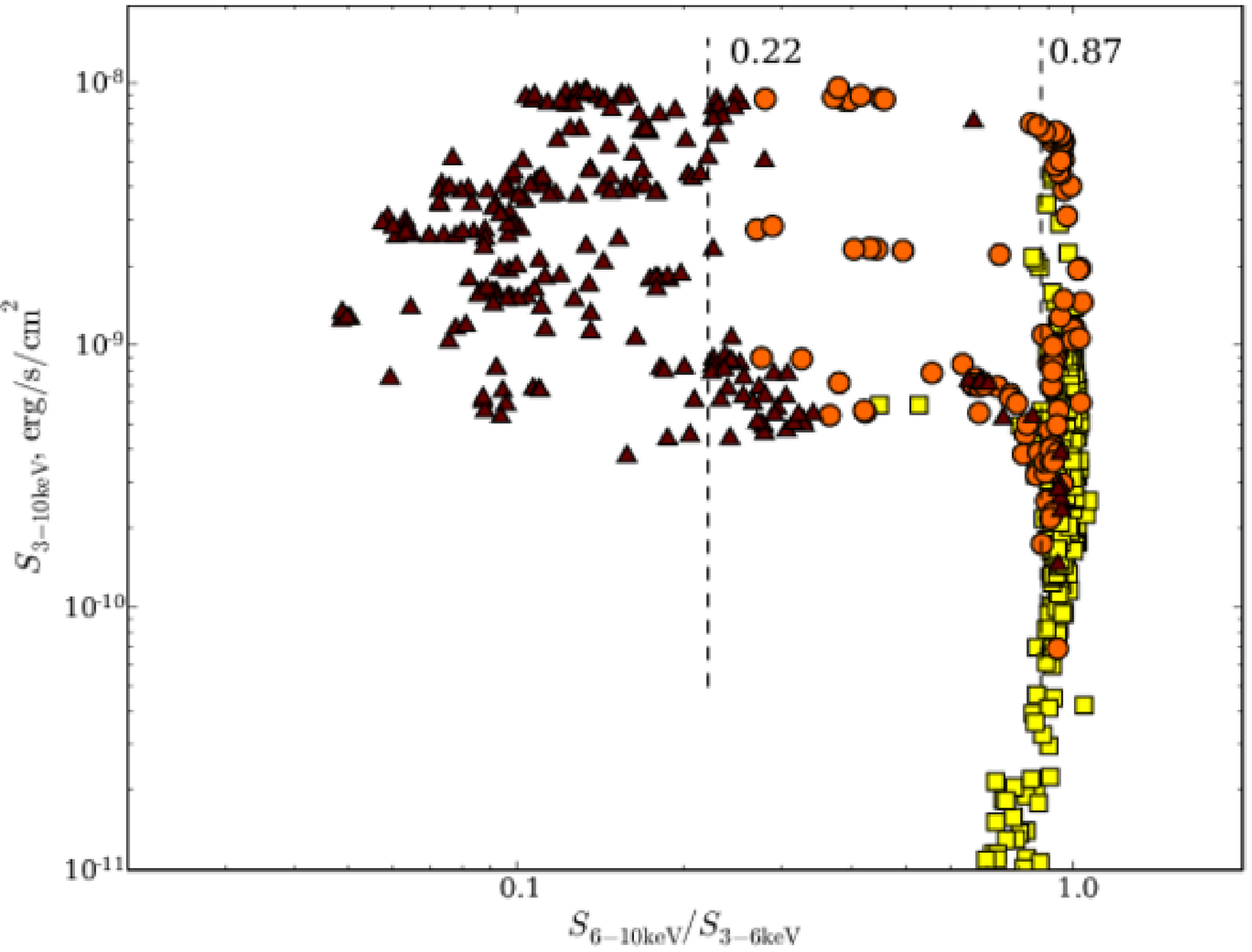}
\includegraphics[width=0.325\columnwidth]{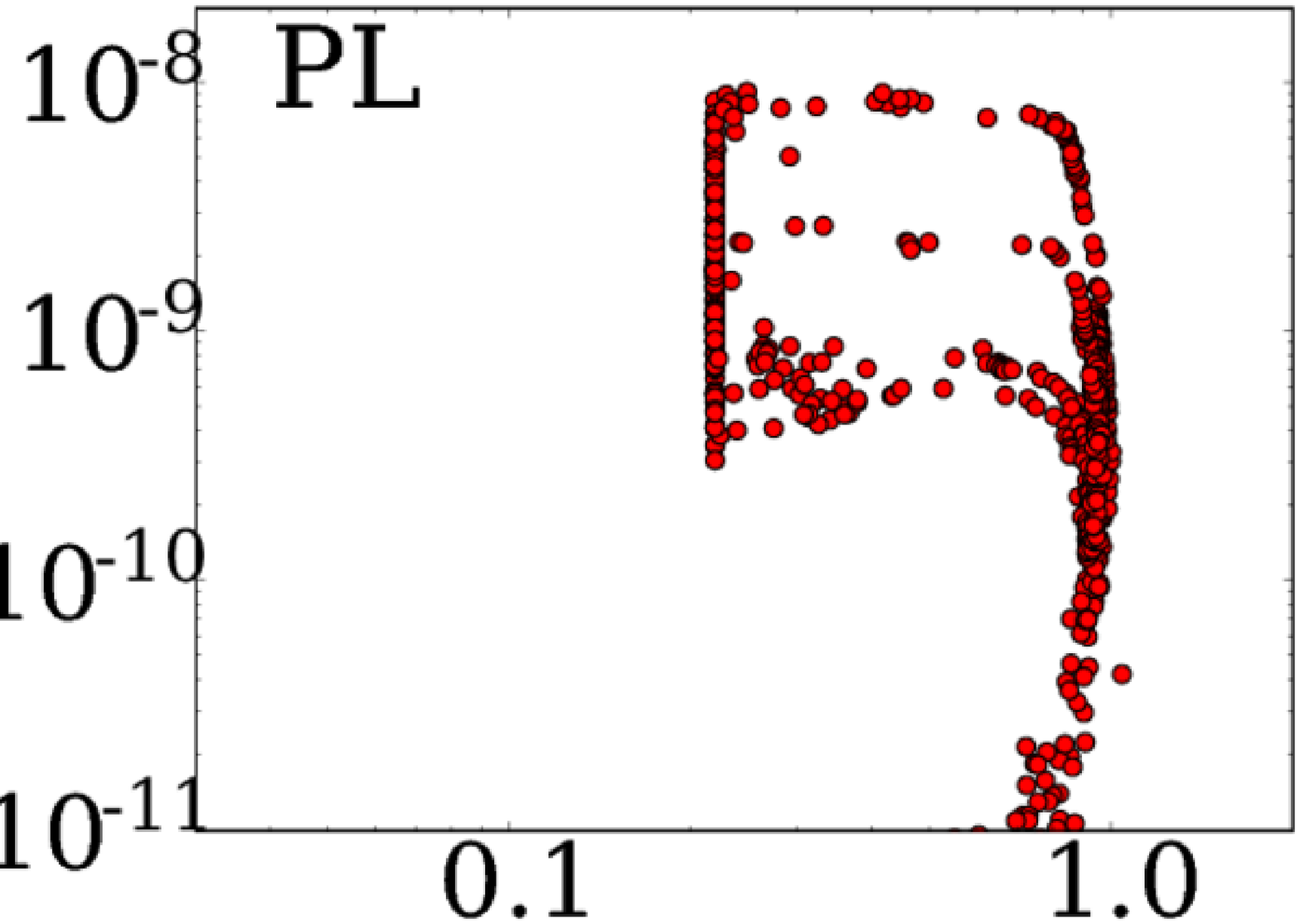}
\includegraphics[width=0.325\columnwidth]{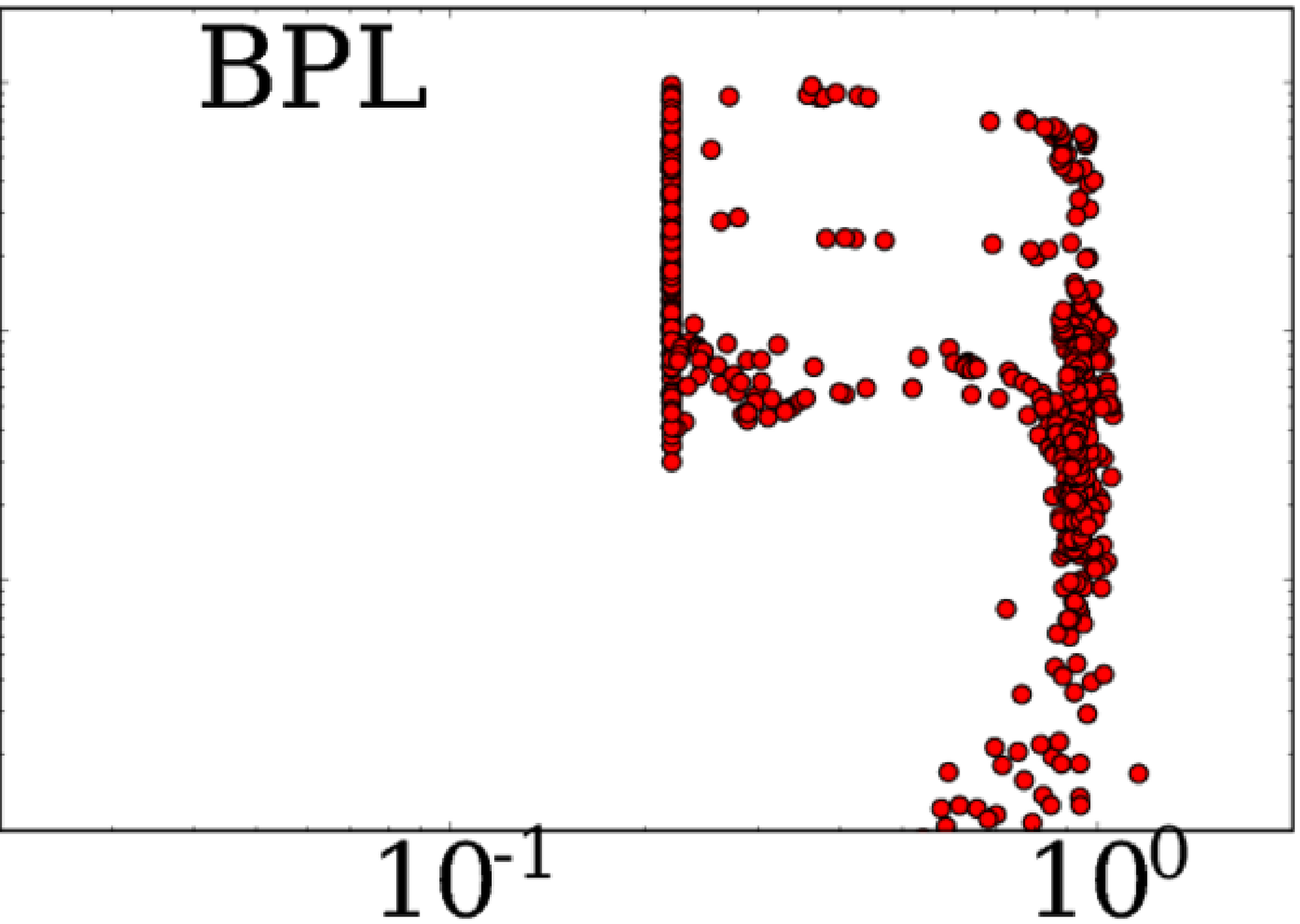}
\includegraphics[width=0.325\columnwidth]{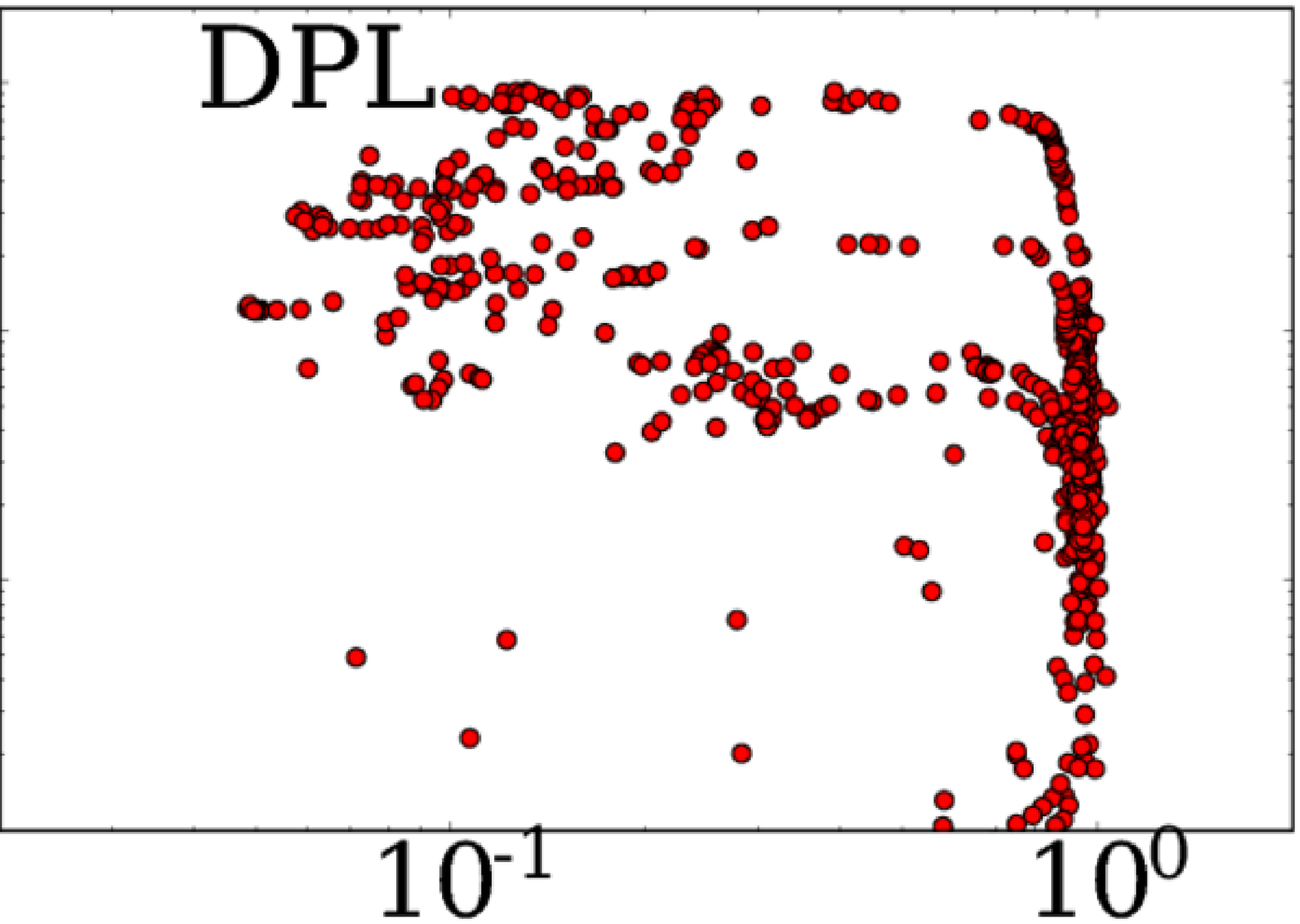}
\caption{\label{fig:Turtle-model} {\scshape top:} The hardness-intensity diagram
  from all the observations.  Powerlaw models in squares, Broken Powerlaw
  models in circles, Powerlaw + Disc models in triangles -- i.e. all
  triangular points require discs.  X-ray colours of 0.225, 0.41 and
  0.85 define the soft--soft-intermediate, soft-intermediate--hard
  intermediate and hard-intermediate--hard state transitions.{\scshape
    bottom:} The HIDs using a single continuum model only powerlaw (PL),
  broken powerlaw (BPL) and disc + powerlaw (DPL), left to right.  The axes are
  the same scale as the main figure.}
\end{figure}

Following the analysis in \eg \citet{Homan01, Belloni04, Fender04}, we plot the X-ray observations
in a Hardness-Intensity Diagram (HID) to show the state changes of
\gx339\ during its outbursts.  We extract the $3-6$, $6-10$ and
$3-10\kev$ fluxes from the spectra.
The X-ray colour was calculated from $S_{6-10\kev}/S_{3-6\kev}$ and plotted
against $S_{3-10\kev}$ in Fig. \ref{fig:Turtle-model}.  

In Fig. \ref{fig:Turtle-model} we show the X-ray colours of \gx339\ where the state transitions between the
high-soft and soft-intermediate states, as well as the hard-intermediate and low-hard states
occur.  The transitions have been determined by changes in the timing
properties of the source as calculated in \citet{Belloni06b} using
data corresponding to our Outburst 3 (see Section \ref{sec:majorOBs}).  Both
transitions towards and away from the soft state occurred between
observations such that a single X-ray colour could be used to
delineate the state changes. 

\citet{Belloni05} study our Outburst 2 and the transitions
from the hard state and into the soft state are very similar to those
in Outburst 3 and so match those we use in this work. However, the
transition between the hard- and soft-intermediate
states occurs at a different X-ray colour.  Also, there is no observation in
soft-intermediate state on the return to the low-hard state in Outburst
2.  The X-ray colour for the transition from the hard-intermediate to
the soft-intermediate state in Outburst 3 is
$S_{6-10\kev}/S_{3-6\kev}<0.37$.  

There is no clear reason why the transition should occur at the same
X-ray colour for any two outbursts.  The X-ray colour is determined by the
temperature and photon index, and also by the relative strength of the
two components.  The transitions towards the soft state are very
different between the two outbursts, and so we should not expect the same X-ray
colour to work for all three transitions.  As a single X-ray colour
could be determined for the transition between the two intermediate
states in Outburst 3 we show this on the appropriate figures, but do
not show an equivalent no figures for Outburst 2. 

Initial investigations into
the HID for the most recent outburst (Outburst 4) whose transition
into the soft state occurred at a similar flux to Outburst 2, shows
transitions between states calculated from X-ray timing, at similar
X-ray colours to Outburst 2 (Del Santo et al., in prep.).

From now on we concentrate mainly on the
soft and the hard states as these are the best sampled (e.g. \citealp{Belloni06}).  For the
remainder of this work we define the soft state when the X-ray 
colour, $S_{6-10}/S_{3-6}<0.22$, and the hard state when
$S_{6-10}/S_{3-6}>0.87$ unless stated otherwise (see
Fig. \ref{fig:Turtle-model}), and the intermediate state is not split in
any discussions.

We also show the HIDs obtained for \gx339\ using each of the thee continuum
models individually in Fig. \ref{fig:Turtle-model}.  The basic shape
of the HID is recovered in all three cases, with the most similar
being the one for the disc + powerlaw.  There is a clear ``pile-up''
of observations with an X-ray colour around 0.2, which is close to our
adopted transition into the soft state.  This is mainly the
result of our restriction on the slope 
of the powerlaw to have $\Gamma<4.0$.  This limit was set to aid the
fitting of a disc.  As the \pca\ is only sensitive down to $\sim
3\kev$ the curvature of the disc component can be difficult to fit
when the disc is not dominating the spectrum.  Therefore a fit might
result in an extremely steep powerlaw rather than a disc.  Also, if
there are insufficient \hexte\ counts then the powerlaw slope cannot
be well determined from the hard X-rays, which would allow an
nonphysical steep powerlaw fit with the disc in the soft state.  Powerlaws of $\Gamma>4.0$ are
thought to be unphysical.  The best fitting models just harder of this
X-ray colour are ones with a disc component, and observations in the
intermediate states which have broken powerlaw models as their best
fit have well determined slopes not pegged at $4.0$.  Therefore we do not believe
that this restriction is influencing our results (see Section
\ref{sec:majorOBs} for disc models in the intermediate state).  There is little difference between the
powerlaw and broken powerlaw models as the break usually occurs at
around $10\kev$ and so has little effect on the X-ray colour adopted here.

\subsection{Comptonisation Models}\label{sec:model:comptt}

We note that the powerlaw models we use here have limited physical
meaning.  We therefore performed a quick investigation into
fitting the {\scshape comptt} model combined
with a {\scshape diskbb} and gaussian line component to the data.  Using this
combination of models would result in a disc component being fitted
whatever the state of the source and so would be a method for
investigating the disc parameters outside of the soft state.
Selecting the four hard state spectra with the greatest
number of \pca\ counts resulted in a good fit
with well defined values for the $\tau$ and $kT$ parameters, which
were reasonably similar between different observations ($\sim 2.5$ and
$\sim 25\kev$ respectively).

However, turning to the soft state, again using four spectra with the largest number
of \pca\ counts, we found that the lack of high energy counts from
\hexte\ meant that we could not determine both $\tau$ and $kT$ for
every observation.  Fixing $kT=50\kev$, the average value from those
observations where both parameters could be estimated, allowed a value of $\tau\sim
0.5$ to be estimated.

Using a single analysis scheme and script for all the observations is
vital to remove any systematic differences between the analysis of
different observations.  As the values of both $\tau$ and $kT$ varied
between the soft and hard states, and the fact that we were not able
to determine independent values for both parameters in the observations of the
soft state with the highest \pca\ counts, we did not investigate any further.  The observations
investigated with {\scshape comptt} had between 0.7 and 11 million
\pca\ and 60 and 350 thousand
\hexte\ background subtracted counts, whereas our cut-off for the number of counts in an
observation is 1000 \pca\ and 2000 \hexte\ background subtracted counts.  We do not believe
that many well defined parameters will be extracted from fitting
{\scshape comptt} models to these shorter, low counts observations.

\subsection{Best Fitting Models} \label{sec:model:best}

The description of the method used to select the best fitting model is
outlined in Section \ref{sec:model}.  Fig. \ref{fig:Turtle-model}
shows which was the best fitting model through the HID.  Models
without the disc component are the best fitting for the majority of the 
hard  and intermediate state observations.  The soft state observations are exclusively best fit by
models which include a disc
component.

\begin{figure*}
\centering
\includegraphics[width=0.325\textwidth]{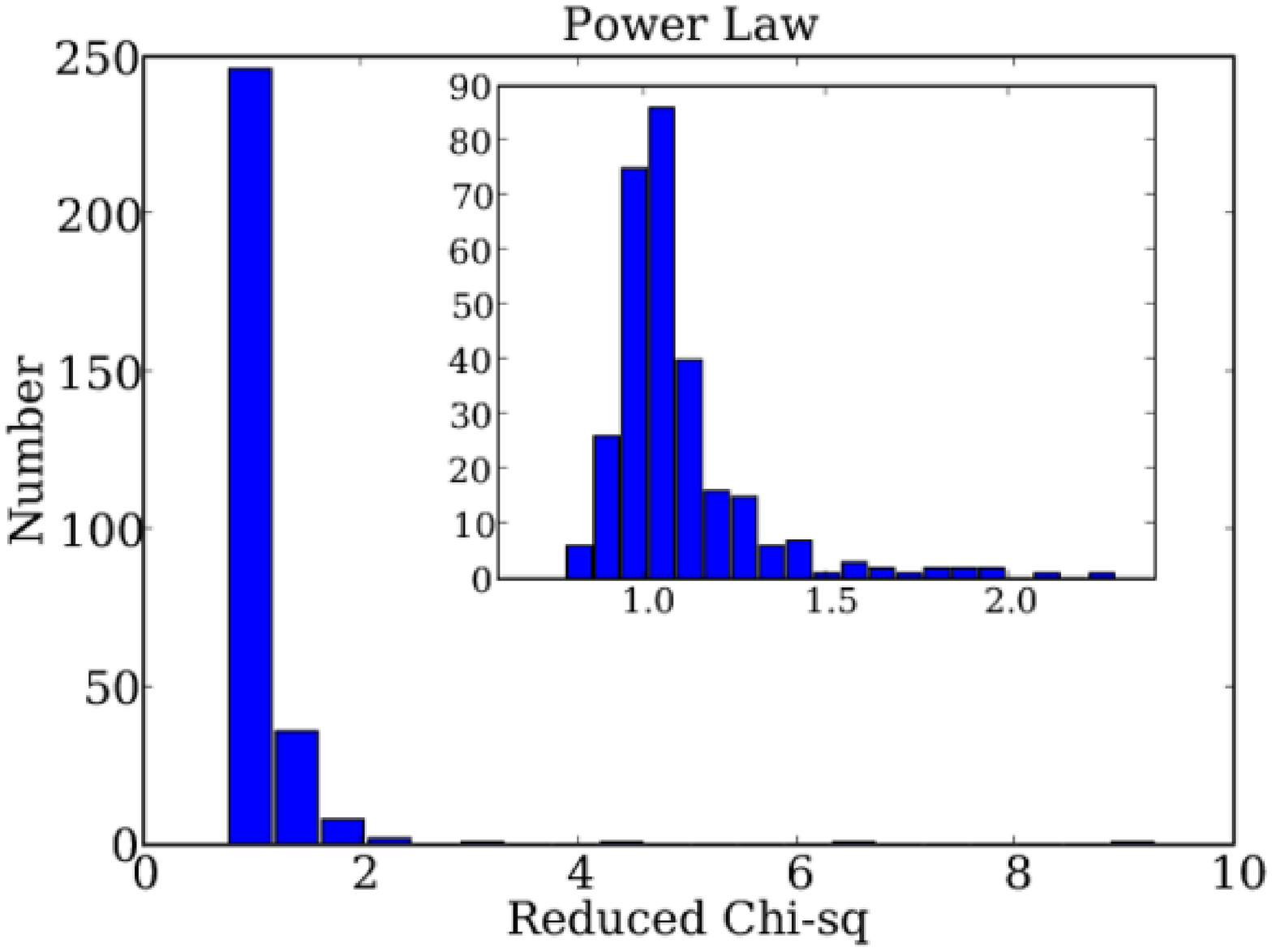}
\includegraphics[width=0.325\textwidth]{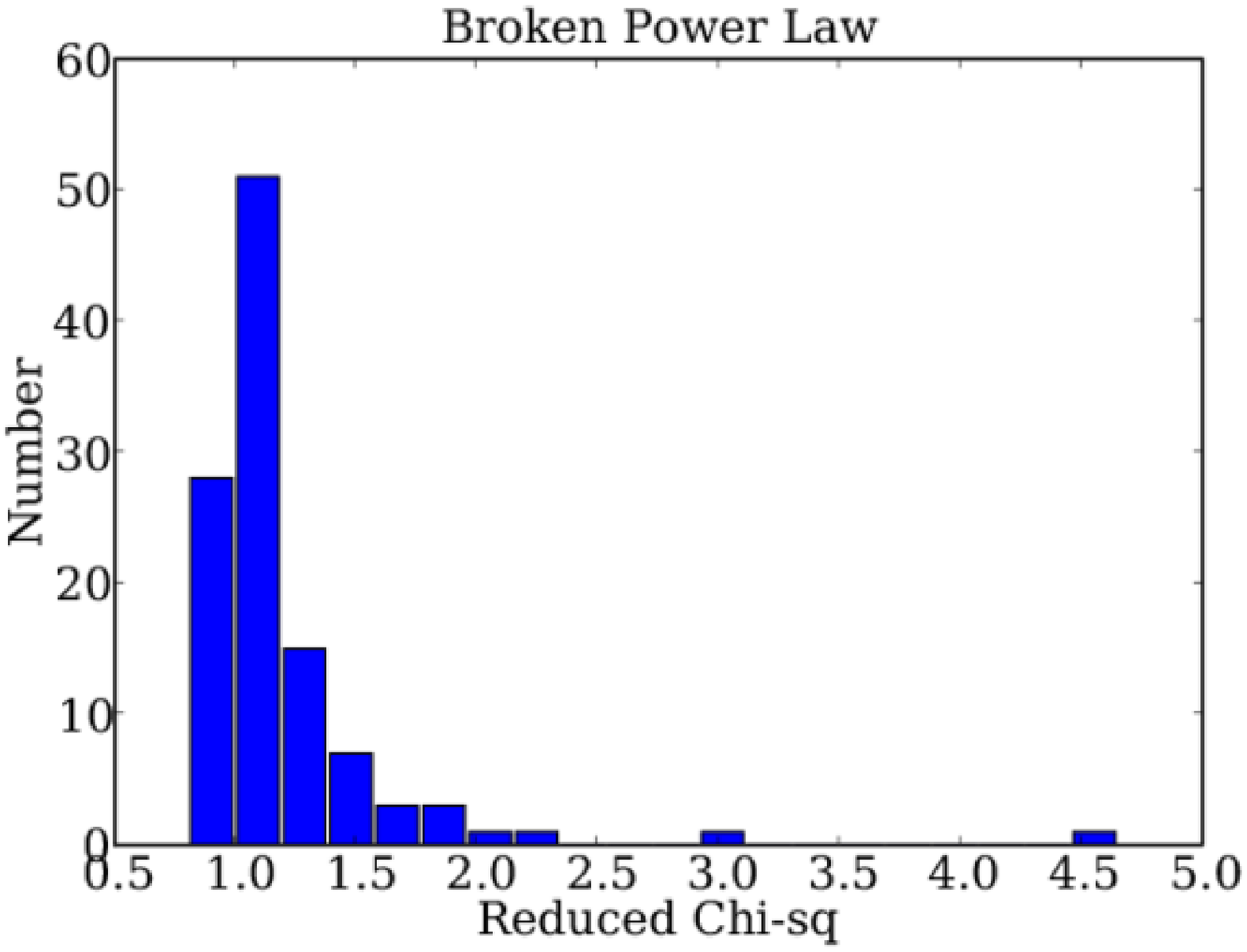}
\includegraphics[width=0.325\textwidth]{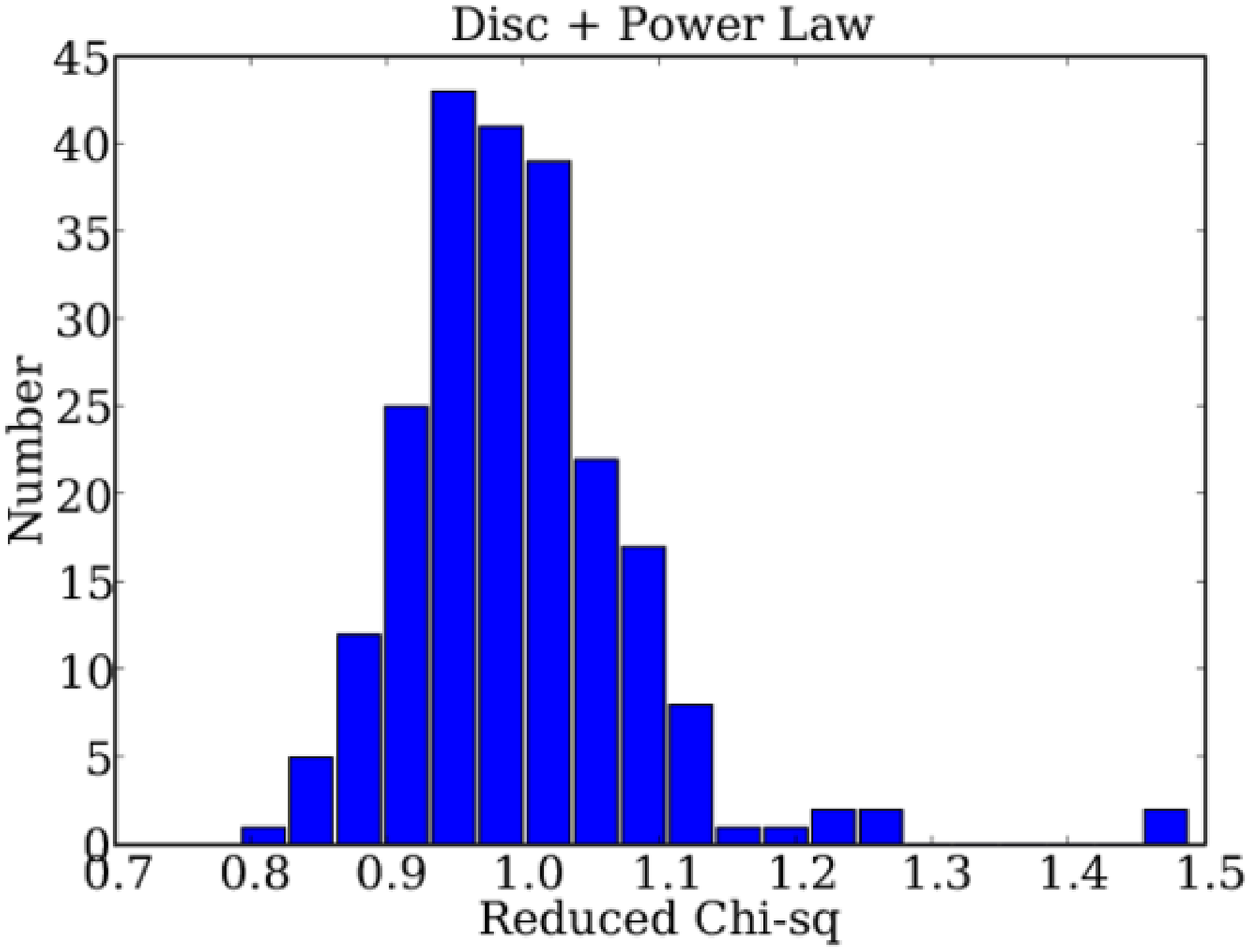}
\caption{\label{fig:chisq_dist} The distributions of the reduced
  $\chi^2$ for the three different continuum models selected as best
  fitting using the method presented in the text.  The inset in the
powerlaw model is an enlargement of the peak of the distribution.}
\end{figure*}

However, some of the observations in the hard state have
best fitting models which include a disc component.  It is expected
that the disc emission should be very low compared to the power-law emission
when \gx339\ is in the hard state.  As a result it is surprising
that for some hard state observations the best fit model includes a disc
component.  It is likely that, although these models have the lowest
$\chi^2$ from the ones fitted, their disc parameters are not
accurate.  These disc fluxes are one to two orders of magnitude below
those from observations in the soft state.  The errors on the disc temperature from these fits are
larger than those obtained from observations in the soft state.  A
combination of a broken power-law and the galactic absorption may have
combined to emulate the emission from a disc sufficiently that a disc
model is the best fitting.

Other X-ray missions (\eg {\it Swift}, {\it XMM-Newton}) have a
response at low energies such that, if a disc were present in the hard
state, then they would be able to detect it in a suitable length
observation.  In \gx339\, Cabanac et al. (in prep.) detect discs
outside of the soft state, as do \citet{Miller06}.  However, in the
closer source {\it XTE
  J1118+480} discs have been seen in the hard state \citet{Chaty03,
  McClintock01}.  

The distributions of the reduced
$\chi^2$ for the three best fitting continuum models are shown in
Fig. \ref{fig:chisq_dist}.  The range in $\chi^2$ for the three
continuum models is on the whole between $0.8 \to 1.5$ with some
outliers.  Most outliers are in the simple power-law model fits as
this is the default model, but they are few in number.

\begin{figure*}
\centering
\includegraphics[width=0.495\textwidth]{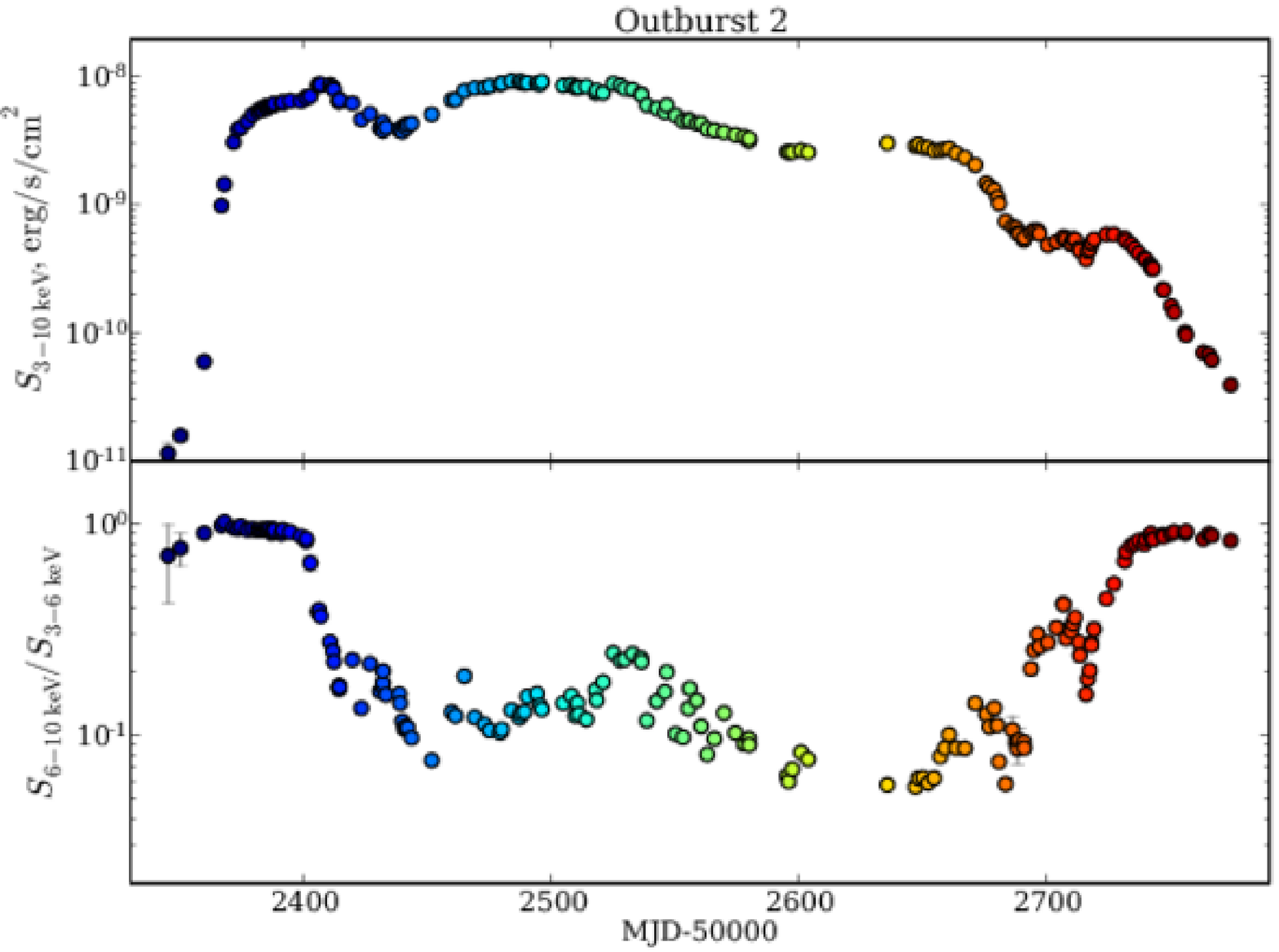}
\includegraphics[width=0.495\textwidth]{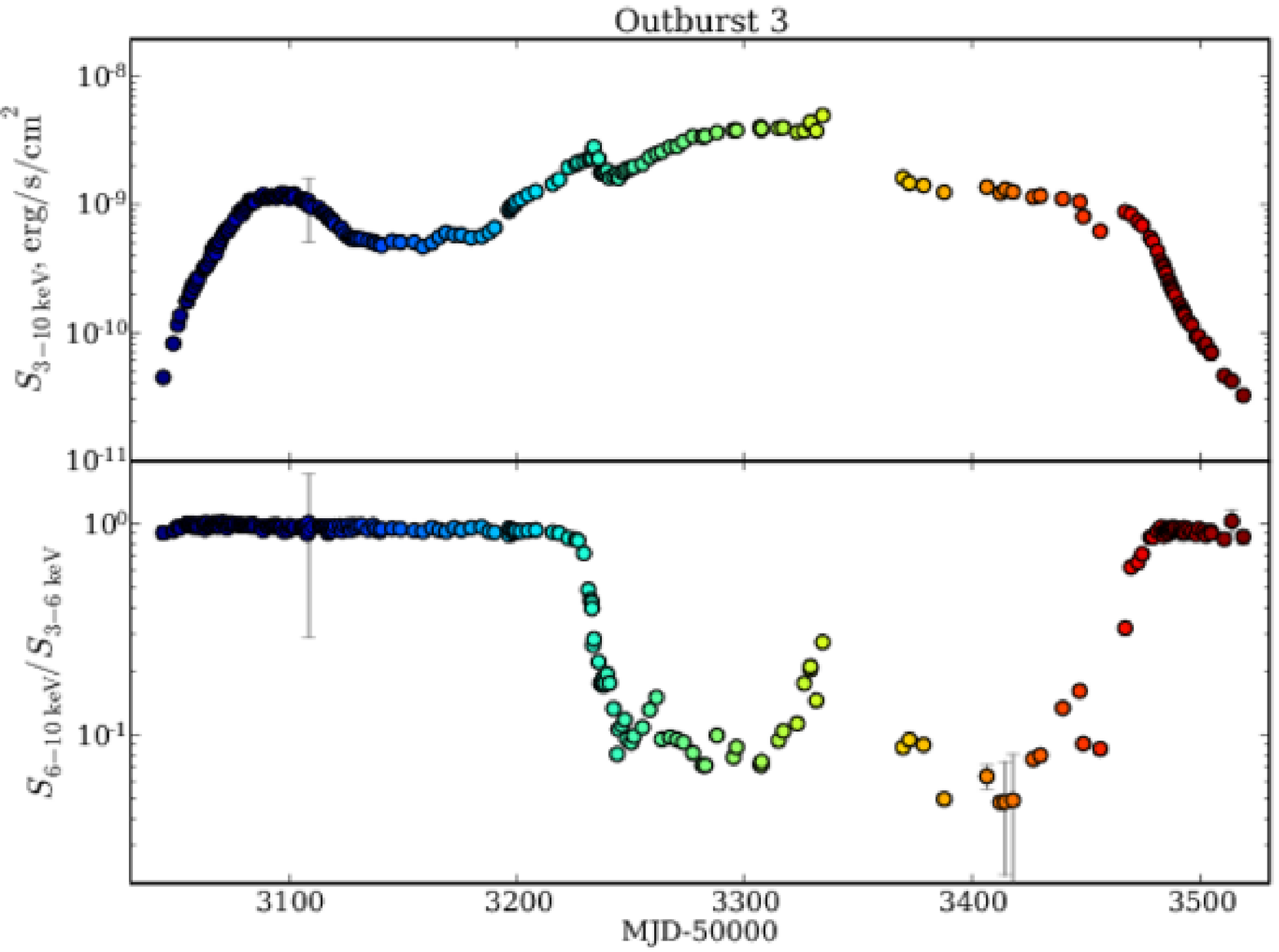}
\includegraphics[width=0.495\textwidth]{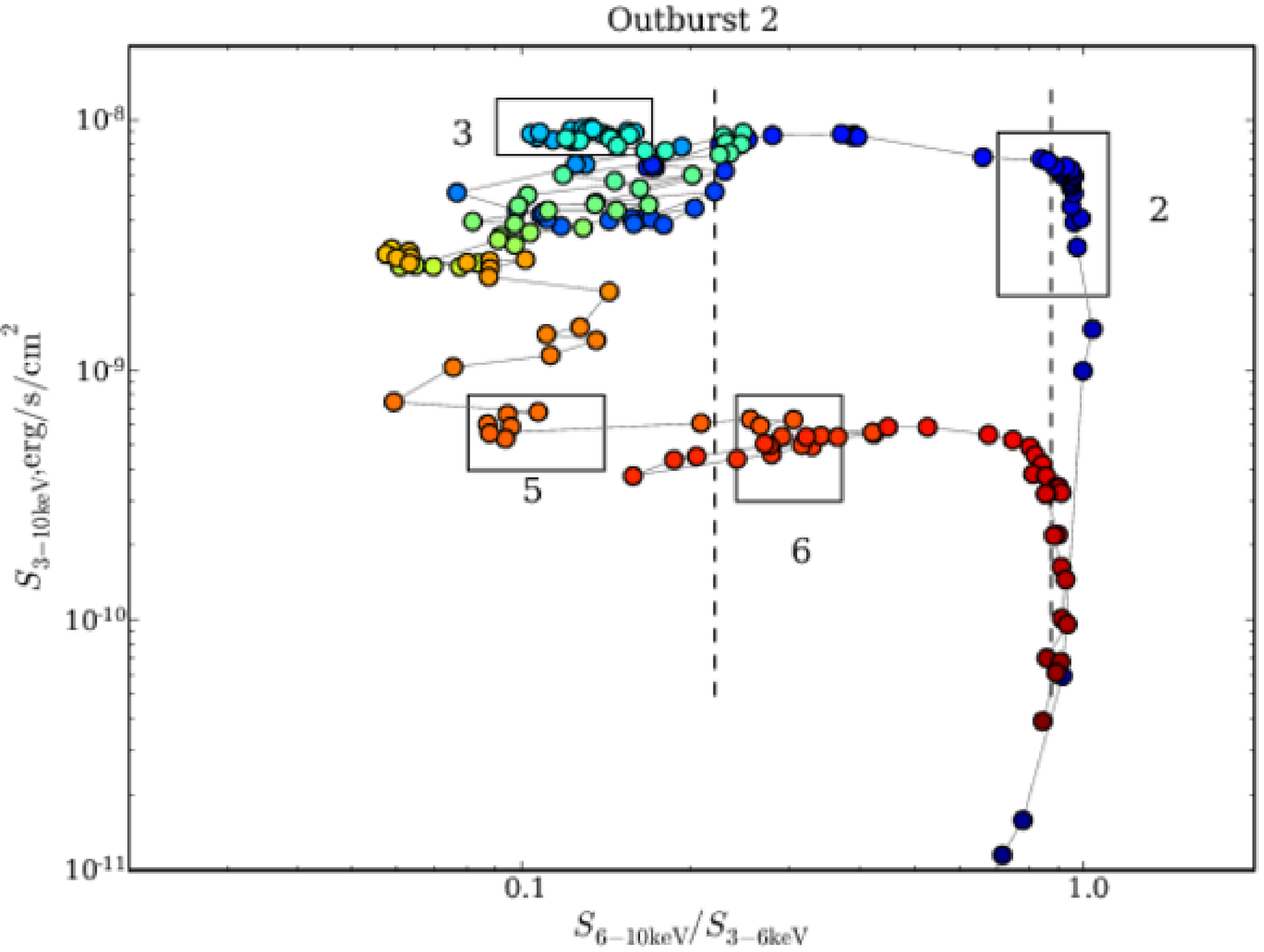}
\includegraphics[width=0.495\textwidth]{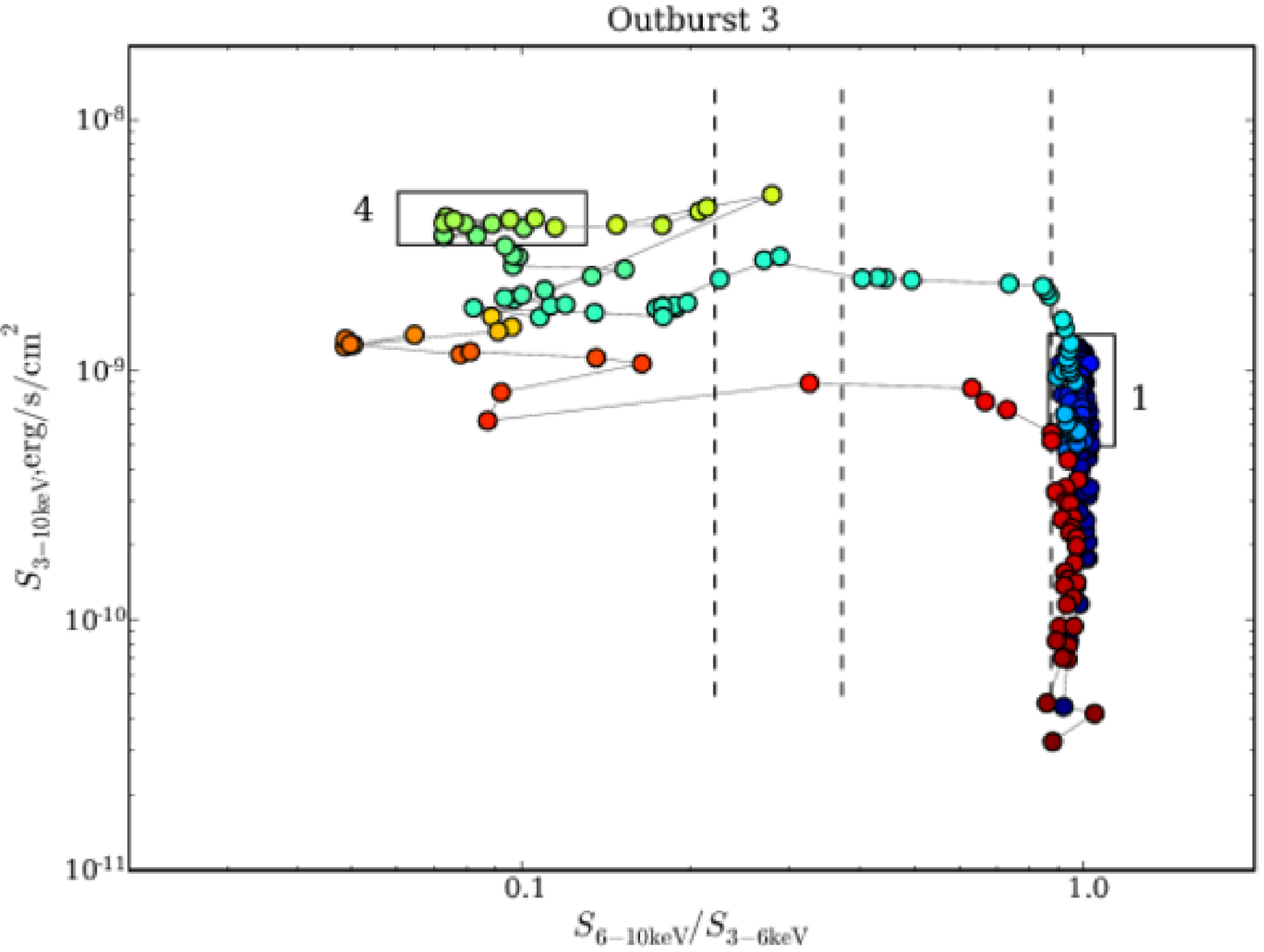}
\caption{\label{fig:Outbursts} The two best covered outbursts of
  GX339-4.  {\scshape top:} Detailed light curves
  and X-ray colour plots; {\scshape bottom:} HIDs of the two main outbursts.  For clarity the errorbars are
  only shown in the flux and X-ray colour light curves rather than on
  the HID.  On the HIDs we show the line tracking \gx339\ from observation to
  observation.  The colour scale is the date
  since the beginning of the outburst, blue is early and red is late.
The boxes in the bottom diagrams show the selected outbursts for the
spectral stacking described in Section \ref{sec:DFLD} and shown in Fig
\ref{fig:spectra}.  The number of the box corresponds to the number of
the spectrum shown in Fig.\ref{fig:spectra}. The dotted vertical lines
show the X-ray colours adopted for the state transitions outlined in
Section \ref{sec:model}.}
\end{figure*}

\begin{figure*}
\includegraphics*[angle=-90, width=0.495\textwidth]{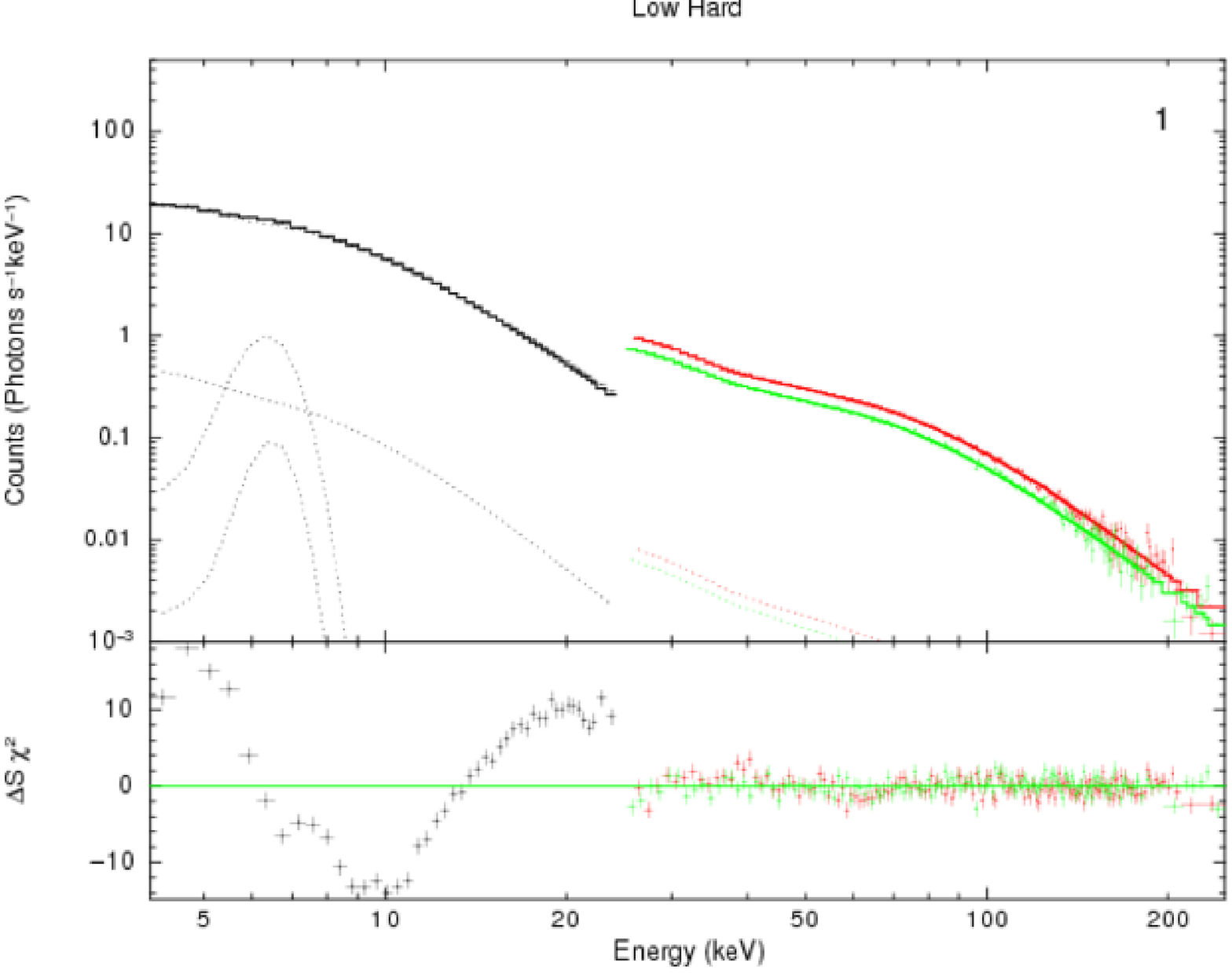}
\includegraphics*[angle=-90, width=0.495\textwidth]{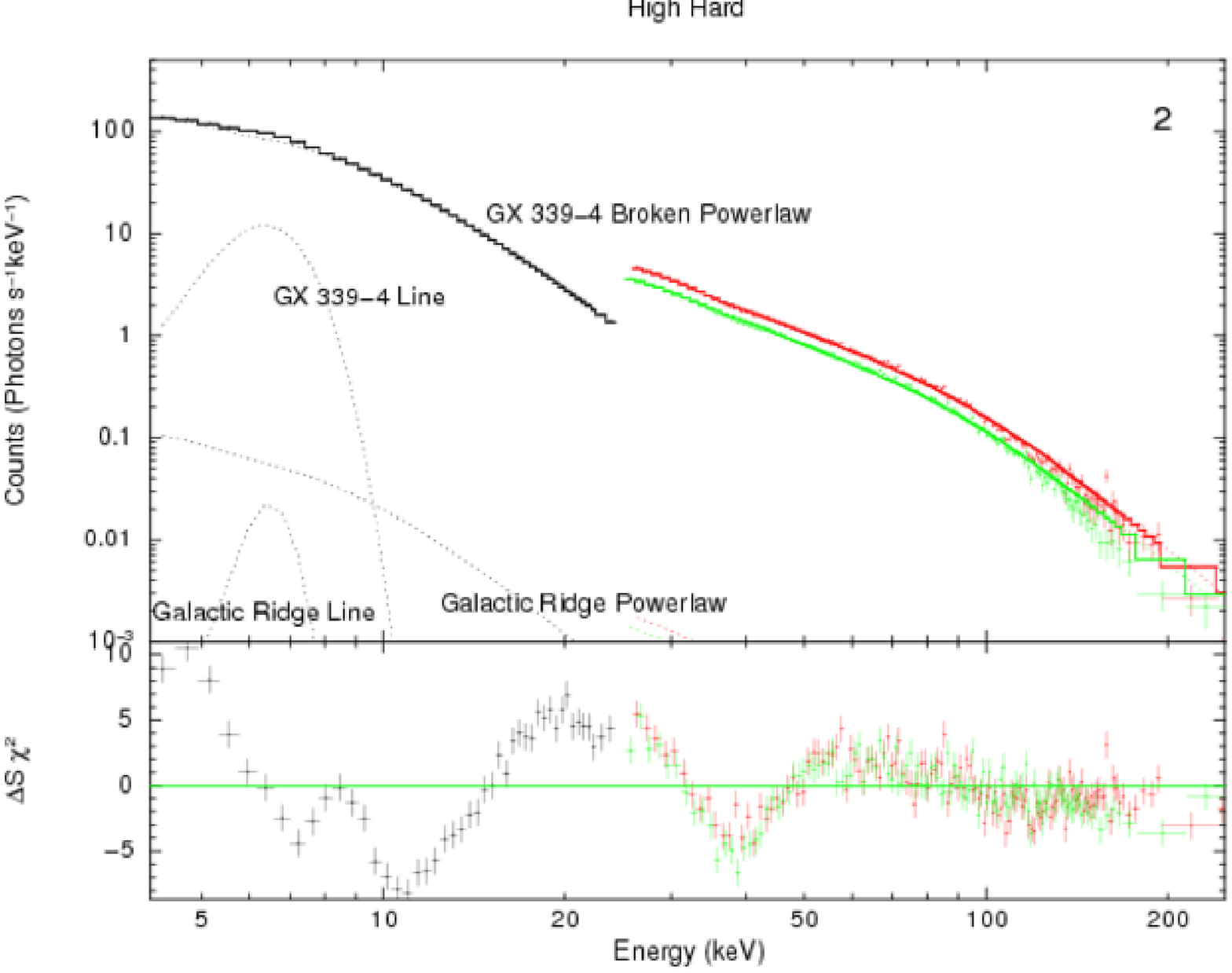}
\includegraphics*[angle=-90, width=0.495\textwidth]{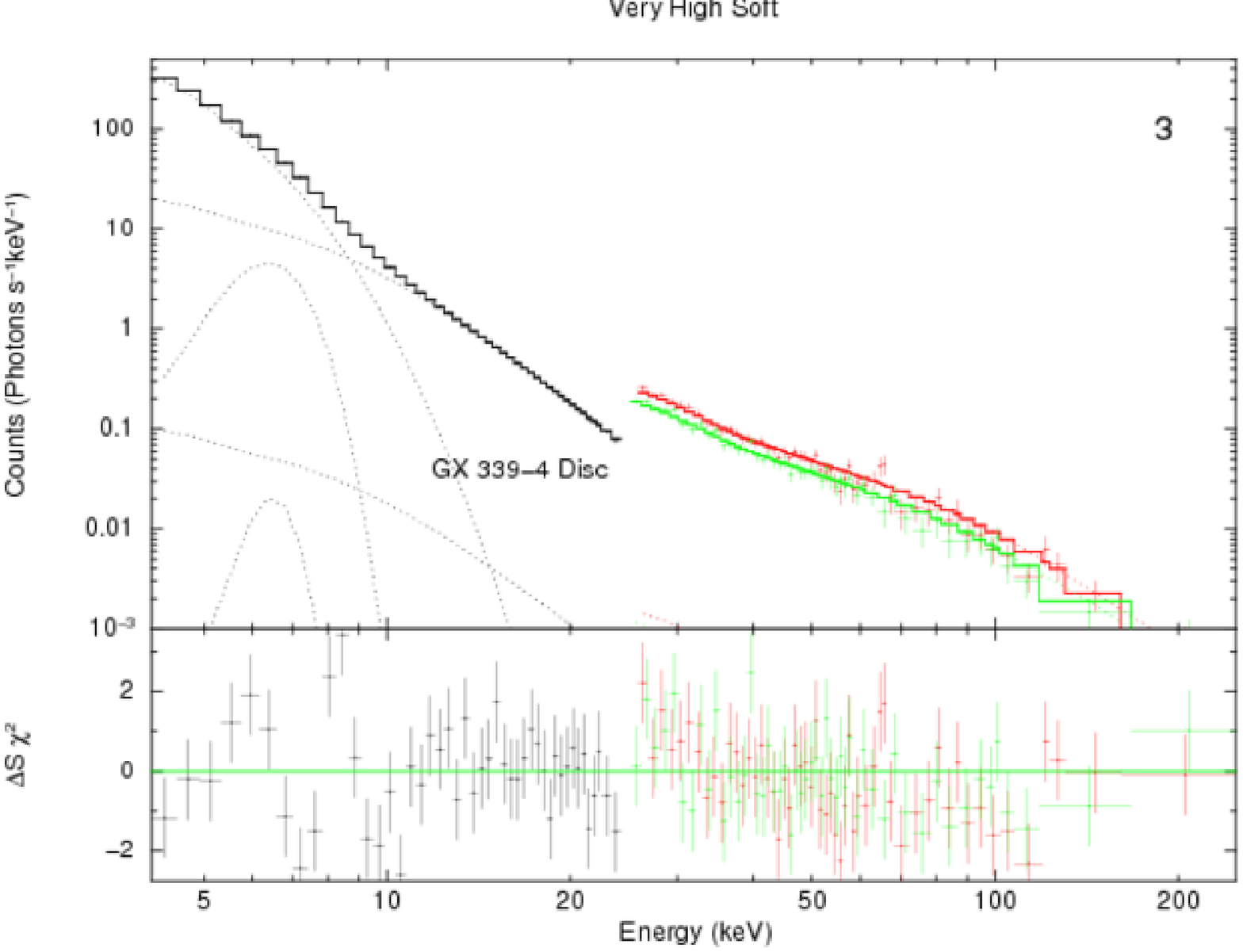}
\includegraphics*[angle=-90, width=0.495\textwidth]{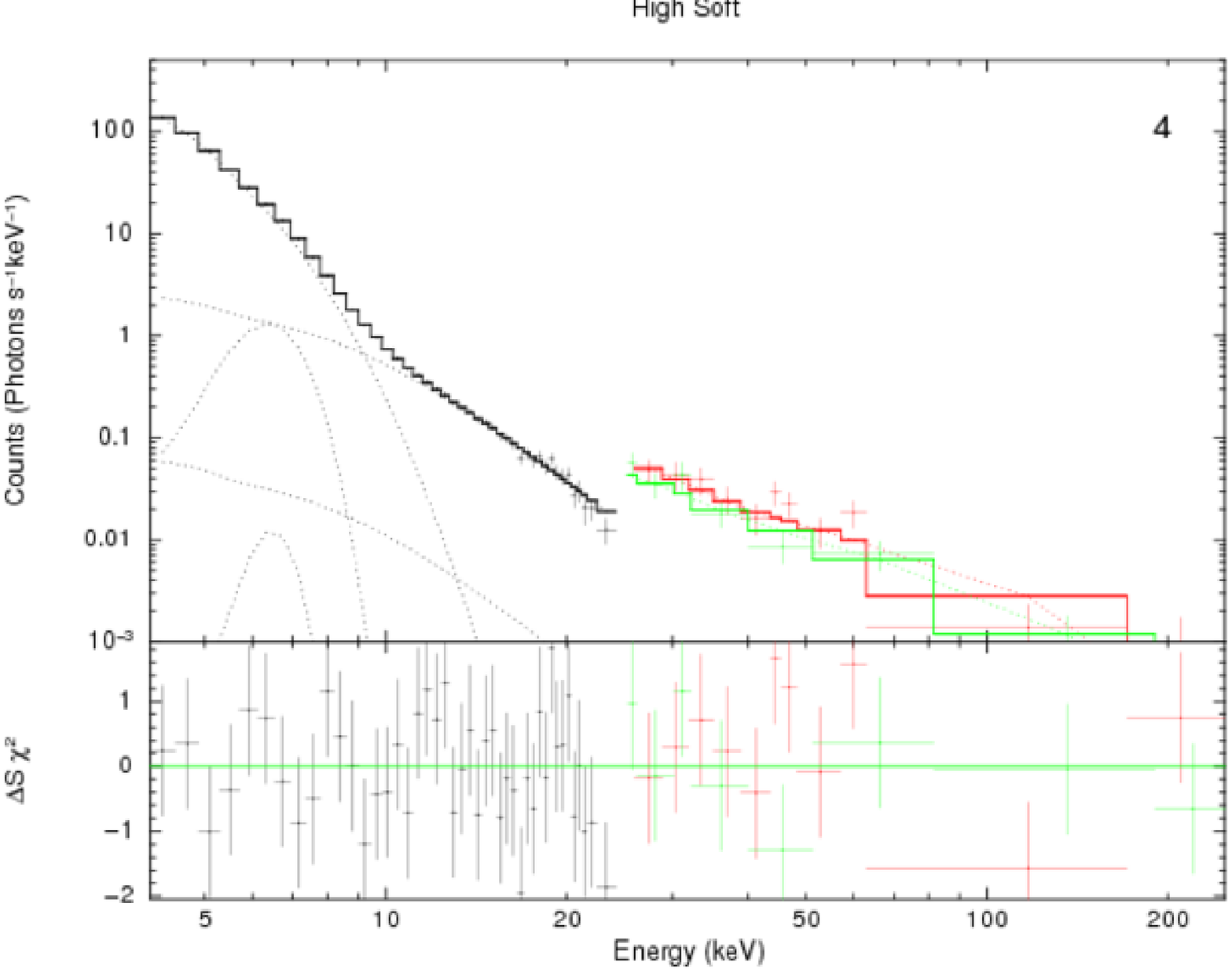}
\includegraphics*[angle=-90, width=0.495\textwidth]{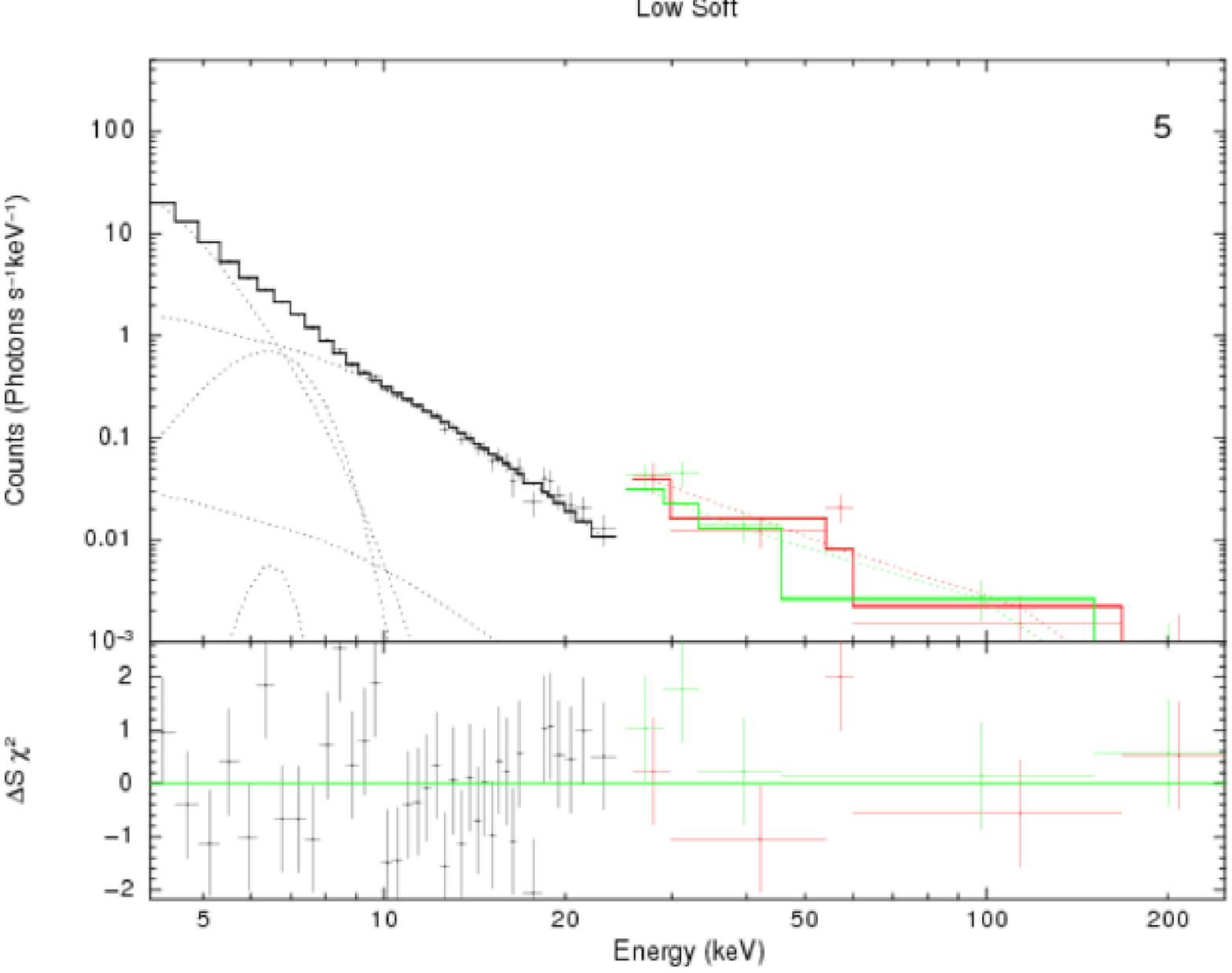}
\includegraphics*[angle=-90, width=0.495\textwidth]{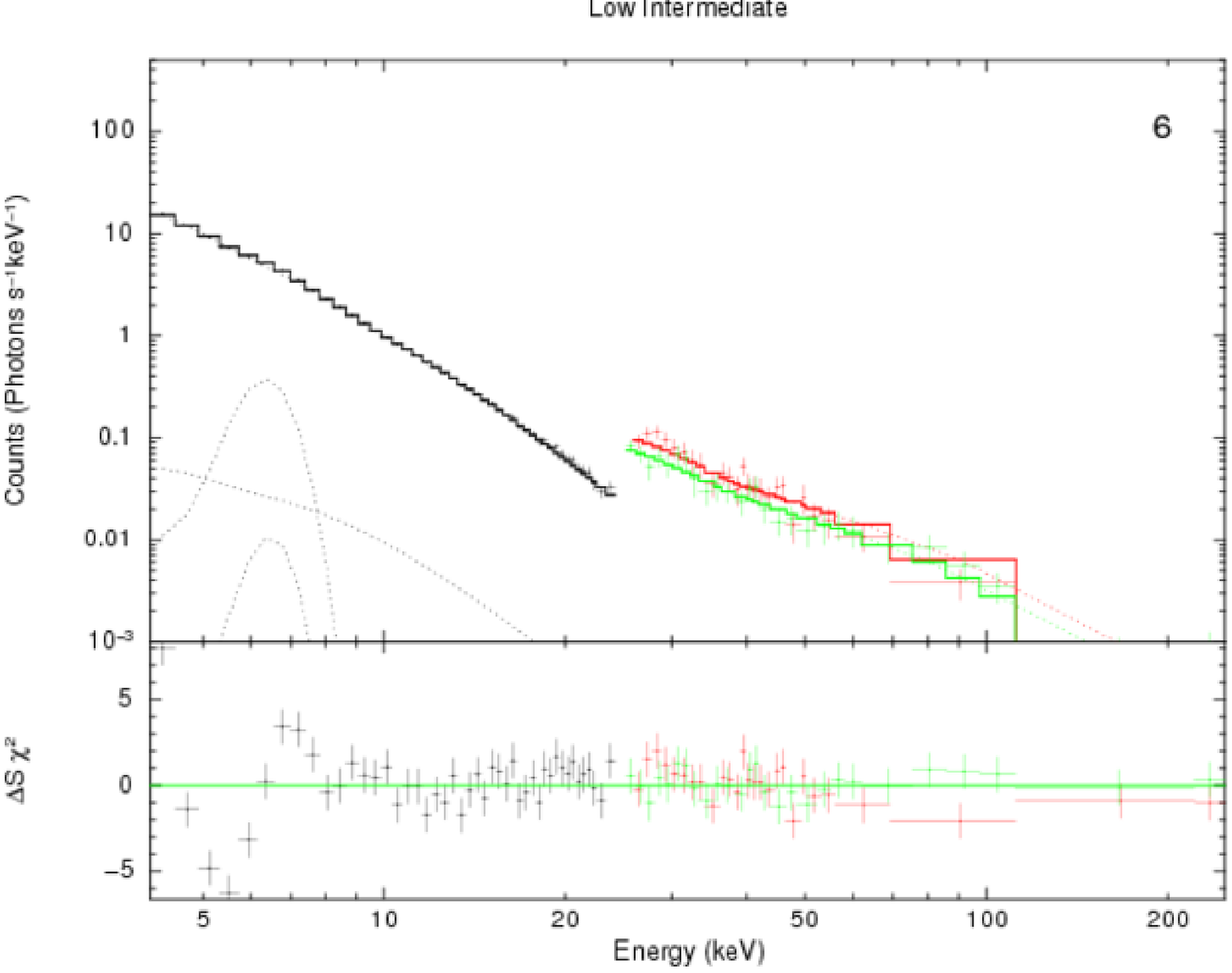}
\caption{\label{fig:spectra}The spectra and model fits to the summed
  spectra.  The fits are for those models which have been used in the
  automated fitting of the individual spectra, except for the Low
  Intermediate state, where the {\scshape discbb+powerlaw} model is
  shown.  The more complicated models detailed in Table \ref{tab:fits}
are not shown here.  The locations in the HID from where the spectra
were taken are shown in Fig. \ref{fig:Outbursts}, and are also
indicated by the numbers in the top right corner of each spectrum.  The data have been
binned for the clarity of the plots alone (no binning was done prior
to fitting) so that the significance of each data point is at least
$3\sigma$.  The effect of the binning is most obvious at the high
energy end of the \hexte\ spectra as well as the shorter exposure time data.}
\end{figure*}

\section{The Two Major Outbursts} \label{sec:majorOBs}

We give a short summary of the major features of the two main
outbursts of \gx339.  Most of the observations in the
rest of this study (424/634) come from these two outbursts.  The detailed lightcurves, along with
the X-ray colours and HIDs are shown in Fig. \ref{fig:Outbursts}.
Outburst 2 reaches its peak flux on the initial rise.  There is a
small decay as \gx339\ evolves into the soft-state, and then the
flux rises again to the same maximum flux. After this there is a gradual decay
in flux. A detailed analysis of this outburst is given in \citet{Belloni05}.

In Outburst 3 the initial peak in the flux is almost an order of
magnitude below the eventual maximum of the outburst.  \gx339\ remains in
the hard state for over 100 days after this first peak, with the flux
falling by a factor of two, then \gx339\
rapidly moves into the soft-state.  The flux continues to increase
once in the soft state, and spectrum hardens fractionally.  After the
break in the light curve, by which time \gx339\ is back in the soft
state, the spectrum hardens before the flux falls and \gx339\ rejoins
the hard branch.  This motion can be traced from the
change in colour of the points in the HIDs shown in
Fig. \ref{fig:Outbursts}.  A detailed analysis of the early part 
of this outburst is given in \citet{Belloni06b}.

It is clear that Outbursts 2 and 3 reach different peak fluxes on the
hard branch (see Fig. \ref{fig:Outbursts}).  Even though there are only a few observations, Outburst
4 also appears to peak at similar fluxes to that of Outburst 2 (see
Fig.  \ref{fig:lc}.  Outburst 1 is
insufficiently well sampled to be able to comment on its peak flux.
On the return to the hard branch, all of the outbursts are
at very similar fluxes, as noted by \citet{Maccarone03} for black hole X-ray
binaries in general.  It is possible that Outburst 3 is a factor 2
higher, but the offset is small compared to offsets in the flux of the
intermediate states for each outburst.

\begin{figure}
\centering
\includegraphics[width=1.0\columnwidth]{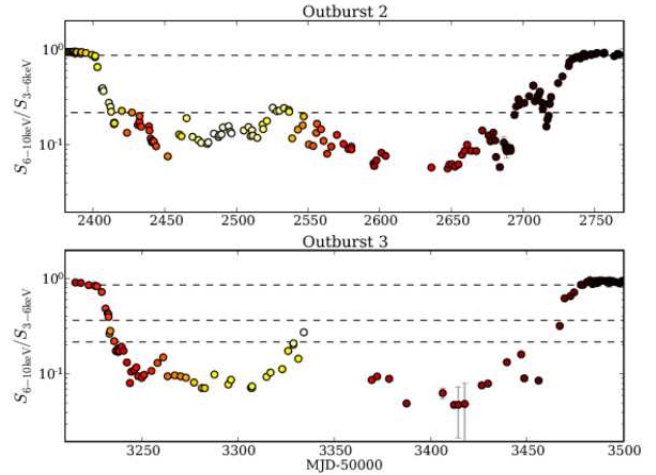}
\caption{\label{fig:XRCL_time} The variation in X-ray colour as the
  outbursts progress for the two major outbursts observed by \rxte.
  The date scale as been adjusted to align the beginning and end of
  the outbursts.  The colour scale is that of the $3-10\kev$ flux of
  \gx339, white for high flux and black for low flux.  The colour
  scales have not been adjusted to match between the two plots.  The
  dotted lines show the X-ray colours adopted for the different state
  transitions outlined in Section \ref{sec:model}}
\end{figure}

The variation of the X-ray colour during Outbursts 2 and 3 is similar
even though they occur two years apart.  The change in X-ray colour with
the outburst is shown again in Fig. \ref{fig:XRCL_time}, rescaled so that the
beginning and end of the transition to and from the soft state are
aligned.  The ``stretching factor'', the relative lengths of the two outbursts, is around $4:3$ for Outburst 2 to
Outburst 3.  The two curves are remarkably similar - the rise in flux and
associated hardening just before halfway through an outburst occurs
in both.  There is also a suggestion that the temporary softening of the spectrum as \gx339\
hardens again, which is clearly seen in Outburst 2, is also present in Outburst 3.  This dip is also seen in the tail end of Outburst 4
when matched to the previous outbursts.

We investigated whether disc components could be present in
observations where individual spectra were unable to indicate any
significant disc component in the intermediate and hard states.  We
selected six areas of similar X-ray colour and flux from the HIDs of
these two well observed outbursts corresponding to low hard, high
hard, very high soft, high
soft, low soft and low intermediate states (see boxes in Fig. \ref{fig:Outbursts}).  These areas were selected
to cover a range in states, but also where there were a concentration
of observations at a similar epoch.  We sum the \pca\ and \hexte\ spectra and
associated backgrounds and refit the same sets of models as before in {\scshape xspec} (see
Fig \ref{fig:spectra} for the spectra and the Appendix for the
numerical results of the fitting).  

The soft states are, unsurprisingly, well fit by a {\scshape discbb + powerlaw}
model.  The best fit model to the intermediate state spectrum is also a
{\scshape discbb + powerlaw} model.  We note that this is different to the best fit
model of some of the individual spectra which make up this summed spectrum
(compare Figs. \ref{fig:Turtle-model} and \ref{fig:Outbursts}).  The
improvement in $\chi^2$ is small, but significant as the degrees of
freedom remain the same.  We therefore conclude that the improved
signal-to-noise of the summed spectrum allows an accurate fitting of
the disc and so this model wins out over a purely broken powerlaw model.
The low intermediate summed spectrum was constructed from observations
of \gx339\ as the outburst decayed, and as such a faint disc would be
expected in this state.  The disc temperature does, however, appear to
rise relative to the soft state rather than fall as would be expected
(see Section \ref{sec:DiscT} for more details). 

Both summed spectra from the hard state are poorly fit by the simple
absorbed broken powerlaw model.  The high flux of \gx339\ during the
observations or the long equivalent exposure of the summed spectra
has resulted in very high signal-to-noise.  Consequently the simple
model is no longer a good fit to the data.  An improved, but by no means
good, fit to the data is achieved by introducing another break in the
powerlaw slope as there is indication for this in the \hexte\ data.
We also added in a {\scshape discbb} model just in case a disc were
present and detectable in these high quality spectra.  There was no
improvement to the fit in the low hard state (which is taken from
observations on the rise of the outburst).  In the high hard state,
however, adding a disc improved the fit (see notes for Table
\ref{tab:fits}).  

The summed spectra do show disc components when they are not
significantly detected by all the individual observations.  This study can
therefore not shed much light on the rise or fall of the disc in the
hard state of \gx339.  However, in long exposure observations in the
high hard state, the fitted disc is at higher disc temperatures than
what is observed in the soft state.

\citet{Yu07} show that the peak hard state flux of an outburst depends on the length of
time since the previous outburst of \gx339.  Using the detailed
light- and X-ray colour-curves from the four outbursts observed by
\rxte.  The duration of Outbursts 2 and 3 (their Outbursts 6 and 7) appear to correlate with the
length of time since the previous outburst (see Table \ref{tab:OBs}).  However Outburst 4 lasts
longer than would be expected from the previous two and also reaches
the same peak flux as Outburst 2 without such a long period of
quiescence preceding it.  Under the
assumption that the duration of the outburst is linked to the
``waiting time'' by a power-law
relationship, then Outburst 4 would be expected to last around 280
days (Outburst 4 was not in the study of \citealp{Yu07}).  However, the end of the outburst is well monitored by
\rxte\ and there are sufficient observations at the beginning to rule
out a much earlier start (a mis-measurement of 100 days would be
required to match the relation obtained for the other outbursts) or a
lower peak flux.

The length of the outburst, however, would depend on how quickly the
material in the disc was drained during the outburst as well as the amount of material
which has built up in the disc since the last outburst .  Under the
assumption of a constant deposition rate of
material into the disc and that the disc is completely emptied during
each outburst, then the interval is a measure of the amount 
of material which could contribute to the outburst. In reality the
disc is unlikely to be totally emptied during each outburst.

As is clear from Table \ref{tab:OBs}, the length of the outburst,
$t_{\rm outburst}$, does
not correlate well with the interval between outbursts, $t_{\rm
prior}$.  However, the duration of an outburst depends not only on
the amount of material in the disc, but also on the rate at which it
is depleted.  The peak X-ray flux is likely to be a good measure of
the depletion rate within the outburst.  If we correct the
interval between outbursts by dividing by the X-ray flux, we obtain values which do
correlate with the length of the outburst, $\mathcal{L}$.  We note that we only have
four outbursts and hence three estimate on the relative magnitudes of
the durations of the outbursts to use in this study.

\begin{table}
\centering
\caption{\label{tab:OBs} {\sc Outburst Durations}}
\begin{tabular}{cllllll}
\hline
\hline
OB & Start & Stop & $t_{\rm outburst}$ & $t_{\rm prior}$ & Peak Flux & $\mathcal{L}$\\
&\multicolumn{2}{c}{MJD-50000}&days&days&($\ergpspcmsq$)\\
\hline
1 & 800 & 1240 & 440 & - & $2.7times 10^{-9}$ & -\\
2 & 2400 & 2730 & 330 & 1160 & $9.4\times 10^{-9}$ & 123\\
3 & 3225 & 3475 & 250 & 495 & $5.0\times 10^{-9}$ & 99\\
4 & 4130 & 4240 & 110 & 655 & $9.7\times 10^{-9}$ & 68\\
\hline
\end{tabular}
\begin{quote}The start and stop times for the
  outbursts are taken as the departure and return to the hard state
  (X-ray colour = 1).  The separation is then the time between the
  return to the hard state for one outburst on its decline and the
  departure from the hard state on the subsequent outburst on its rise.
\end{quote}
\end{table}

\section{Iron Line} \label{sec:FeLine}

An $F$-test in combination with the line normalisation (see Section
\ref{sec:model} for more details) were used to test for the presence of an emission line
(assumed to be the iron $K_\alpha$ fluorescence line) at
$6.4\kev$.  However  out of 628 spectral
fits, 400 require a line\footnote{This is using both the $F$-test and
  the line normalisation test to check that the gaussian component is significant.}, and these occur all over the HID, in all
states.

We did investigate the effect of allowing the line energy to be free,
between $6-7\kev$.  However this caused some erroneous fitting, mainly in
cases where there is some slight curvature to the low energy powerlaw.
The curvature is a broad feature usually at lower energies
than $6\kev$.  In trying to fit this feature, the gaussian component became very broad and
``pegged'' at the lower energy limit with a large uncertainty in the
line energy, even though the residuals showed clear excesses in the $6-7\kev$ range.  Selecting
these models with a significant ``line'' detection as the best fitting
would not give information about any line present as this has not been
fitted by the gaussian.  Therefore, any line component, even if
deemed significant by the $F$-test which had an energy uncertainty
of $>1\kev$ was taken not to be representing a line, but another
continuum component, and so the line ``detection'' was discarded from
further analysis.  In some cases a line was well and sensibly fit at
energies different from $6.4\kev$, however these were few.  However,
this method discards observations where a line may be present but
{\scshape xspec} has preferentially fitted the curvature in the
powerlaw instead.  We therefore present the results on the analysis
with a gaussian component fixed at $6.4\kev$ in this
section.  We note that using the most stringent cut-off in the
$F$-test when selecting lines when their energy was free did show a
variation in the energy across the HID.  The lowest energy was at the
top of the low-hard state and the highest energy was in the
intermediate state\footnote{Given the problems with fitting a line
  whose energy was free, we do not split the intermediate state for
  this comment.} at the end of an outburst, during the return to quiescence.

\begin{figure}
\centering
\includegraphics[width=1.0\columnwidth]{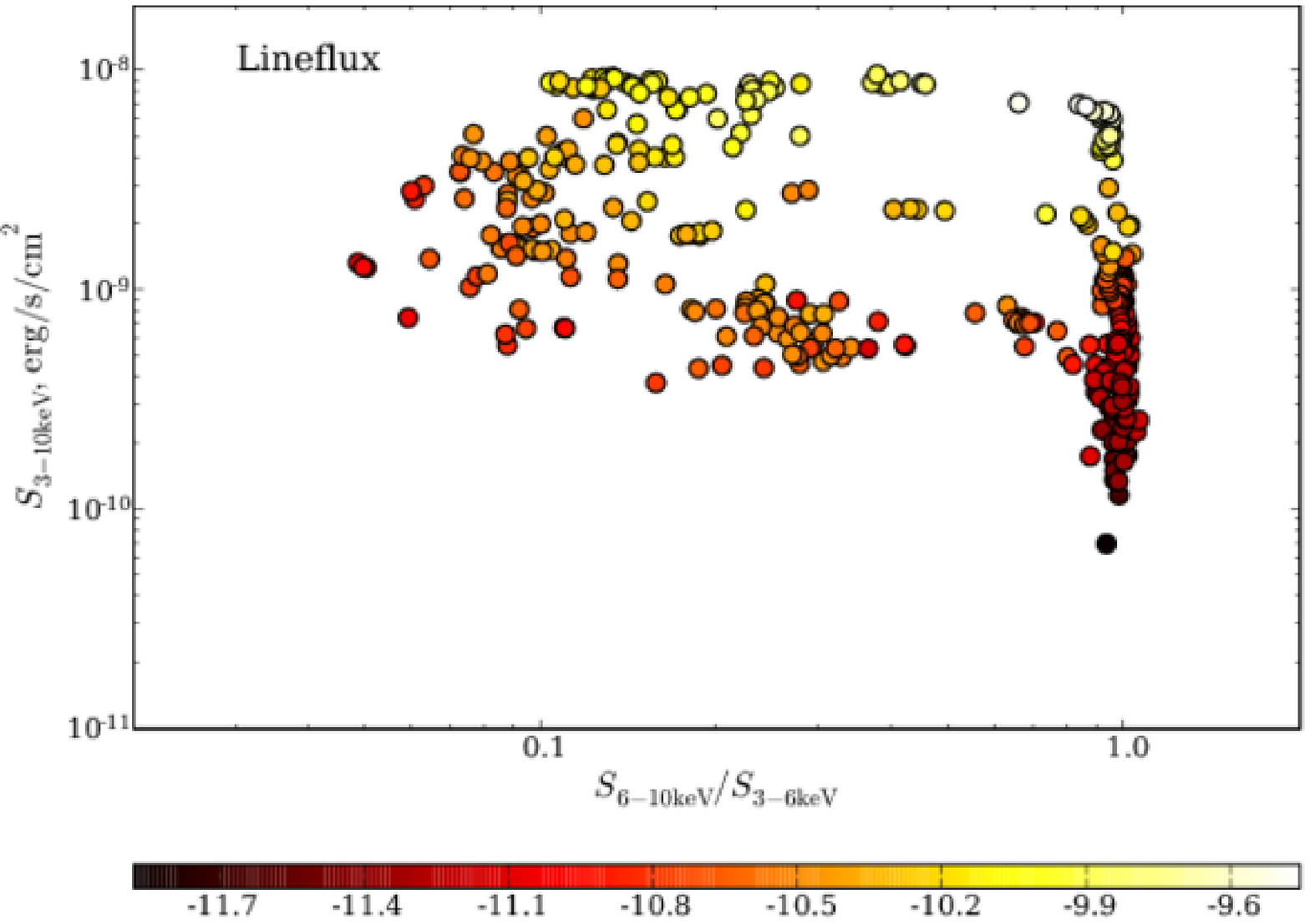}
\includegraphics[width=1.0\columnwidth]{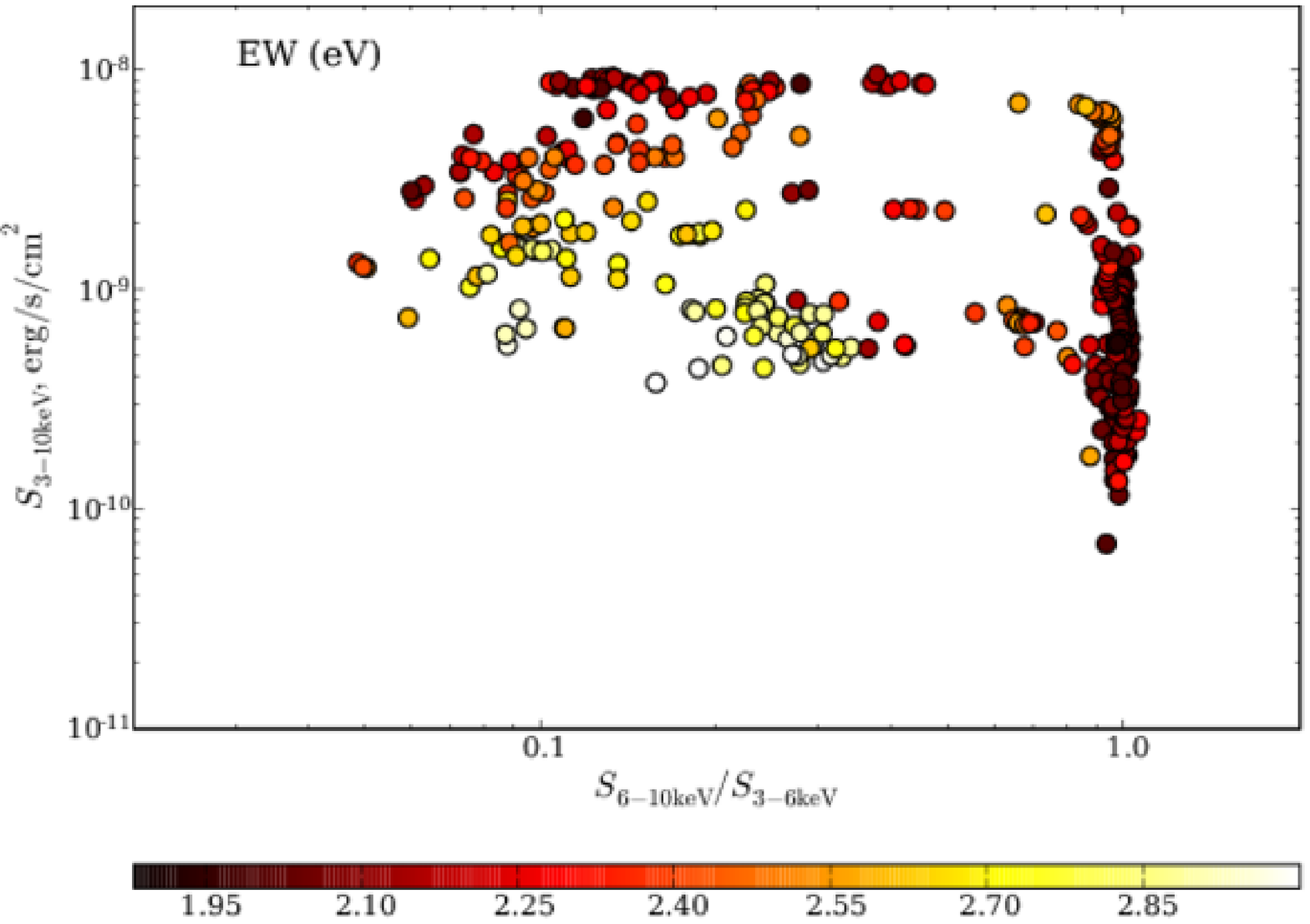}
\includegraphics[width=1.0\columnwidth]{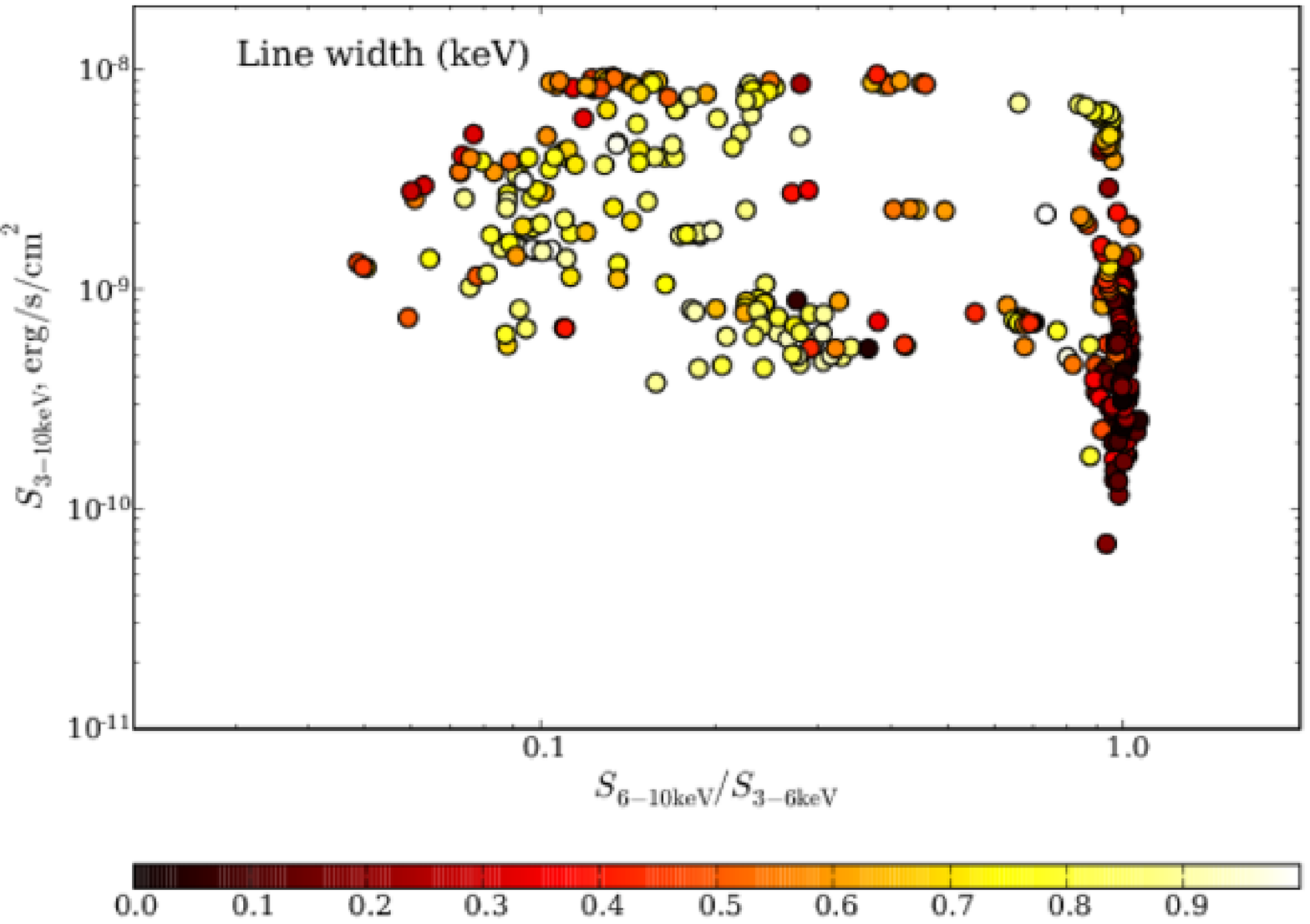}
\caption{\label{fig:Turtle-EW} The hardness-intensity diagram
  from all the observations where a significant line is detected.  {\scshape top} The
  colour scale is from the line flux.  The decrease in line flux
  during the course of the outburst is clearest at the low soft
  states.  {\scshape middle} The colour scale is for the equivalent
  width of the iron line (logarithmic, eV). In some cases the calculated
  equivalent width was greater than $1\kev$ and these have been
  truncated in the colour scale to aid clarity. {\scshape bottom} The
  colour scale is for the line width in $\kev$.}
\end{figure}

\begin{figure}
\centering
\includegraphics[width=1.0\columnwidth]{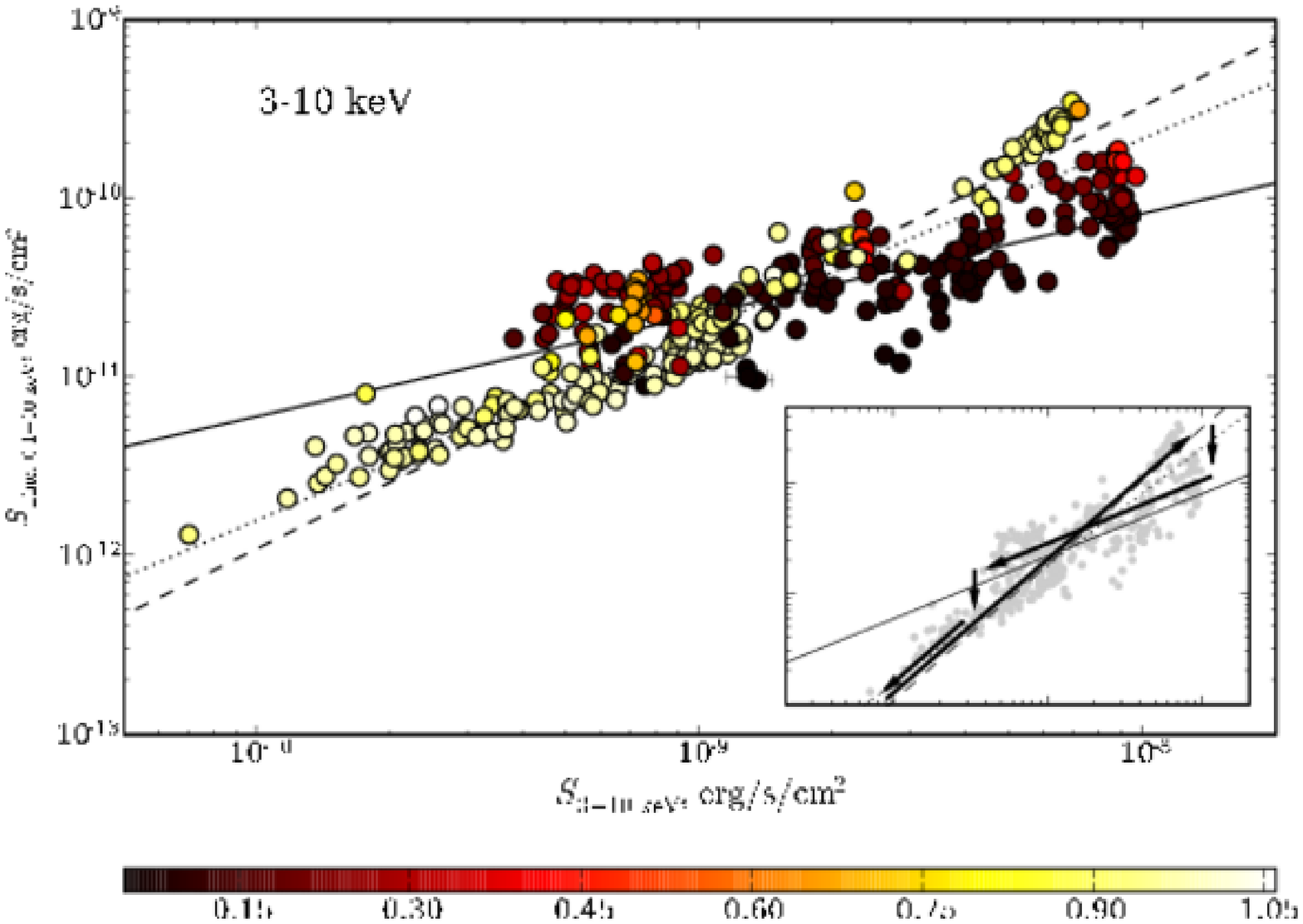}
\includegraphics[width=1.0\columnwidth]{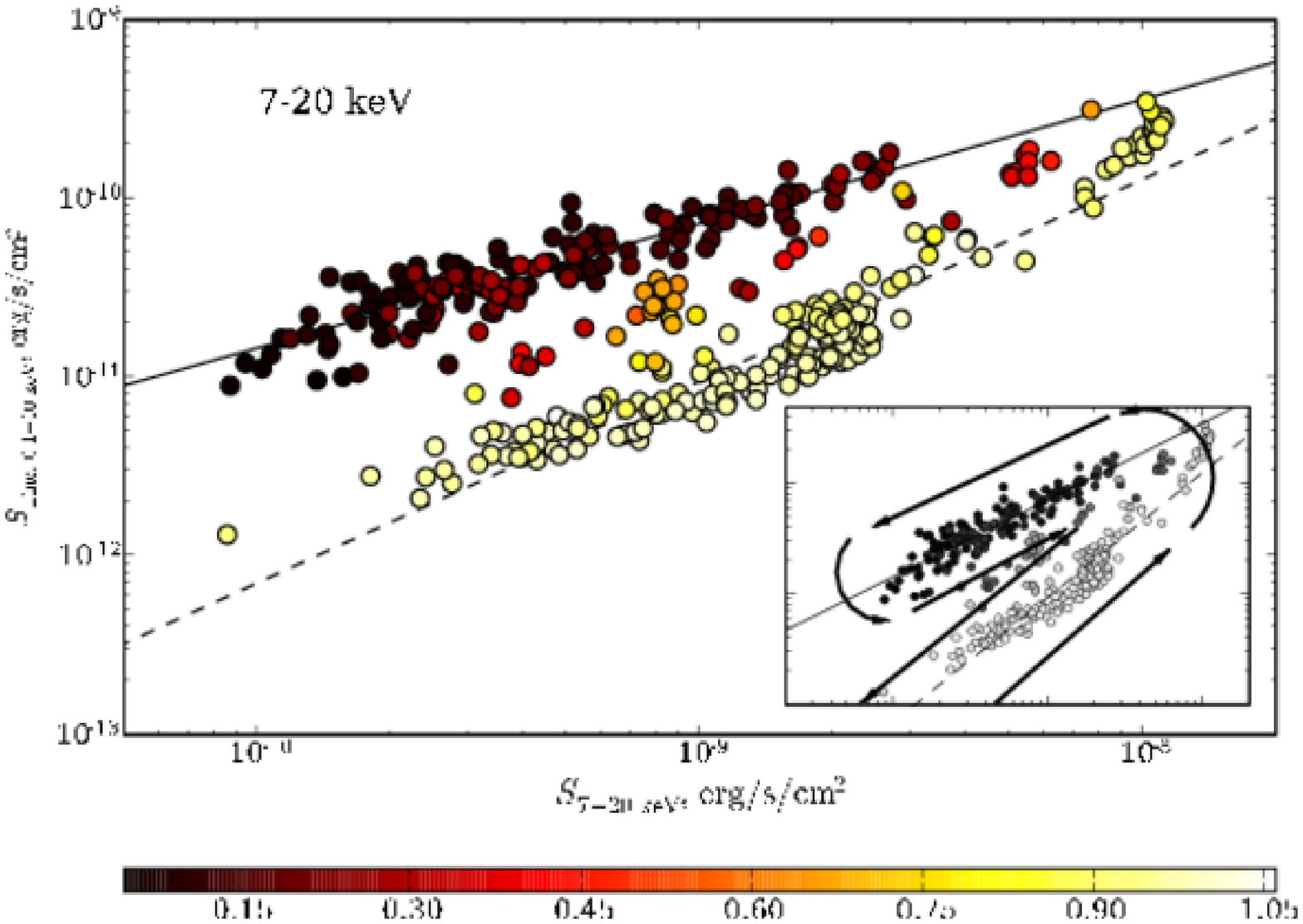}
\caption{\label{fig:lineflux} The flux of the $6.4\kev$ iron line as a
function of {\scshape
  top} the $3-10\kev$ total flux and {\scshape bottom} the $7-20\kev$ total flux.    The colour scale on the plot 
is the X-ray colour of the observation.  The best fit relation is to
all the data is shown by the dotted line.  The best fit to the
soft and hard states are shown by the black solid and dashed lines
respectively.  The insets show the approximate motion of
\gx339\ through the diagrams during an outburst.}
\end{figure}

We extract the line flux over a $0.1-10\kev$ range.  The line is
strongest at the highest fluxes in the hard state, see
Fig. \ref{fig:Turtle-EW}, whereas the EW is largest at the lowest flux
soft states.  The line width is also largest in the soft state (Fig. \ref{fig:Turtle-EW}).   Fitting the line flux against the
$3-10\kev$ total flux shows a relationship for all of the data of
\[
S_{\rm 0.1-10,  line}\propto S_{\rm 3-10\kev}^{1.07\pm0.02},
\]
however there is a large
scatter of around the best fitting line (see Fig \ref{fig:lineflux}).  The
Spearman Rank correlation coefficient is $0.860$.  We also calculate
the Kendall's Tau correlation coefficient, $0.678$.  The variance of
the latter\footnote{Kendall's Tau is more non-parametric than the
  Spearman Rank correlation as it uses only the relative ordering of
  the ranks.  The variance is calculated as
  $\rm{Var}(\tau)=\frac{4H+10}{9N(N-1)}$, with the significance, in $\sigma$, as
    then $\tau/\rm{Var}(\tau)$.} is $0.0011$ and so the significance of the
correlation is $20.2\sigma$.  As the
continuum increases, then the line does in an almost linear
relationship.  However, the X-ray colour of \gx339\ for each
observation shows that the hard and soft states have different
behaviours.

We split the data into extremely soft (X-ray colour $<0.22$) and hard
(X-ray colour $>0.87$) states and refit the correlation.  The hard and
soft state correlations for both the $7-20\kev$ and $3-10\kev$ fluxes are then 
\begin{eqnarray}
S_{\rm 0.1-10,  line}&\propto& S_{\rm 3-10\kev}^{1.24\pm0.02}\nonumber\quad{\rm hard}\\
S_{\rm 0.1-10,  line}&\propto& S_{\rm 3-10\kev}^{0.57\pm0.06}\nonumber\quad{\rm soft}\\
\nonumber\\
S_{\rm 0.1-10,  line}&\propto& S_{\rm 7-20\kev}^{1.13\pm0.03}\nonumber\quad{\rm hard}\\
S_{\rm 0.1-10,  line}&\propto& S_{\rm 7-20\kev}^{0.70\pm0.03}\nonumber\quad{\rm soft},
\end{eqnarray}
respectively (see Fig \ref{fig:lineflux}).  We also show the
correlations of the $7-20\kev$ total flux with the line flux,
following  \citet{Rossi05}, as only photons with
energies $\gtrsim 7\kev$ are be able to ionise iron.  We do not,
however, fit all observations with a single line as there are clearly
two separate relations in the $7-20\kev$ flux in Fig. \ref{fig:lineflux}

In both flux bands, the slope of the relation of the line flux is
different between the soft and hard states.  In the 
$7-20\kev$ flux band, the powerlaw is the dominant spectral component being
probed.  The powerlaw flux falls away during the soft state and, noting
that the lineflux remains approximately constant across all X-ray
colours in Fig. \ref{fig:Turtle-EW}, the
soft state points have been displaced to the left.  However using the
lower energy $3-10\kev$ band, the disc is the dominant spectral
component in the soft state, but the powerlaw takes it's place in the
hard state.  Therefore there is no sideways displacement of the soft
state points relative to the hard state.

The difference between the behaviour of the line fluxes in the hard and the soft state
could arise from changes in the ionization parameter as well as from
differences in the spectral slope of the incident radiation. The X-ray
spectrum above $7.1\kev$ of a soft state object is steeper than that of
a hard state object. Thus, there are more lower energy photons
available to ionize the iron in the soft state compared to the hard
state. As the cross-section of the ionization process decreases with
increasing photon energies, a soft state is therefore more efficient
to produce the iron line than a hard state for a given 7-20 keV
flux. Additionally to this effect a soft state may have a smaller
ionization parameter, which would also increase the observed line
flux (for more details see e.g. \citealp{Reynolds03}).  

\subsection{X-ray Baldwin Effect}\label{sec:FeLine:baldwin}

Having detected lines in most of observations at different
states during out an outburst we investigated whether there is an
X-ray Baldwin effect in \gx339.  \citet{Baldwin77} found that the
equivalent width of the C{\scshape iv} line decreased with increasing
UV luminosity in a sample of AGN.  An X-ray Baldwin effect has been
reported in AGN by a number
of authors \citep{Iwasawa93, Nandra97, Page04, Bianchi07, Mattson07}.  They find
that the equivalent width depends on the luminosity in the following
range
\[
EW\propto L^{\beta},
\]
where $-0.20\leq\beta\leq-0.14$.

\begin{figure}
\centering
\includegraphics[width=1.0\columnwidth]{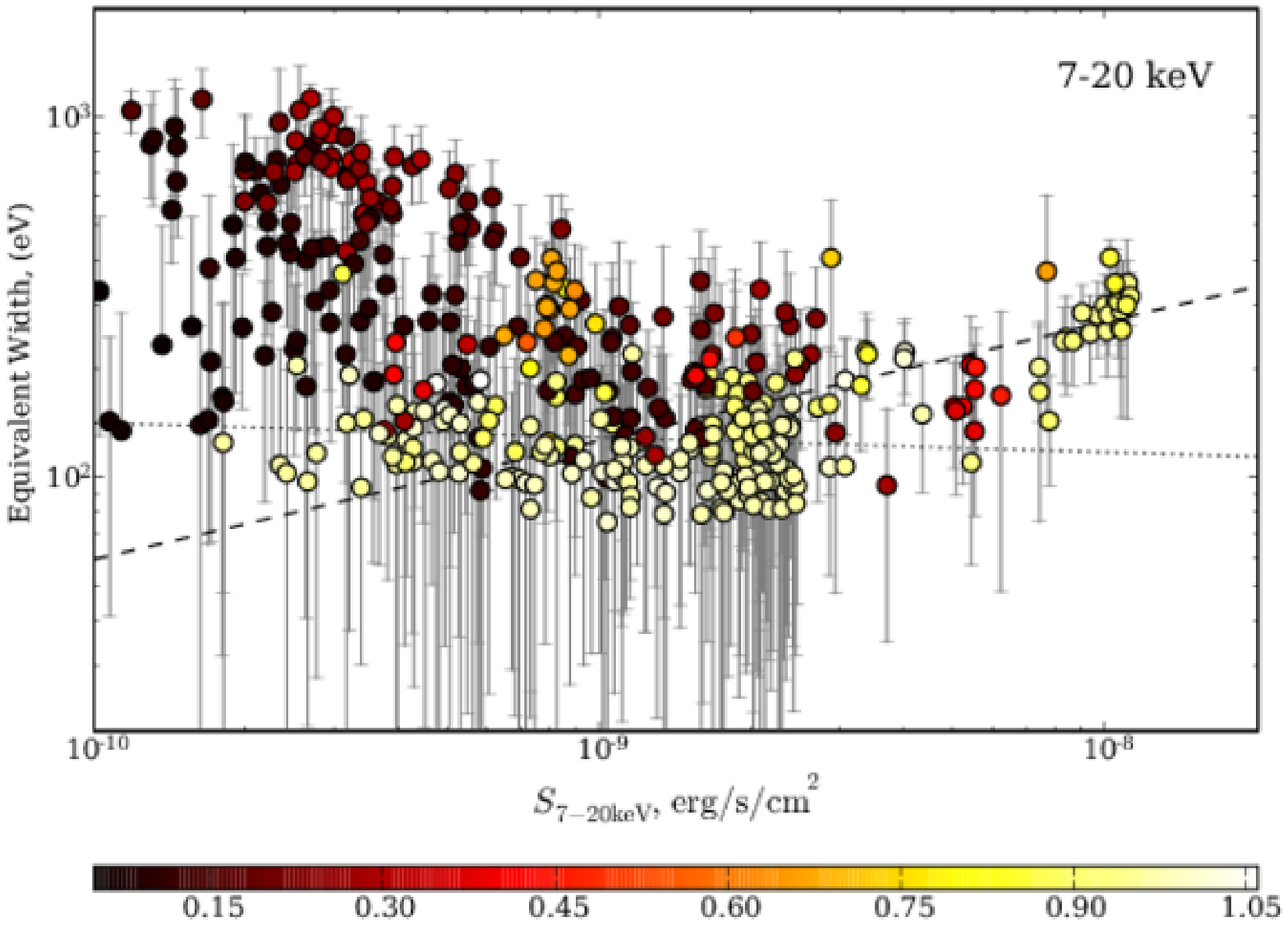}
\includegraphics[width=1.0\columnwidth]{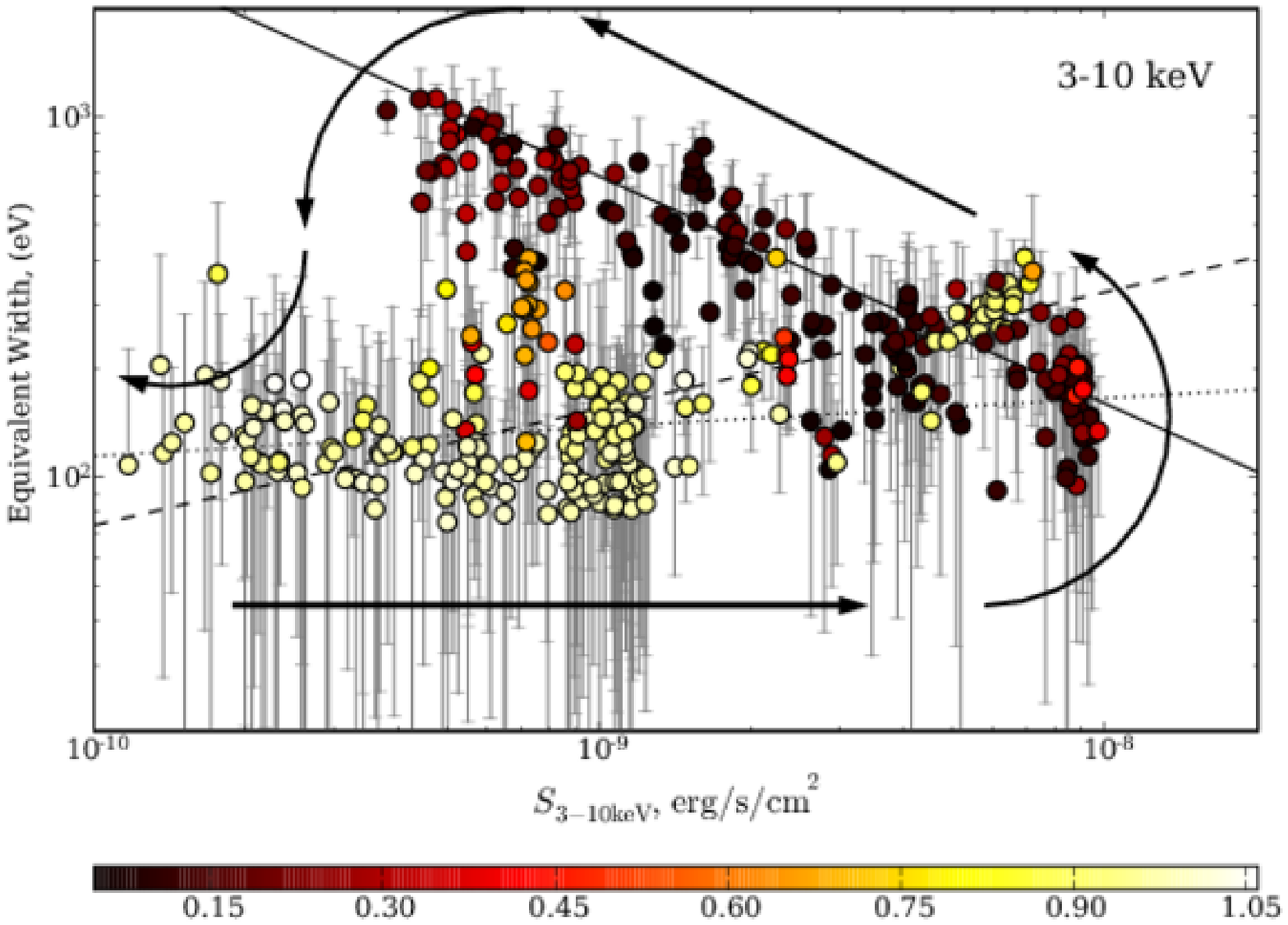}
\caption{\label{fig:Flux-EW} The Equivalent Width of the Iron
  line against the flux during the outbursts, {\scshape top:}
  $7-20\kev$ flux, {\scshape bottom:} $3-10\kev$ flux.  The colour scale shows the
  X-ray colour of the spectrum for each observation, black being soft
  and white being hard.  The solid line shows the best fit to the soft
  state, the dashed line the best fit to the hard state, and the
  dotted line to the low-flux hard states.  The arrows indicate the
  motion of \gx339\ through the diagram during an outburst.}
\end{figure}

The equivalent width was calculated during the course of the fitting,
and is shown on the HID in Fig. \ref{fig:Turtle-EW}.
The HID clearly shows that, on average, the equivalent width is highest in
the softest and lowest flux states.  Fig. \ref{fig:Flux-EW} shows the
equivalent width of the iron line against the flux of the observation,
with the X-ray colour of the spectrum as a colour scale.  The two
states (hard and soft) appear to have different correlations between
the flux and equivalent width.  Extracting those observations which
have an X-ray colour of $<0.22$ (the extreme soft state), the best fit
to the anti-correlation is,
\begin{eqnarray}
EW\propto S_{3-10\kev}^{-0.61\pm0.09}\nonumber\\
EW\propto S_{7-20\kev}^{-0.55\pm0.05}\nonumber,
\end{eqnarray}
with Spearman Rank correlation coefficients of $-0.840$ and $-0.430$ respectively.  The Kendall
Tau are $-0.644$ and $-0.308$, with significances of $10.6\sigma$ and
$5.07\sigma$ respectively.  We also take the hard state observations (those with X-ray
colours $> 0.87$) we fit the observations with
\begin{eqnarray}
EW\propto S_{3-10\kev}^{0.32\pm0.07}\nonumber\\
EW\propto S_{7-20\kev}^{0.33\pm0.02}\nonumber,
\end{eqnarray}
with Spearman Rank correlation coefficients of $0.262$ and $0.237$ respectively.  The best
fitting line has a positive index.  However this is the result of the
combination of the large scatter in the majority of the hard state
points and the small cluster at high fluxes, separate from the rest.
If we ignore those points with $S>2\times 10^{-9}\ergpspcmsq$ and
$EW>400\ev$, and fit the bulk of the hard state observations, we find $EW\propto S_{3-10\kev}^{-0.09\pm0.14}$ and $EW\propto S_{7-20\kev}^{-0.04\pm0.04}$ , shown as the
dotted line in Fig. \ref{fig:Flux-EW}.  These are consistent with no
change in the EW with flux.  No errors in the $7-20\kev$ fluxes could
be obtained from {\scshape xspec} and so the uncertainties in the
slopes for the $7-20\kev$ fluxes only include the errors in the EWs.
The uncertainties in the slope of the $3-10\kev$ fits include the
errors from the EWs and the fluxes.

If the lineflux decreases at the same rate as the continuum then the
measurable equivalent width of
the line decreases as the flux decreases.  This effect would therefore
lead to a lower cut-off for the minimum measurable equivalent width with flux
with a negative slope.  This would result in a small slope in the data
points, whereas it should be flat.  We therefore do not believe that
there is an X-ray Baldwin effect in \gx339\ in the hard state.

The motion of \gx339\ through the flux-equivalent width plane is that
at the beginning of the outburst the flux increases at constant
equivalent width.  At the peak flux, when the source moves into the
soft state, the equivalent widths start to increase as the flux
decreases.  When the source returns to the hard state, the equivalent
widths reduce suddenly and then become constant as the flux falls (see
arrows in Fig. \ref{fig:Flux-EW}).
This is another example of hysteretical behaviour from \gx339.

There is an X-ray Baldwin effect in the $6.4\kev$ iron
line in the soft state of \gx339.  The slope of the relation, however, is much
steeper than that observed in AGN.   \citet{Iwasawa93} studied Seyfert 1s and
quasars, whereas in a sample of 53 AGN from the {\it XMM-Newton}
archive, \citet{Page04} used 14 radio loud and 39
radio quiet sources.  \citet{Bianchi07} excluded
radio loud objects\footnote{Quasars with $\log(R)>1$ and Seyferts with
$\log(R)>2.4$ and $\log(R_X>-2.755)$ where $R$ is the radio-loudness parameter from
  \citet{Stocke92} and $R_X$ is the X-ray radio loudness parameter
  \citet{Terashima03}\citep{Bianchi07}} from their complete study of radio quiet Type
1 AGN in the {\it XMM-Newton} archive.  These studies all show an X-ray
Baldwin effect between a variety of sources, rather than an intrinsic
effect within one source.  Most of the AGN in studies of
the X-ray Baldwin effect have been radio quiet AGN, analogous to the soft
state of black hole X-ray binaries.   Scaling the effect described
here for \gx339\ up to AGN would result in an effect which occurs on
very short timescales.

\subsection{Links to the Powerlaw Slope}\label{sec:FeLine:PLSlope}

\begin{figure}
\centering
\includegraphics[width=1.0\columnwidth]{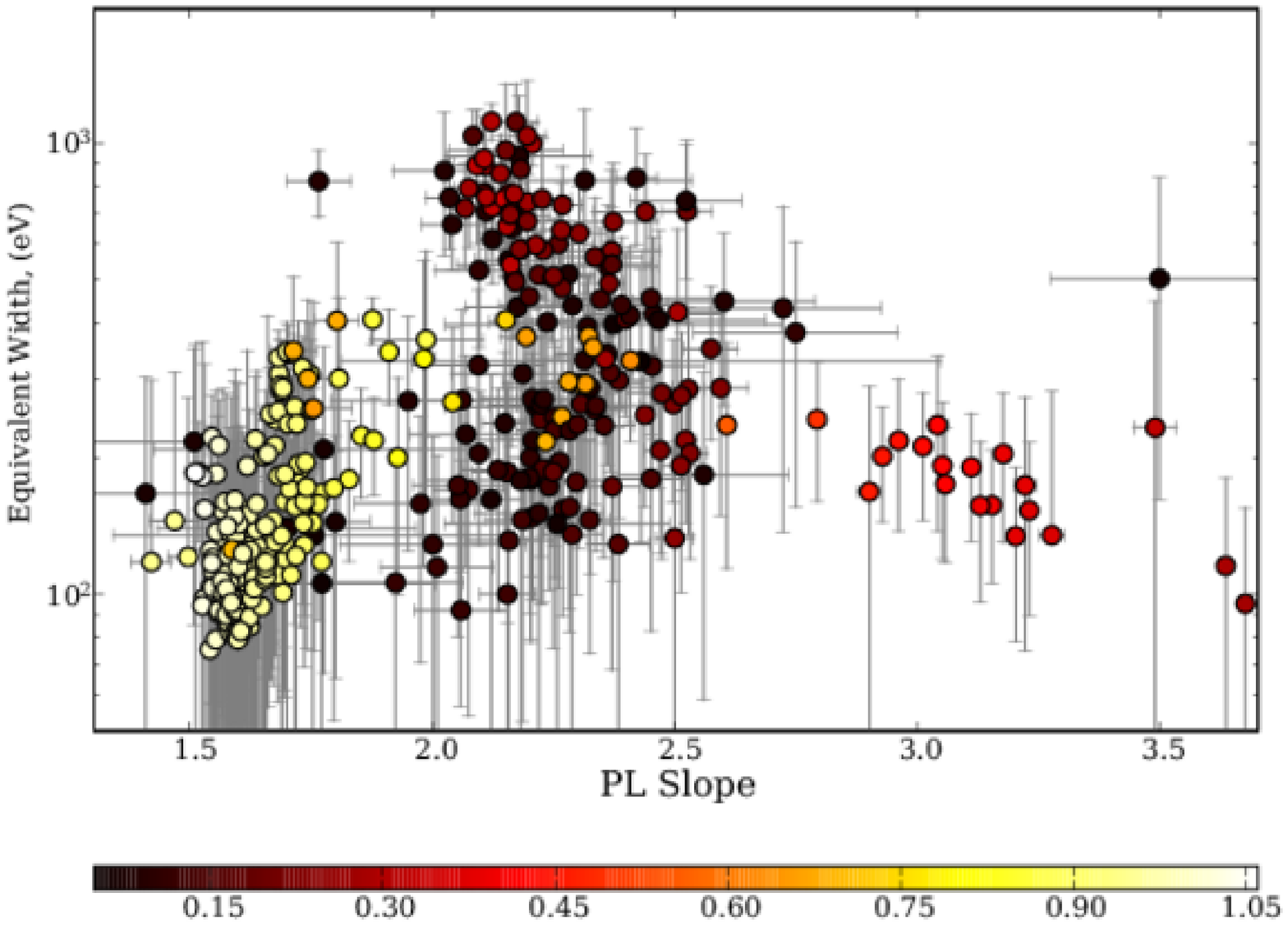}
\includegraphics[width=1.0\columnwidth]{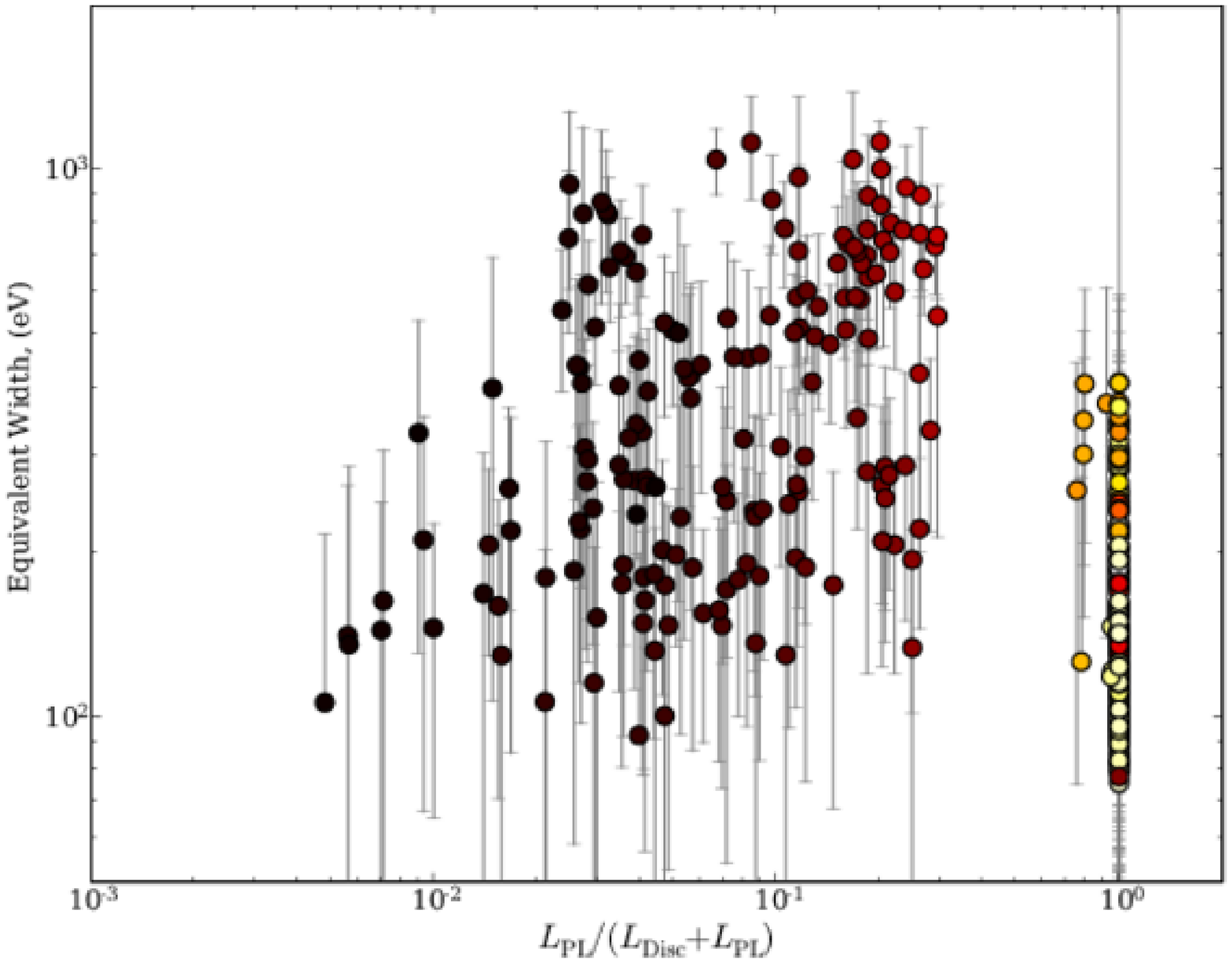}
\caption{\label{fig:PL_EW} {\scshape top:} The equivalent width against the power law
  slope.  The colour scale shows the
  X-ray colour of the spectrum for each observation, black being soft
  and white being hard.  The plot has been truncated at $\Gamma=1.4$
  and $\Gamma=3.5$, and $\Gamma$ outside of this range is unlikely to
  be an accurate description of the data. {\scshape bottom}: The disc
  fraction of \gx339\ against the equivalent width with the colour
  scale representing the X-ray colour.  See Section \ref{sec:DFLD} for more
  details about the disc fraction.}
\end{figure}

In a recent study of 12 Seyfert 1 and 1.2 galaxies, \citet{Mattson07} fitted \rxte\
data with a {\scshape pexrav} model, which simulates an exponentially
cut-off power-law reflected by neutral matter, and a Gaussian line.  They find a positive
correlation between the line equivalent width and the underlying power
law slope for $\Gamma<2$.  For $\Gamma>2$ they found an anti
correlation, with a peak equivalent width of $\sim250 \ev$, see their
Figs. 1a and 3.  We plot the powerlaw slope (below the break in the
case of a broken powerlaw) against the
equivalent width for \gx339\ in Fig. \ref{fig:PL_EW}.  We note that
superficially the shape of the distribution is very similar, however
there is a larger scatter in the shape of the distribution.  The
colour scale corresponds to the X-ray colour of the observation.  The
correlation corresponds to the hard state whereas the anti-correlation
corresponds to the soft state.

In \gx339\ the hard state, which usually associated with the presence of a
steady, mildly relativistic jet, is found at the low $\Gamma$s.  In
\citet{Mattson07} the low $\Gamma$ end of relation is dominated by
points from 3C~273.  The positive correlation is composed of
observations of radio loud Seyfert 1 galaxies, the peak by radio quiet
Seyfert 1s, and radio quiet Seyfert 1.2s comprise the anti-correlation
part of the relation.  In \gx339, the soft, disc dominated state is
mostly at the peak of the relation with the intermediate state at the
high $\Gamma$ end.

The model of an X-ray binary outburst presented in \citet{Fender04}
indicates that the jet is quenched as the disc becomes dominant in the
soft state.  As the jet is the source for the radio emission, when
this quenches, the X-ray binary becomes radio quiet, in the soft state,
as shown in Fig. \ref{fig:DFLDradio}.  We therefore would expect to see
observations of the soft, radio quiet state of \gx339\ on the negative
correlation and the hard, radio loud state on the positive correlation
from this model.

\citet{George91} simulated the spectrum expected from an X-ray source which
illuminates a slab and showed that it should include an iron line and
a ``Compton hump.''  As the spectrum softens ($\Gamma$ increases)
there are fewer photons with enough energy to photo-ionise iron, and so
the iron line EW decreases.  This is observed in the soft state of
\gx339\ as well as in the sample in \citet{Mattson07}.  

\citet{Mattson07} suggest that for $\Gamma$ to increase as the jet
dominance decreases the jet would have a hard X-ray component
associated with it.  On the other side of the relation, observations
of MCG--6-30-15 have shown that the reflection component remains
relatively constant \citep{Miniutti07}, so when $\Gamma$ increases the
EW decreases.

As the relations between the power-law slope and the equivalent width
of the iron line are similar for \gx339\ and the sample of AGN it is
reasonable to assume that the processes at the centre of an AGN and an
XRB are similar.  As XRBs have radio (jet) dominated quiescent periods and
disc dominated outbursts, then this similarity in the relations points
towards to AGN also having outburst phases.  As the black hole masses
in AGN are $\sim 10^5-10^8$ times larger, their outburst phases are
much longer.  However, this result indicates that the path of an AGN
through an HID would be similar to XRBs, though their direction is
currently unknown.  The AGN HID would have to be constructed so that ultra-violet
emission from the disc would be included.   This is further discussed in
Section \ref{sec:DFLD}.

As the hard, intermediate and soft state points appear in different
locations in the diagram, we reform the relation using the disc
fraction (see Section \ref{sec:DFLD} and Fig. \ref{fig:PL_EW}).  The X-ray colour is indirectly
related to the disc temperature and luminosity.  The disc fraction,
however, is a more physical quantity describing the relative
luminosities of the disc and the powerlaw components of the spectrum.
The hard and intermediate
state appear with no disc fraction as the best fitting model contains
no disc component.  The soft states occur over a range of disc
fractions and equivalent widths.  We split the observations into two
sets; those with a disc (disc fraction $<0.5$) and those without (disc
fraction $\ge0.5$) and perform a t-test to indicate whether the means
are significantly different and also a K-S test to see if the two
distributions of equivalent width are drawn from the same population.
The probabilities for both tests show that the two data-sets have the same mean/were drawn
from the same population are very low ($P_{KS}=1.29\times10^{-25}$ and
$P_{t}=1.3\times10^{-46}$ respectively).  We therefore conclude that,
although the separation of the two populations is small, the
behaviour of the equivalent width of the line is different when
\gx339\ is in outburst to when it is in quiescence/the hard state.
However the difference in behaviour is less clear in this presentation
of the variation of the equivalent width with the state of \gx339.

\section{Disc Temperature} \label{sec:DiscT}

\begin{figure}
\centering
\includegraphics[width=1.0\columnwidth]{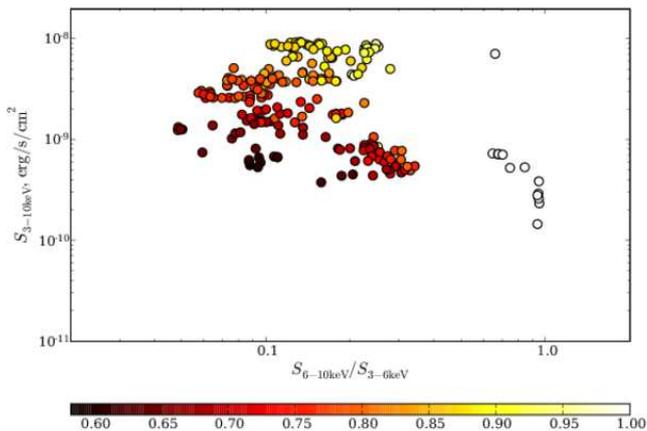}
\caption{\label{fig:Turtle-DiscT} The hardness-intensity diagram
  from all the observations where a disc is detected.  We do not
  consider the apparent detections of the disc in the hard state as
  true detections.  The colour scale is for the disc temperature in keV, and has
been truncated at $1\kev$ to emphasise the variation in disc
temperature during the soft state.}
\end{figure}

Selecting those observations where a disc is present we extract
the disc temperature.  The variation in disc
temperature through the HID is shown in Fig. \ref{fig:Turtle-DiscT}.  The highest disc
temperatures are observed at the highest fluxes in the soft state, and
the lowest temperatures at the lowest fluxes.  The variation through
the soft state is smooth, even though we show data from all four
outbursts.  The observations with
an X-ray colour $>0.4$ show a disc temperature much higher than those
in the soft state (though the colour scale is truncated at $1\kev$).  In fact they appear to indicate that the disc
temperature is rising as the source is fading down the hard branch.
As stated in Section \ref{sec:model:best} it is unlikely that these
disc ``detections'' are true.  Without spectral information below
$3\kev$ we are limited to how accurately we can determine the presence
or absence of a disc.  If there is any curvature in the spectrum, this
will be better fit by a higher temperature disc combined with a
powerlaw rather than a simple powerlaw.  However, the curvature may
not be the result of a disc component, and as we have on purpose
restricted the choice of models to a limited range, the disc turns out
to be the statistical best fit.  The temperature of these disc fits is
much higher than would be expected for a decaying disc ($>1\kev$).  If
these are true discs, then the temperature at the end of an outburst
doubles over 10-20 days compared to the gradual $\sim$250 day decay in
the soft state.  Also, as the fluxes of these disc components
is $10^{-11}-10^{-10}\ergpspcmsq$ rather than the
$10^{-9}-10^{-8}\ergpspcmsq$ for the soft-state disc detections, we
conclude that the {\scshape diskbb} model is not fitting the true disc
component.  We do not include any of these points in any
analysis in this section.

The spectrum of an optically thick, geometrically thin accretion disc
is the result of a sum of black-body spectra, one for each radius - a
so-called multi-coloured disc model.  The maximum temperature for such
a disc should occur close to the inner edge of the accretion disc, $R_{\rm in}$.  In a similar
way to the Stefan-Boltzmann law, the resulting disc flux should follow:
\[
S_{\rm Disc}\propto R_{\rm in}^{2}T^{4},
\]
mimicking the emission from a black body whose radius is that of the
inner edge of the disc and whose temperature is that of the disc at
the innermost radius.  In the first instance we assume a constant inner radius for the disc.
We initially assumed that in the soft state, the flux from the disc
would dominate the total flux to such an extent that this relation
would hold for the total flux received from \gx339 and hence expected
that $S_{3-10 \kev}\propto T^{4}$. 

\begin{figure*}
\centering
\includegraphics[width=0.49\textwidth]{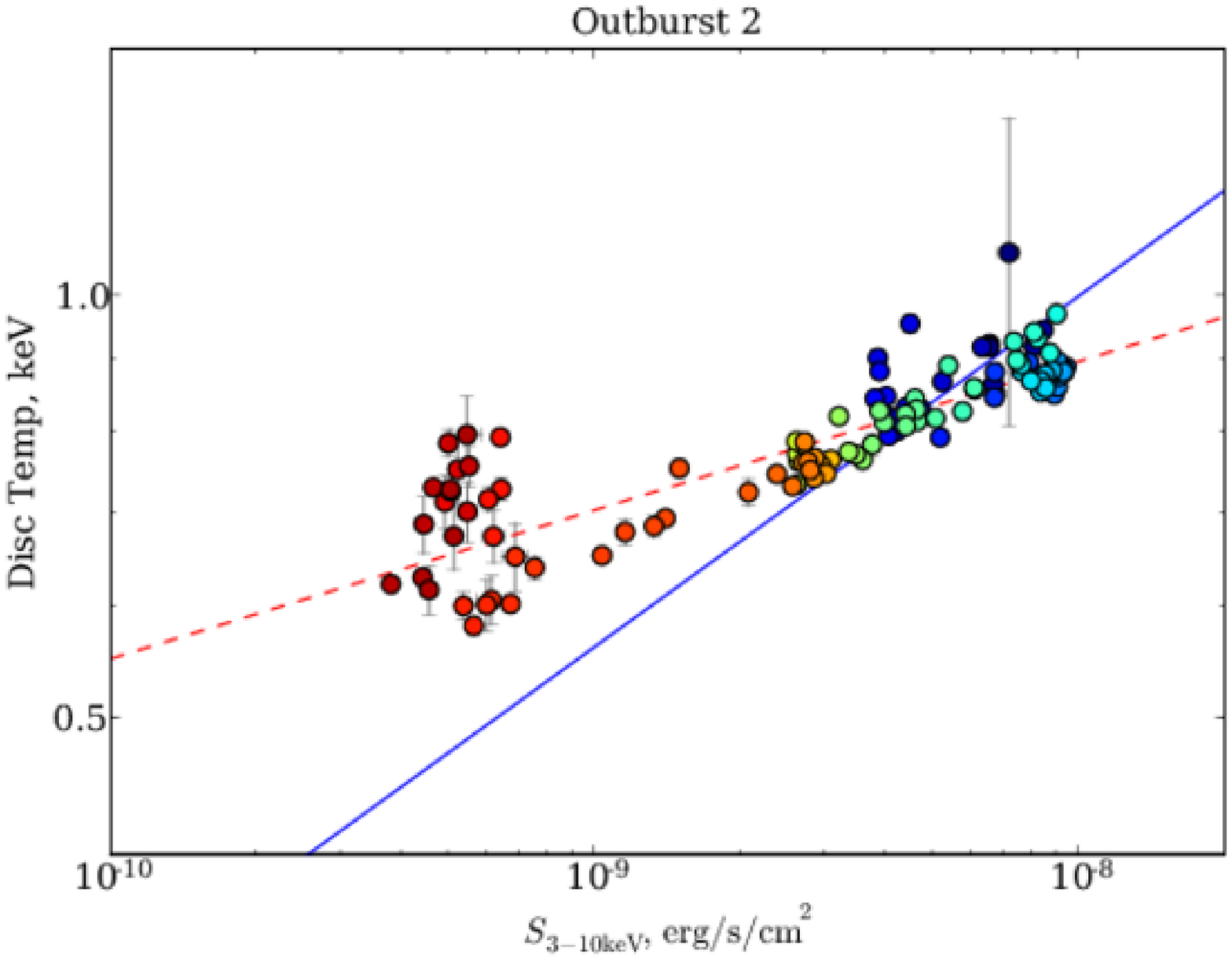}
\includegraphics[width=0.49\textwidth]{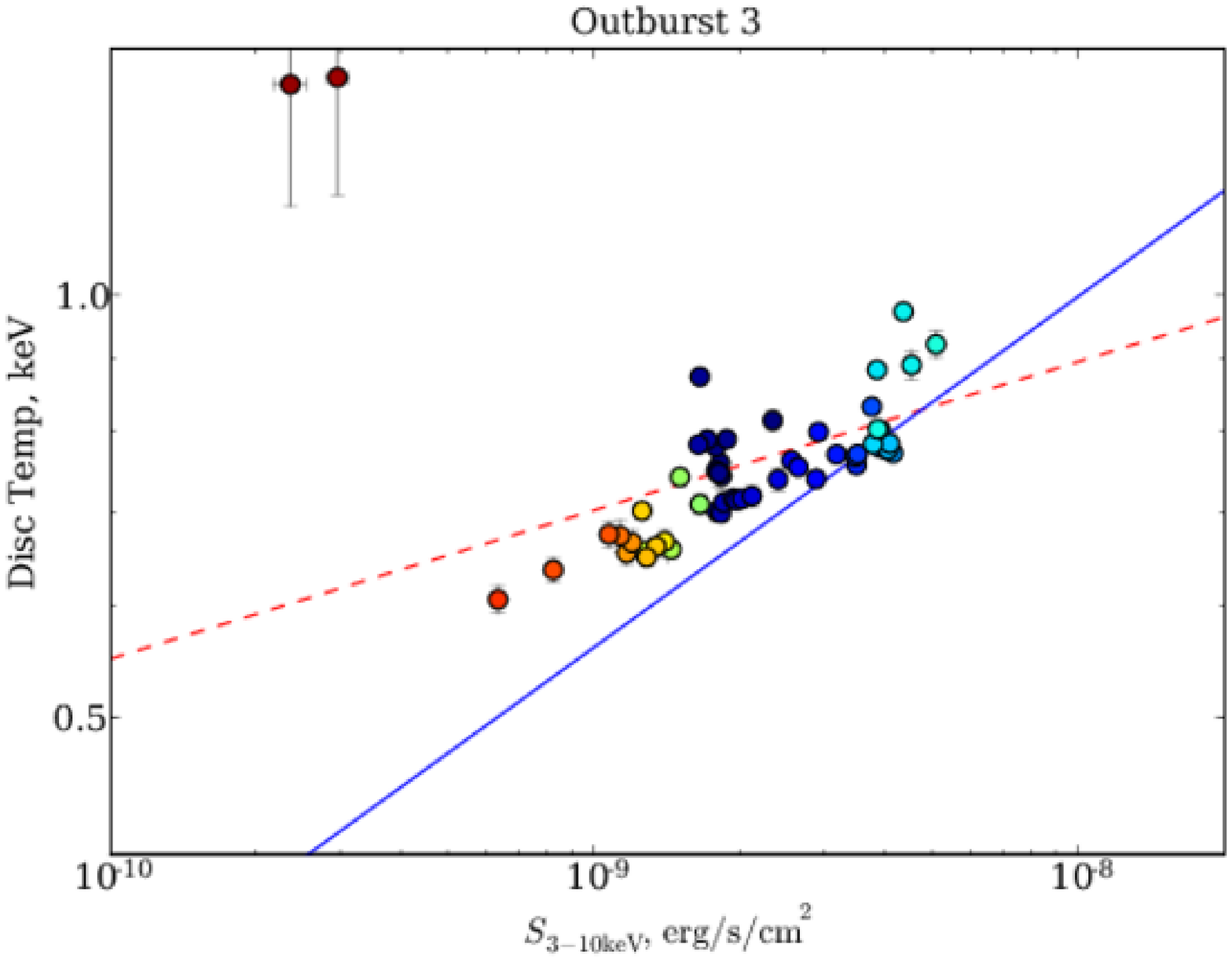}
\includegraphics[width=0.49\textwidth]{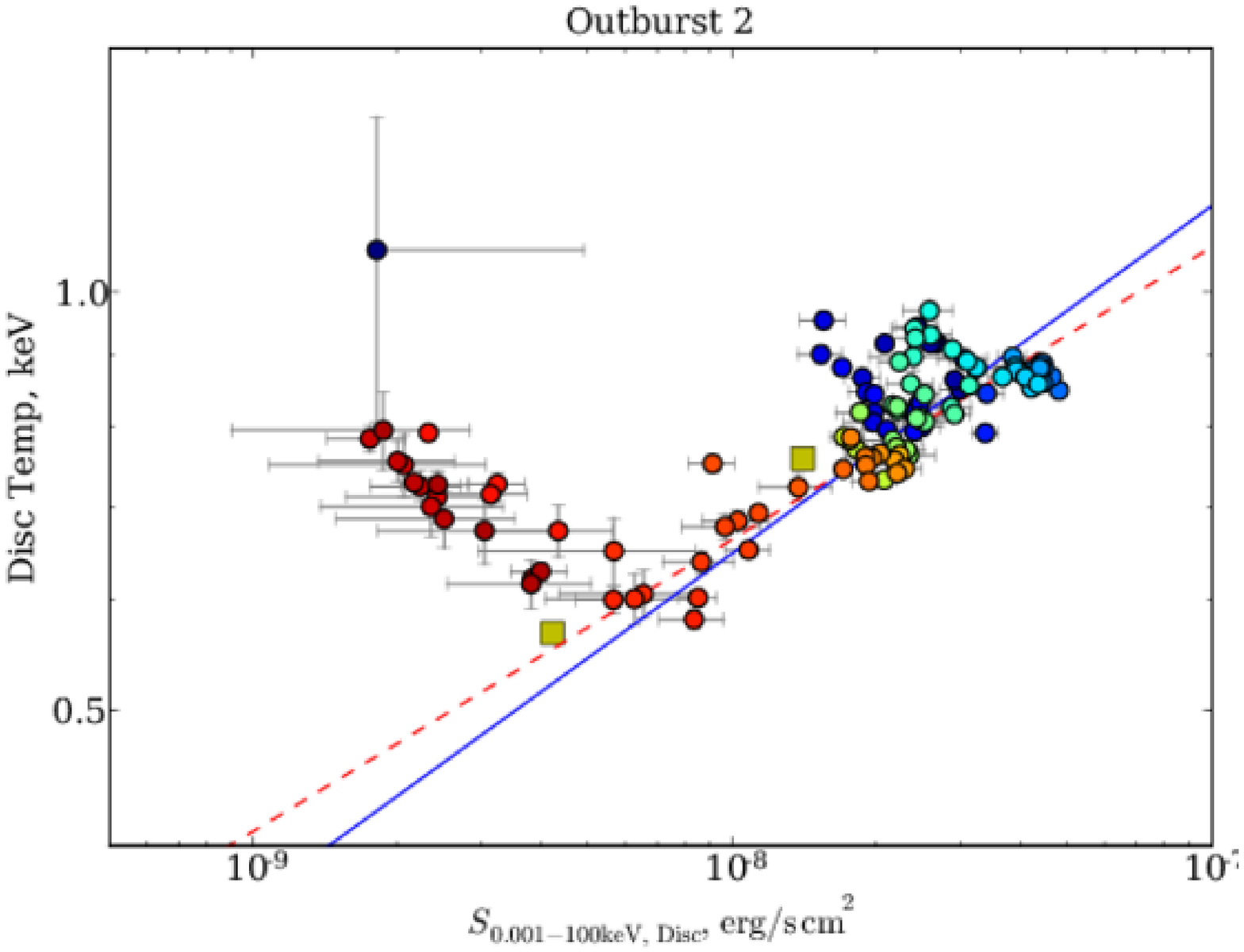}
\includegraphics[width=0.49\textwidth]{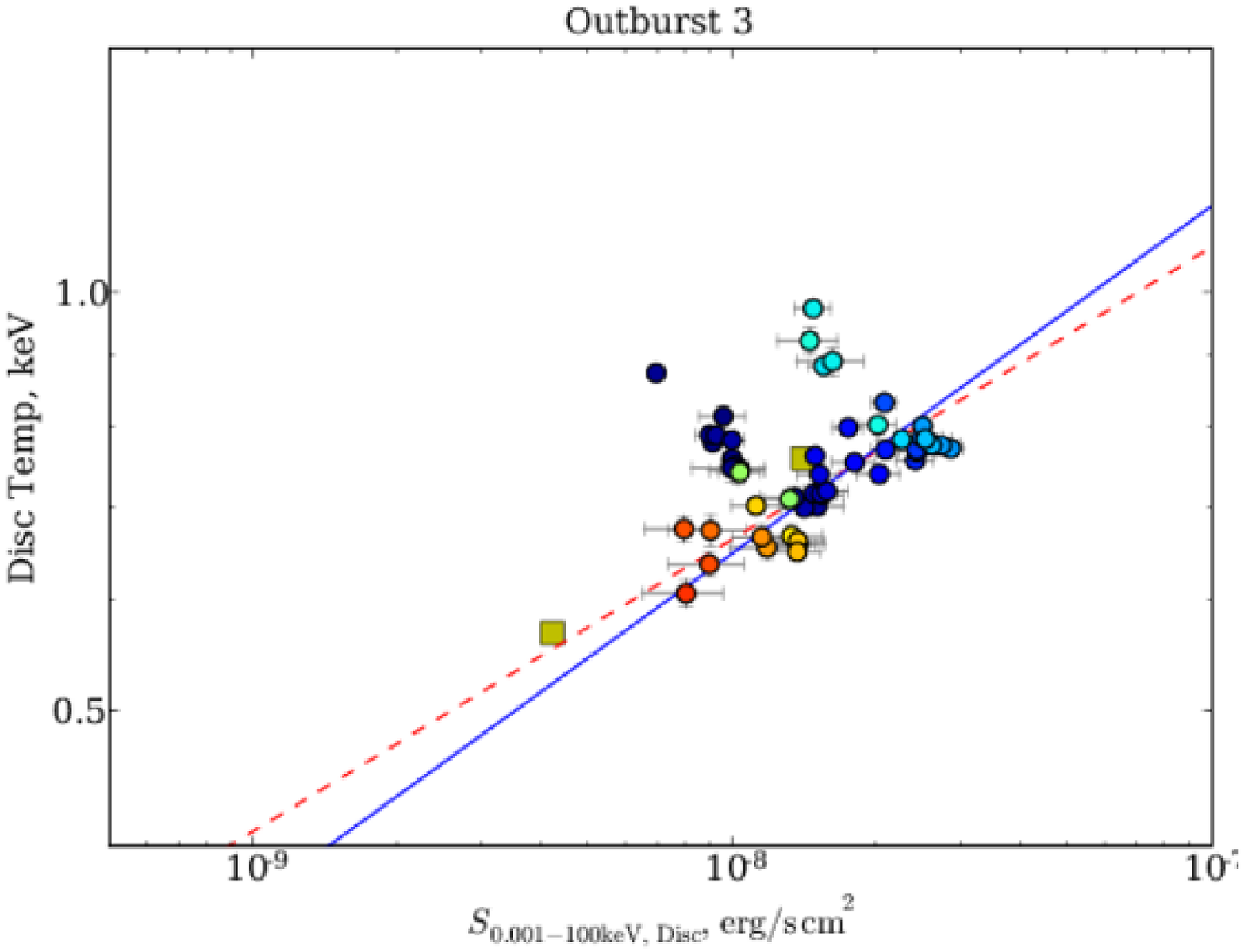}
\caption{\label{fig:FluxDiscTdate} For Outburst 2 ({\scshape left})
  and Outburst 3 ({\scshape right}): The X-ray flux against the disc
  temperature ({\scshape top}) and the disc flux against the disc
  temperature for the two main outbursts of
  \gx339 ({\scshape bottom}).  The colourscale
  shows the time since the transition into the soft state - blue is
  early, red is late.  The best fit lines are shown 
  by the dashed line ($T^{9.44}$ for the flux plots, and $T^{4.75}$
  for the disc flux plots -- from the observations with X-ray
  colour$<0.2$).  The expected correlation of $S_{\rm 
    Disc}\propto T^{4}$ is shown by the solid line. The points
  off the top left of these plots are the observations in the hard
  state which appear to require discs. The square points are
  measurements from \citet{Miller04a, Miller04b, Miller06}.}
\end{figure*}

Fitting the correlation of the total flux from \gx339 with the disc
temperature (for disc temperatures $<1\kev$), we obtain
\[
S_{3-10 \kev}\propto T^{9.44\pm0.12},
\]
which is steeper than what is expected and is shown by the dashed line
in Fig \ref{fig:FluxDiscTdate} {\scshape top}.  Our assumption that the
flux is dominated by the disc component is unlikely to be true.  We
see extra flux above what would be expected from the disc alone at the
lower flux observations.  As a result we need to use the unabsorbed
disc flux.  There are two methods of obtaining the disc flux from the
fit; direct from {\scshape xspec} by setting the normalisations of all
other emission components to zero (including the absorption) or
calculating it from the fitted normalisation and temperature of the {\scshape
  diskbb} model.  The advantage of using the second method is that the
uncertainties on the disc flux are able to be estimated using error
propagation\footnote{A quick investigation into the differences
  between error propagation and a parameter space investigation show
  that the uncertainties from the error propagation are around a
  factor of two larger than those from the parameter space
  investigation.  Newer versions of {\scshape xspec} will have a Monte
  Carlo method for estimating the uncertainties, which will hopefully
  allow the utilisation of first method (Keith Arnaud, private
  communication.)}, and so this method is used for the remainder of
this work.

\begin{figure}
\centering
\includegraphics[width=1.0\columnwidth]{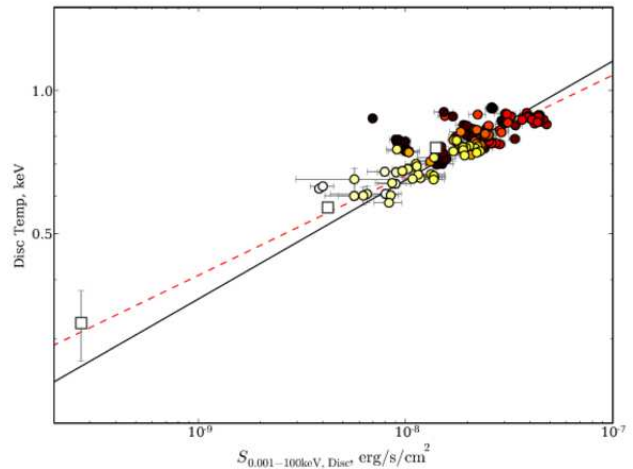}
\caption{\label{fig:WideDiscT} The Disc Flux and Disc temperature for
  observations with an X-ray colour of $<0.2$.  The x-axis has been
  extended to show all the points from \citet{Miller04a, Miller04b,
    Miller06} in white squares.  The black solid line is the theoretical $T^4$ relation and
  the red dashed line is the best fit line to these observations
  $T^{4.62}$.  The colour scale shows the date since the first
  detection of the disc in the continuum emission.}
\end{figure}

However, the decay in the disc temperature as the disc flux decreases is consistent
with the model of $S_{\rm Disc}\propto T^{4}$ as shown by the lines in
Fig. \ref{fig:FluxDiscTdate}.
The square
points are from \citet{Miller04b, Miller04a, Miller06}, some of which
fall outside the limits of this diagram.  These points are consistent
with $T^4$ and the highest temperature one ($0.76\kev$) also matches
the the set of values we obtain (see also Fig. \ref{fig:WideDiscT}).

Although we plot both major outbursts here, the decay is clearest in the
first one as there is a ``clean'' decay in flux with only a small
variation in X-ray colour while in the soft state (Fig. \ref{fig:Outbursts}).   The flux decays
slightly when the \gx339\ reaches the soft state.  There is a small
rise and hardening of the spectrum before the \gx339\ returns to the
soft state and the flux and disc temperature decay gradually.  The
entry and exit from the soft state are seen as ``spurs'' off the $T^4$
relation (Fig. \ref{fig:FluxDiscTdate}).  These may be artifacts of
the fitting process and result from the limitations of the \pca\ bandwidth.

Once the second outburst has reached the soft state, the flux rises at
a fairly steady X-ray colour and then the spectrum hardens.  After a
gap in the light curve the X-ray colour remains 
relatively steady as the flux decays.  Both the entry into and
departures from the soft
state (in the middle and at the end of the outburst) are seen as ``spurs'' off the rest of the
points, which scatter neatly around the $T^4$ relation.  At face
value, the
implication of these ``spurs'' is that the disc {\it cools} onto 
the $T^4$ relation, and at the end of the outburst heats up again.
This increase in the disc temperature as \gx339\ returns to the hard 
state is also seen in the summed spectra in Section \ref{sec:majorOBs}.  

All of the points outside of the general scatter around the $T^4$
relation are from the observations which have a disc but are in the
intermediate state.  Selecting observations with X-ray colours $<0.2$
and disc fluxes of $S_{\rm Disc}>5\times 10^{-9}\ergps$
retains only those which scatter close to $T^4$ (see Fig. \ref{fig:WideDiscT}).  Fitting these we
obtain a best fit of 
\[
S_{\rm Disc}\propto T^{4.75\pm0.23}.
\]
with a Spearman Rank correlation coefficient of $0.904$ and a Kendall
Tau of $0.740$ with a significance of $23.6\sigma$ (taking into
account the errorbars in the disc
temperature and the disc flux).  Although this is
still steeper than the theoretical expectation, the theoretical
expectation still is still a good ``by-eye'' fit.  The range of disc
temperatures is only just under a factor of 2 and the flux a factor of
10, and so the correlation can easily be masked by the scatter.

\subsection{Disc Radii}\label{sec:disc:radii}

\begin{figure*}
\centering
\includegraphics[width=0.49\textwidth]{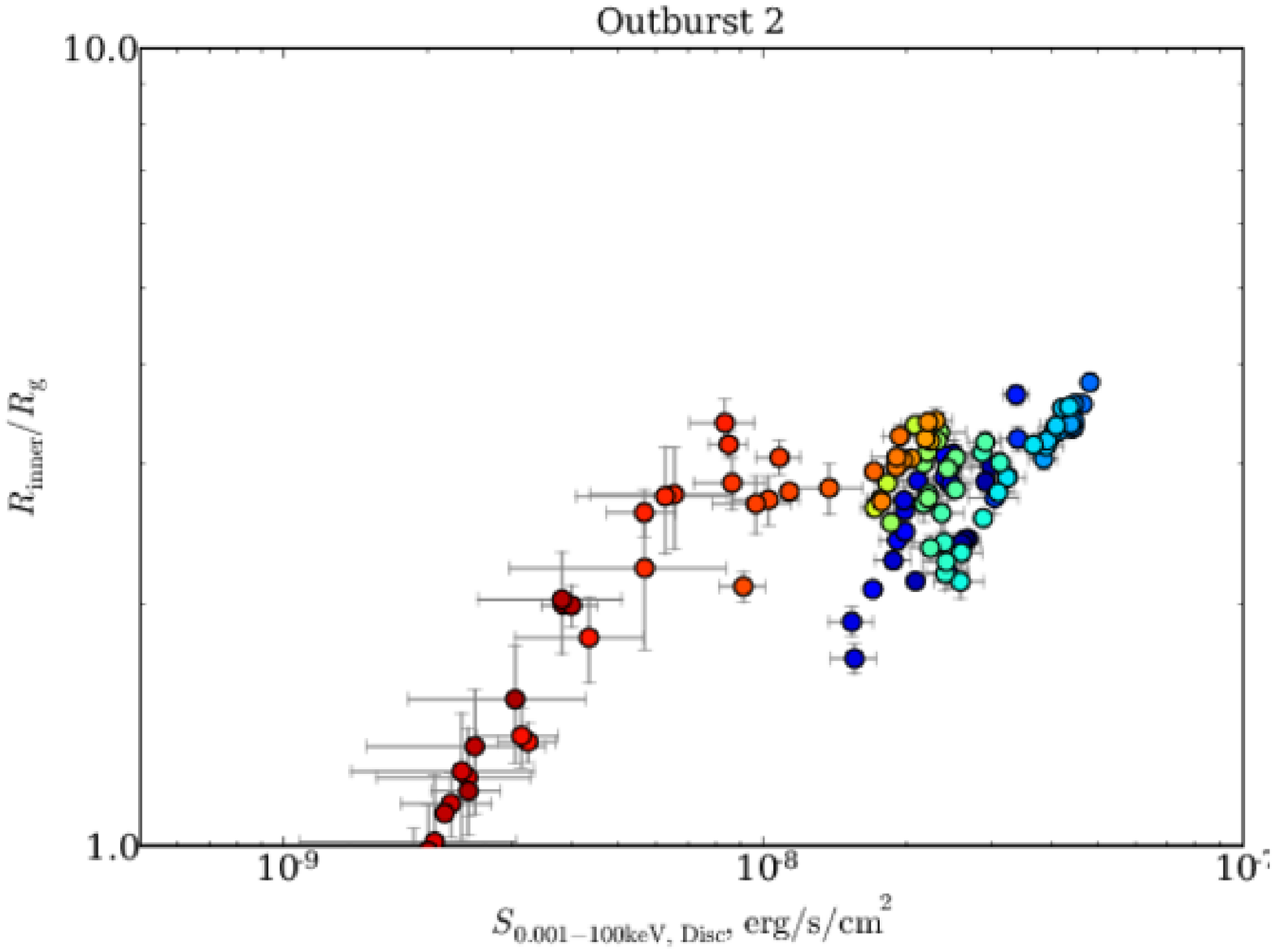}
\includegraphics[width=0.49\textwidth]{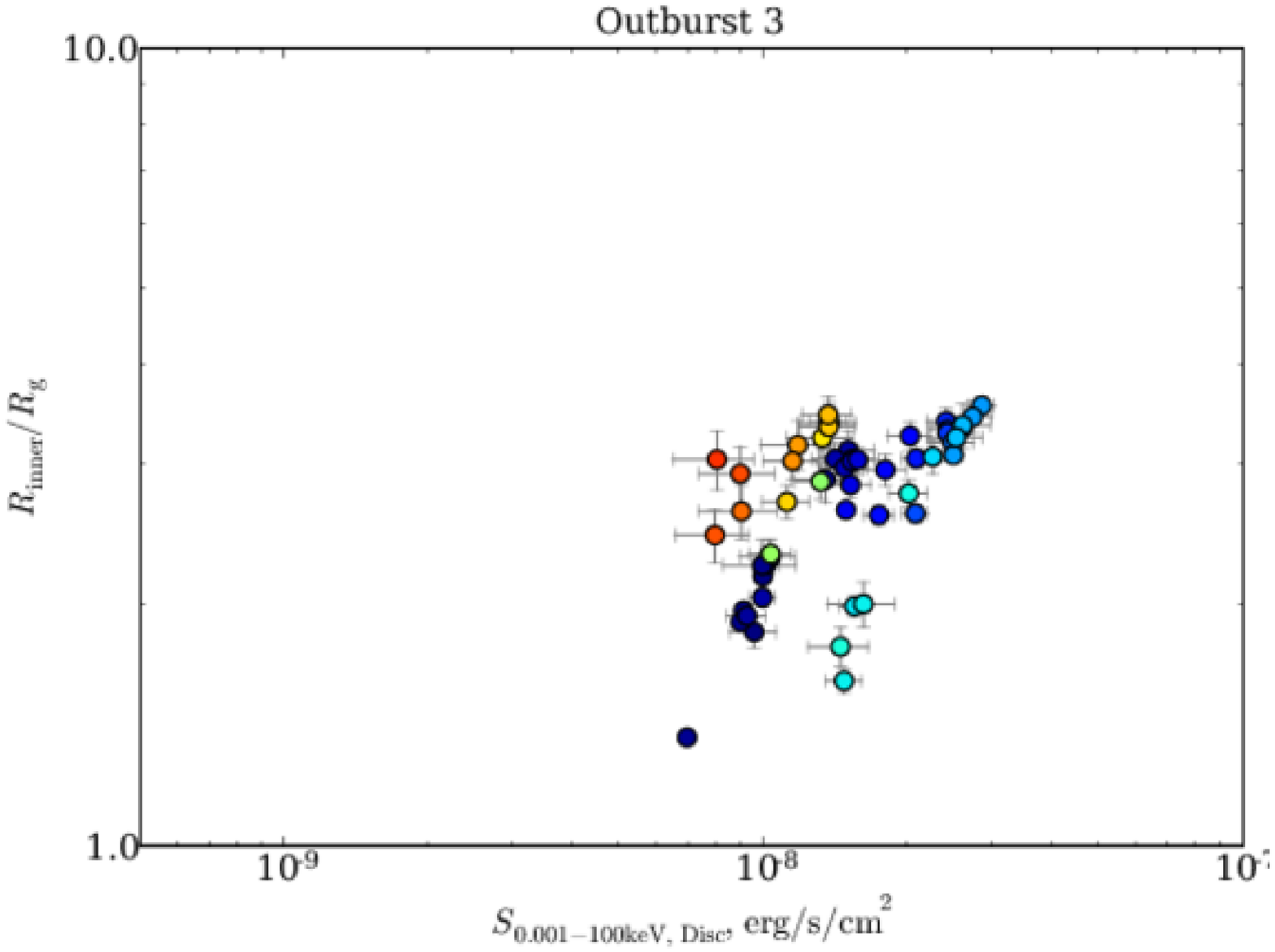}
\includegraphics[width=0.49\textwidth]{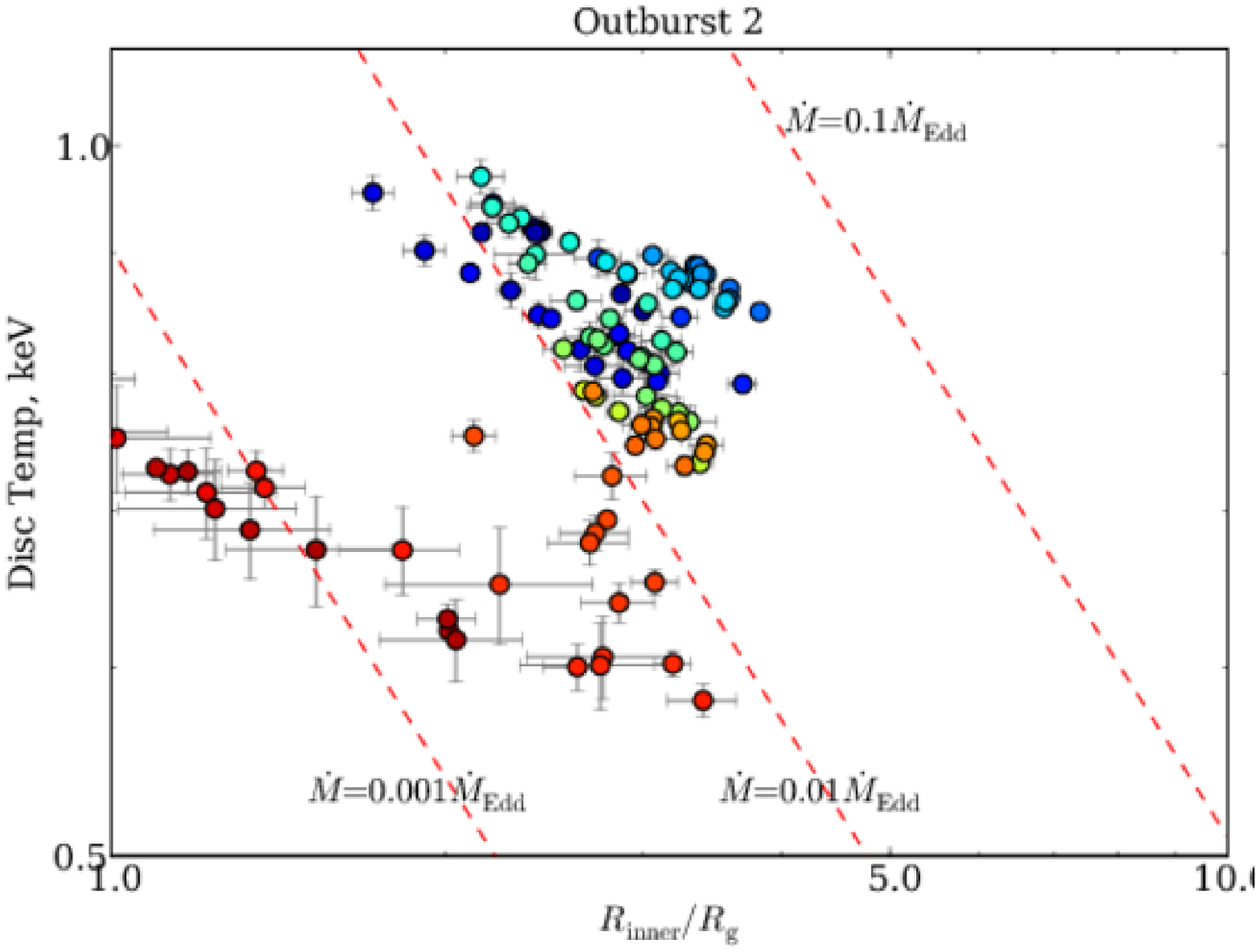}
\includegraphics[width=0.49\textwidth]{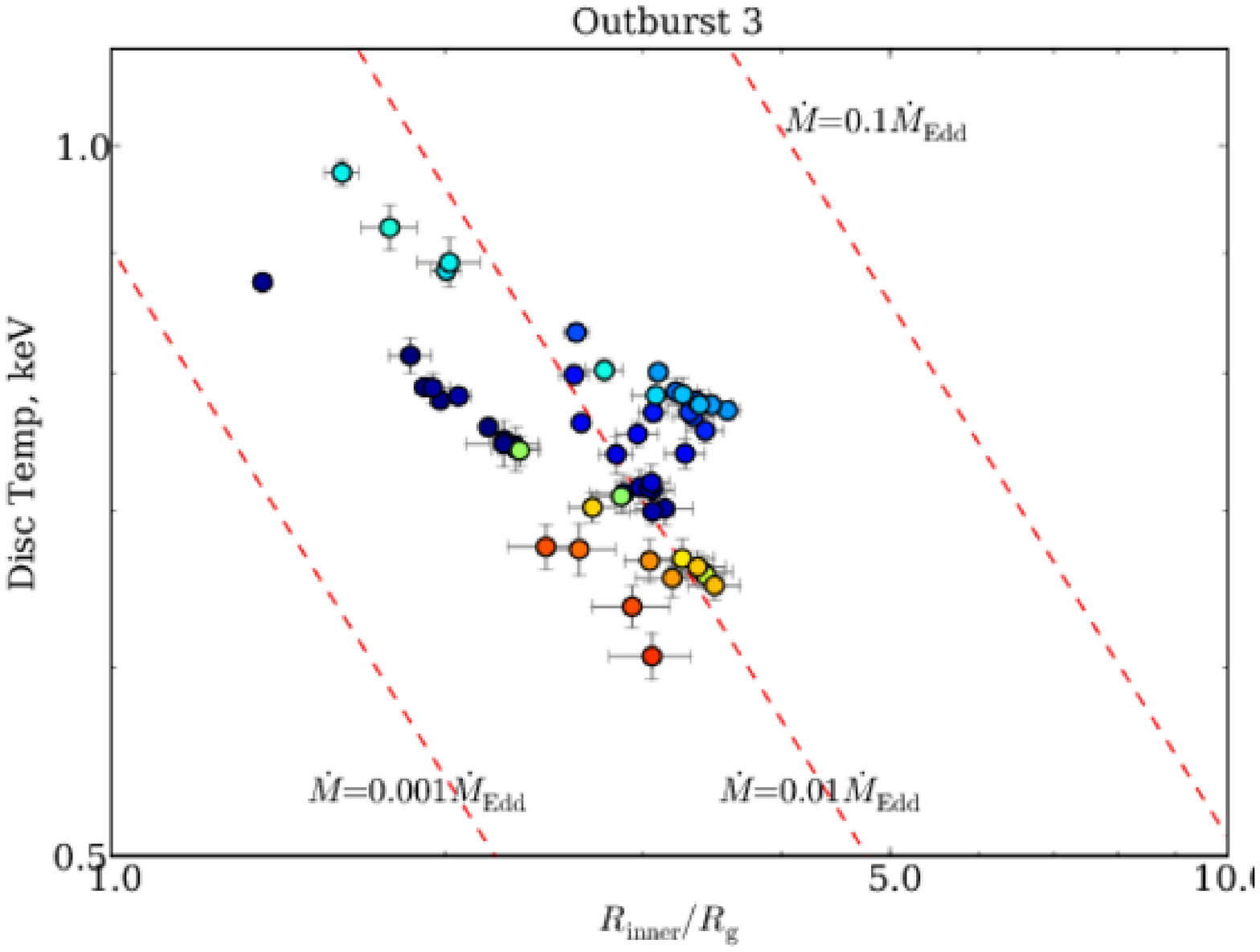}
\caption{\label{fig:Rinner} For Outburst 2 ({\scshape left})
  and Outburst 3 ({\scshape right}):  {\scshape
    top}: The inner radius of the disc in units of $R_{\rm g}$
  through the outburst. {\scshape bottom}: The evolution of the
  temperature of the disc at the inner radius with the inner disc
  radius.  The dotted lines correspond to different constant disc
  accretion rates.  The colour scale shows the time since the
  transition into the soft state - blue is early, red is late.}
\end{figure*}

If the theoretical relation is applicable, then if the inner radius of
the accretion disc is constant, the temperature should follow a $T^4$
law.  The diagrams show that when in the softest part of the soft
state, the temperature variation is as expected.  Therefore we expect
that the inner radius of the disc is constant.  The inner radius of the
accretion disc can be calculated from the normalisation of the
{\scshape xspec} model;
\begin{equation}
R_{\rm in}/R_{\rm G}=0.677\frac{d_{\rm
    10\kpc}}{M/M_{\odot}}\sqrt{\frac{\kappa}{\cos i}},
\end{equation}
where $\kappa$ is the normalisation of the disc model in {\scshape
  xspec} and $i=30^\circ$ is the inclination of the system.  The distance to
\gx339\ is taken to be $8\kpc$ \citep{Zdziarski04} and the mass as $10M_{\odot}$.  We do not
include any uncertainties from the distance and mass into the
calculation of the uncertainties of $R_{\rm in}$.  The variation in
the inner radius of the accretion disc during the two outbursts are
shown in Fig. \ref{fig:Rinner} {\scshape top}.  During the part of the
outburst where the temperature follows a $T^4$ law, the inner radius
of the disc is constant within the uncertainties.  The disc behaves as a
standard thin accretion disc during these parts of the outburst.

However, the same observations which show a departure from $T^4$ also
show a change in the inner radius of the disc.  The implication of these
points is that the disc recedes outwards to something like the
innermost stable orbit at the beginning of the outburst, and falls
inwards at the end.  This is obviously not a physical explanation or
model for the outburst, and so either the model adopted for the emission from
the disc is not appropriate or the spectral analysis is insufficiently
detailed to investigate the disc in the intermediate states.

We follow the analysis of GRS~1915+105 in \citet{Belloni00} and plot
simple derivatives of the two measured and independent quantities, $T$
and $R_{\rm in}$ (see Fig. \ref{fig:Rinner}).  It is in principle possible to calculate the accretion rate from each
spectrum using the expressions from a standard thin disc,
\begin{equation}
\dot{M}=\frac{8\pi R_{\rm in}^3 \sigma_{\rm T} T^4}{3GM}
\end{equation}
(e.g. \citealp{Belloni97}).  Therefore any given $R_{\rm in}$ and $T$
implies an accretion rate, and lines of constant accretion rate as a
fraction of the Eddington limited rate where we assume an efficiency of the
  accretion flow, $\eta$, of 10 per cent. are shown in
Fig. \ref{fig:Rinner} {\scshape bottom}.  The accretion rate is
observed to rise at the beginning of the outburst, reaches a maximum
when the state is at the highest soft flux, and then falls off.  In
both the outbursts shown here the peak accretion rate is approximately
$0.03\dot{M}_{\rm Edd}$, where $\dot{M}_{\rm Edd}=L_{\rm Edd}/\eta c^2$.  The
observations which formed the spurs off the $T^4$ relation have a more
rapid change in the accretion rate through the disc.  Comparing these
two diagrams with the equivalent one in
\citet{Belloni00}, the slope of the observations in GRS~1915+105 are
similar to that of the spurs in \gx339.

\citet{Gierlinski04} include a colour temperature correction
factor in their analysis of discs in ten black hole binary systems,
following \citet{Shimura95, Merloni00}.  The effect of this correction
is almost always to harden the spectrum.  For the same disc
luminosity, adding in the colour correction increases the disc
temperature.  Therefore our results may be consistent with $\propto
T^4$ if the temperature we extract from {\scshape xspec} includes a
colour correction effect.  To further investigate whether the relation is truly shallower
than expected more data on the disc in the soft state is required.  We
note that \citet{McClintock07} also find a deviation from the expected
$T^4$ law in the X-ray Nova H1743-322 and \citet{Tomsick05} present a
detailed correlation between $L$ and $T$ with deviations from the
expected relation for 4U~1630-47.

Comparing Figs. \ref{fig:Turtle-EW} and \ref{fig:Turtle-DiscT}, as
the disc temperature decreases, the equivalent width increases.
Excluding the observations of the disc in the hard state and fitting the correlation gives
\[
EW\propto T^{-3.3},
\]
with a Spearman Rank correlation coefficient of $-0.704$ and a Kendall
Tau of $-0.494$ with significance of $9.48\sigma$.

\section{Disc Fraction Luminosity Diagram} \label{sec:DFLD}

In a study whose aim was to compare AGN and X-ray binaries,
\citet{Koerding06} construct a more general version of the HID.  The
location of points in the HID depends on the total luminosity of the
system and the strength of the non-thermal to thermal emission. In
binaries, both the power-law and disc components are observed in
X-rays, whereas in AGN the disc peaks in the UV band.  As a result the
HID for AGN as estimated from X-ray data alone would be unlikely to
give any insight into the state of the source, and UV data from a
large number of AGN is scarce.

Therefore \citet{Koerding06} generalised the HID by using the total
luminosity and the disc fraction which is calculated from
\[
{\rm Disc\, Fraction}=\frac{L_{0.1-100\kev,\ {\rm PL}}}{L_{0.01-5\kev,\ {\rm Disc}}+L_{0.1-100\kev,\ {\rm PL}}}.
\]
The Disc Fraction characterises the relative strengths of the disc and power-law
components and also remains finite if either of the components
approach zero.

\begin{figure}
\centering
\includegraphics[width=1.0\columnwidth]{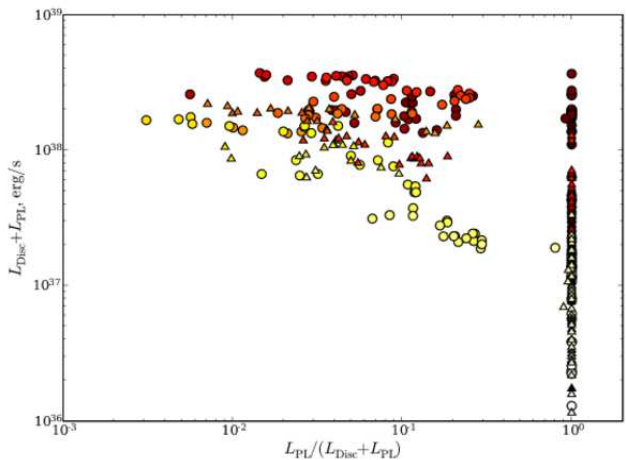}
\caption{\label{fig:DFLD} The Disc-Fraction Luminosity diagram for
  \gx339\ for the two major outbursts.  The colour scale
  shows the date of the observation relative to the start of the
  outburst - black is early, white is late.  The Outburst 2 is shown
  by circles, Outburst 3 by triangles.}
\end{figure}

We adapt this characterisation to the spectrum of \gx339.  The model fits throughout the outbursts of \gx339\ allow the disc and
power-law fluxes to be calculated, $S_{\rm disc}$ from $0.001-100\kev$
and $S_{\rm PL}$ from $0.1-100\kev$\footnote{The disc flux is the
  unabsorbed flux, the powerlaw flux is the absorbed flux.}.  This allows us to calculate a
Disc Fraction Luminosity diagram (DFLD) for an X-ray binary for the
first time.  To calculate the luminosity of the different components
we take the distance to \gx339\ to be at $8\kpc$. The DFLD is shown in
Fig. \ref{fig:DFLD}.  The colour scale tracks \gx339\
through the diagram during the outburst, and is linked to the time
since the beginning of the outburst (similar to Fig. \ref{fig:Outbursts}). 

\citet{Koerding06} simulate a DFLD for 100 objects based on previous
work on the evolution of the disc and power-law components during an
outburst (see their Fig. 10).  Reassuringly, even for only two
outbursts from one object, Fig. \ref{fig:DFLD} looks similar.

We note that as we only use X-ray observations in this study, this
DFLD is limited, as in quiescence a disc is again seen.  However, this
disc is cooler, and so is best observed in the UV and it is not seen
in the X-ray.  This re-appearance of the disc is hinted at by the
softening of the low flux hard state tail in the HID
(Fig. \ref{fig:Turtle-model}).  Future outbursts where the decays are
studied in the Ultra-violet will reveal the shape of the DFLD for all
states that XRBs exhibit.

The shape of the DFLD is subtly different to the HID.  As the disc
flux and temperature decay, the X-ray colour remains almost constant,
before very rapidly hardening at the end of the outburst.  This is a
result of the X-ray colour being an indirect measure of the disc
flux.  The Disc Fraction is a more physical measure, and as a result
the decay of the disc can be easily tracked as the disc fraction
slowly reduces towards the power-law dominate state.

The tracks of the two outbursts through the DFLD is very similar to
that in the HID.  The hardening and softening of the spectrum is seen
along with the decay in flux.  However, the motion is more spread
out.  Whereas in the HID, the range in X-ray colour was from $0.03-0.3$
while the source was in the soft state, the range in the disc
fraction is from $0.007-0.3$.  This allows a more detailed
investigation into the variation of the disc during an outburst. 

There is a large gap in the diagram between the power-law dominated
section and the area where a disc is detected.  This is the result
of the limited spectral coverage of the \rxte\ \pca.  The gradual rise
of the disc may be taken as
the steepening of the lower slope in the broken powerlaw.  Only once the
curve of the black-body is dominant does the disc model become the
best fitting, by which time the disc fraction is already significant.

\section{Radio/X-ray Correlations} \label{sec:Radio}

To investigate correlations of the radio flux within the DFLD the we took the
compilations presented in \citet{Corbel00} and \citet{Gallo04}.
As the radio observations were not coordinated with the \rxte\
observations we match the two data-sets as best possible.  If there is
a radio observation within two days of an X-ray observation, then this
radio observation is linked to the X-ray one.  A two day overlap gave
a reasonable number of X-ray observations which had corresponding
radio flux values.  

There was little data from MOST\footnote{Molongolo Observatory
Synthesis Telescope}, nor from $2.3$ and $1.5\ghz$
ATCA\footnote{Australia Telescope Compact Array} observations.  As a
result we concentrate on $8.5$ and $5\ghz$ data as well as the spectral
index, $\alpha$ (where $S_{\rm Radio}\propto\nu^{\alpha}$).  The data from \citet{Corbel00} cover Outburst 2, whereas the radio data from \citet{Gallo04}
cluster around the peak of Outburst 3.

\begin{figure*}
\centering
\includegraphics[width=0.95\columnwidth]{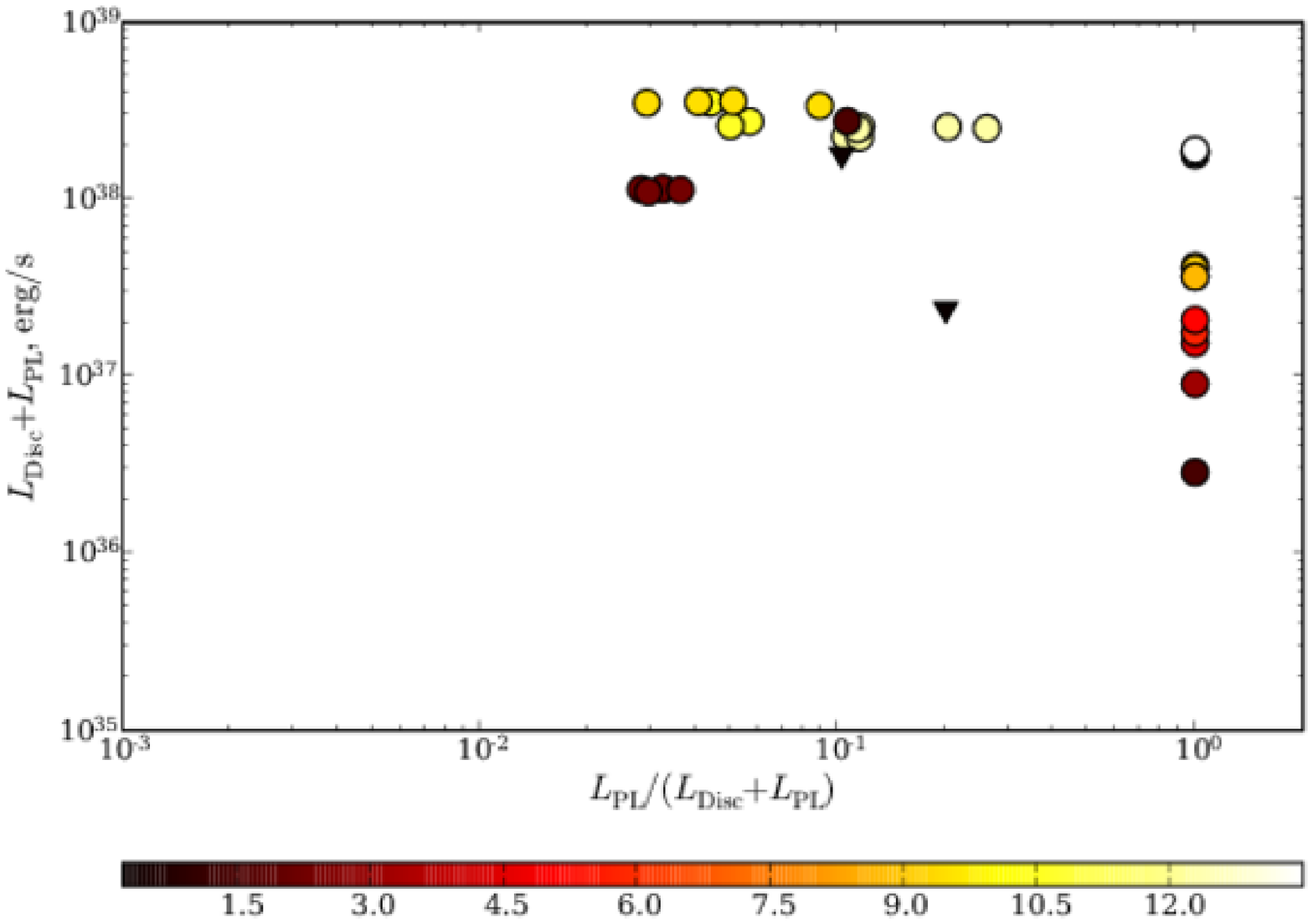}
\includegraphics[width=0.95\columnwidth]{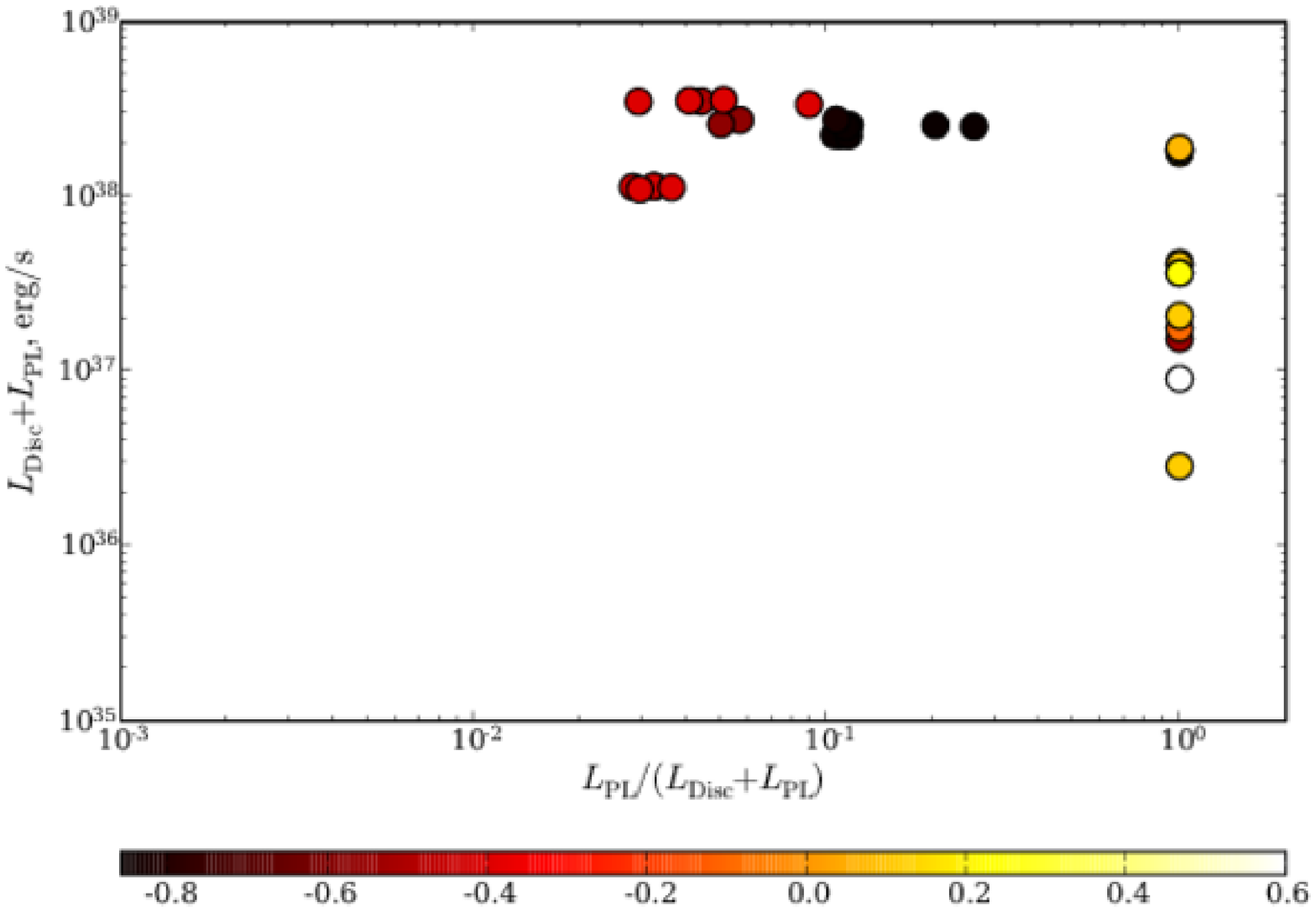}
\caption{\label{fig:DFLDradio} DFLD with {\scshape left} $8.5\ghz$ radio
  flux in mJy and {\scshape right}
  $\alpha$ as the colour scale.  Points which only have upper limits
  on the radio flux are shown by triangles.}
\end{figure*}

As can be seen in Fig. \ref{fig:DFLDradio}, the highest radio fluxes are observed
at the top of the DFLD.  There is a sharp drop in radio flux as
\gx339\ drops to the bottom of the DFLD and heads back to the hard
state.  There is also a possibility that the radio flux decreases as
\gx339\ moves across the top of the DFLD.

In the right-hand panel of Fig. \ref{fig:DFLDradio} the change of the
radio spectrum between the hard and soft states is clear.  The stalk
of the DFLD has a spectral index of $\alpha \sim 0.2$, whereas the
soft state has a spectral index of $\alpha \sim -0.4\to -0.6$.

\section{Summary}\label{sec:conc}

We have performed a comprehensive and consistent investigation into the
disc and iron lines detected in the X-ray emission from \gx339.  All
the 913 public
observations in the 11 year \rxte\ archive were reduced, but only 634
had sufficient number of counts and flux to be analysed further.  Three types
of model were fitted to each observation; single powerlaw, broken powerlaw and a disc +
powerlaw and the best fitting one chosen.  The spectra were also
tested for the presence of a Gaussian line at $6.4\kev$.  There were
four outbursts in the data, but we concentrate on the two best
sampled.  The relative variation in flux and X-ray
colour between these two outbursts are remarkably similar.

A significant iron line was detected in 400 of the 634
observations.  There is an anti correlation between the
flux and the equivalent width of the iron line for observations of the
soft state (X-ray colour $<0.22$) of $EW \propto S_{3-10 \kev}^{-0.621}$.  This is
steeper than the range in slopes found in AGN.  In the
hard state the data are consistent with no correlation.  As such there
is an effect analogous to the X-ray Baldwin effect in AGN present in X-ray
binaries in the soft state.  We compare the powerlaw slope, the line equivalent width and
X-ray colour of the spectra.  The relation obtained is similar to that seen in a
sample of AGN studied by \citet{Mattson07}.  Therefore the behaviour
of the line is hysteretical.

Theoretical arguments indicate that the decay in disc temperature with
flux should mimic that for a black body, $S_{\rm Disc}\propto R^2T^4$.  We find that the disc
flux rather than the source flux needs to be used for this relation to
be valid.  In the soft state the
decay in disc temperature matches the expected $S_{\rm Disc} \propto
T^4$.  The best fit, however, is steeper at $T^{4.75}$ resulting from
the comparatively large scatter and small range in disc temperature.
This implies that during the decay of the disc temperature in the soft
state the inner radius of the disc is constant.  Departures from a
constant inner disc radius are non-physical, implying that the model
for the emission is not appropriate or the spectral analysis is
insufficiently detailed during the the intermediate states.

Following the method outlined in \citet{Koerding06} we construct a Disc
Fraction Luminosity Diagram for \gx339.  We find that the shape
qualitatively matches that produced for AGN.  Linking this with 
the radio emission from \gx339\ the change in radio spectrum between
the presence and absence of the disc is clearly visible.  The large
gap in the DFLD at low disc-high powerlaw fractions is surmised to
result from the lack of spectral resolution at low energies.  Summed
spectra from different parts of the HID show that this is indeed the
case for the low intermediate state.

\section*{Acknowledgements}

We thank the referee for a detailed and helpful report, Mike Nowak for enlightening suggestions and discussions, Dan Summons, Vanessa McBride and Dave Russell for help with
the {\it RXTE} and general data reduction and James Graham for help
with {\scshape matplotlib}. EGK acknowledges funding form a Marie Curie Intra-European fellowship under contract Nr. MEIF-CT-2006-024668. 

\bibliographystyle{mn2e} 
\bibliography{mn-jour,/net/jets/black/r.j.dunn/bibtex/dunn3}

\section{Appendix: Spectral Fits}\label{sec:appendix}

\begin{figure*}
\includegraphics*[angle=-90, width=0.495\textwidth]{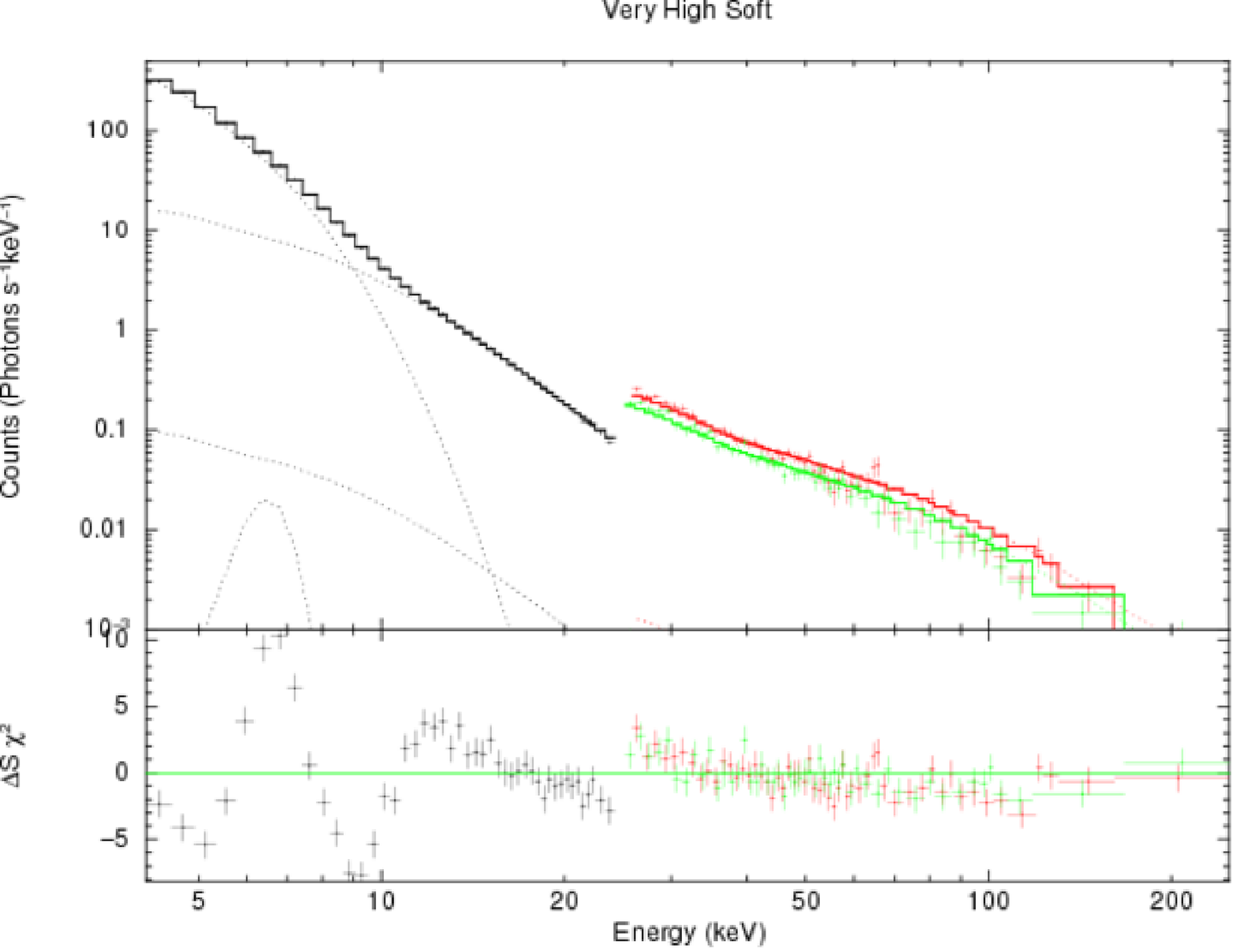}
\includegraphics*[angle=-90, width=0.495\textwidth]{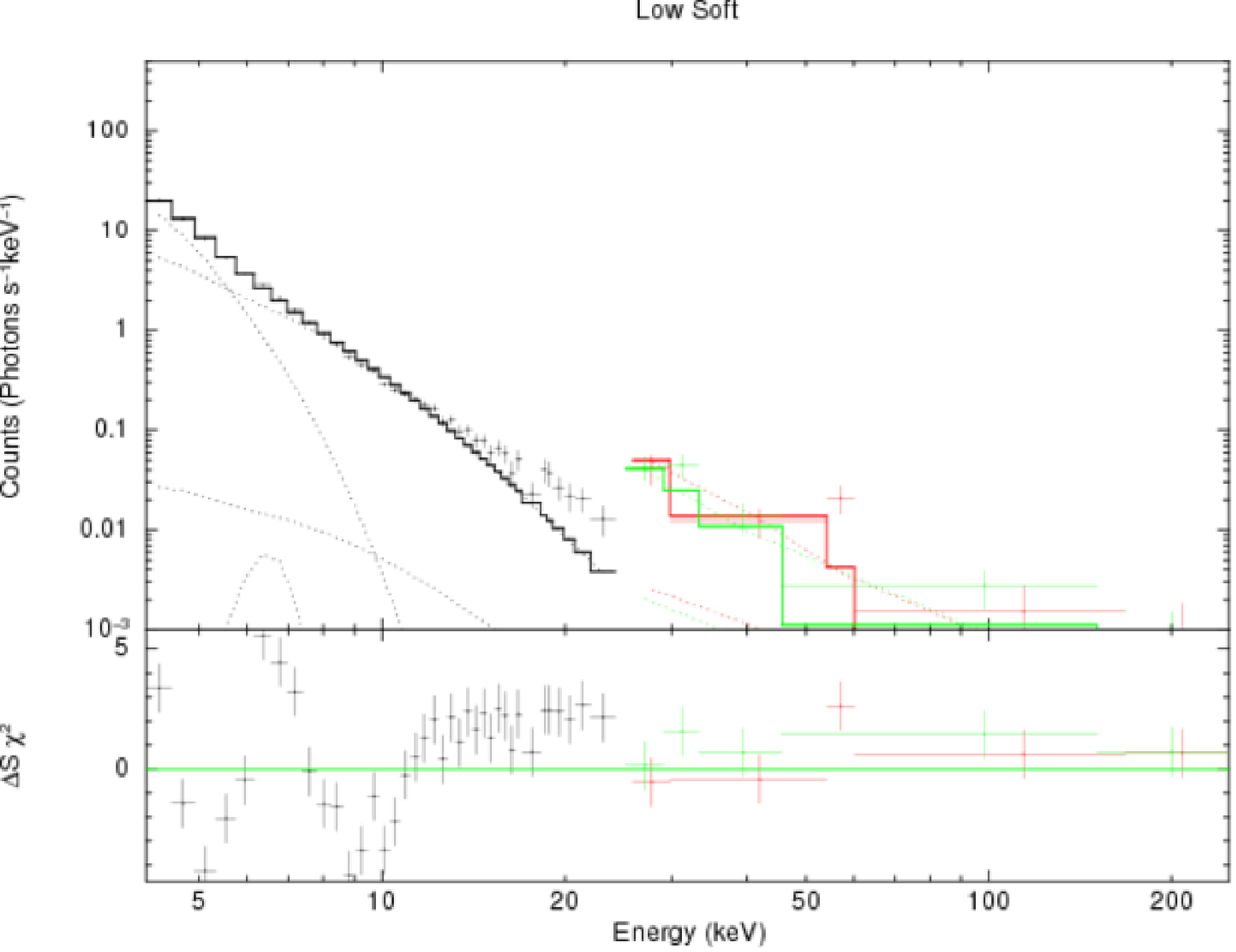}
\includegraphics*[angle=-90, width=0.495\textwidth]{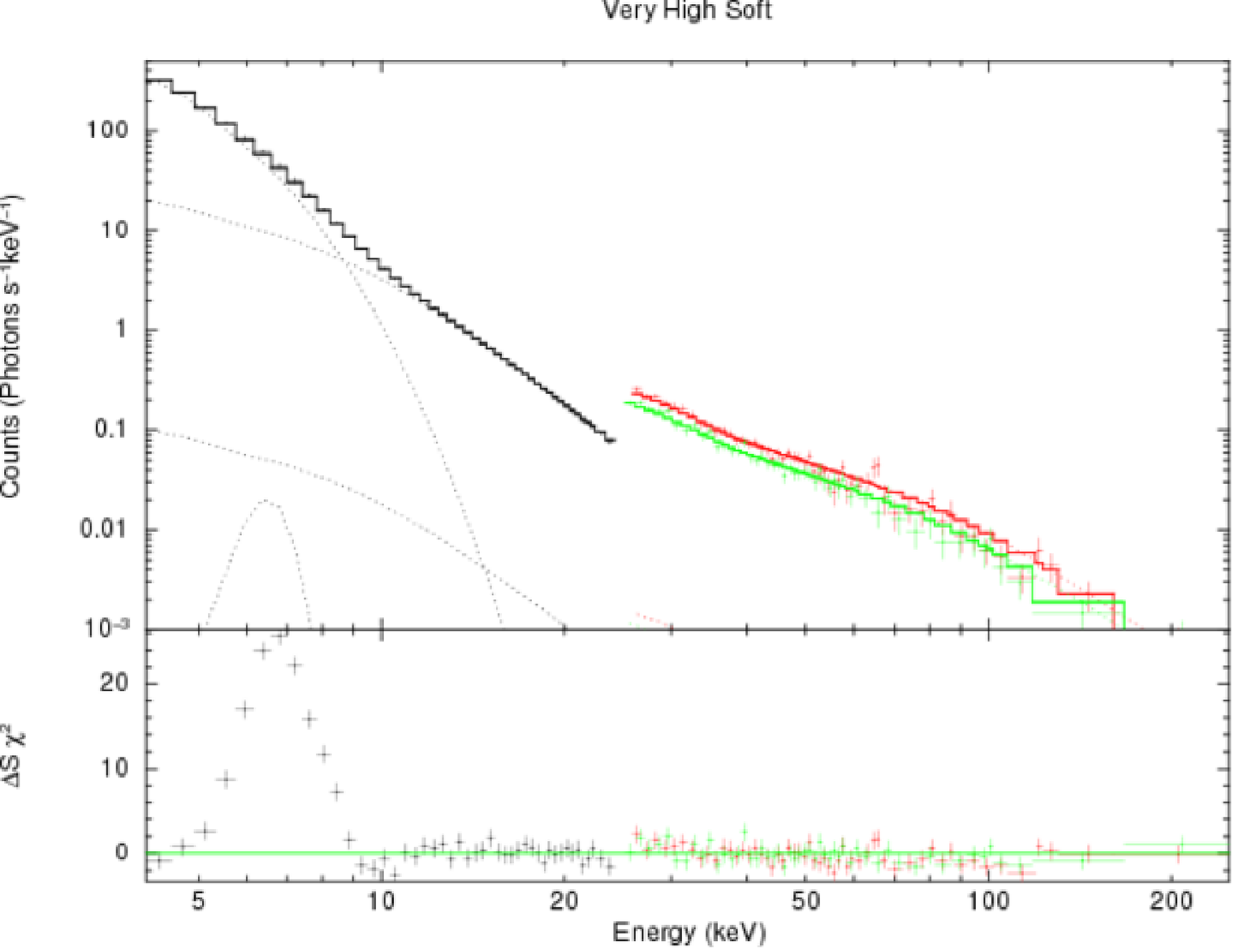}
\includegraphics*[angle=-90, width=0.495\textwidth]{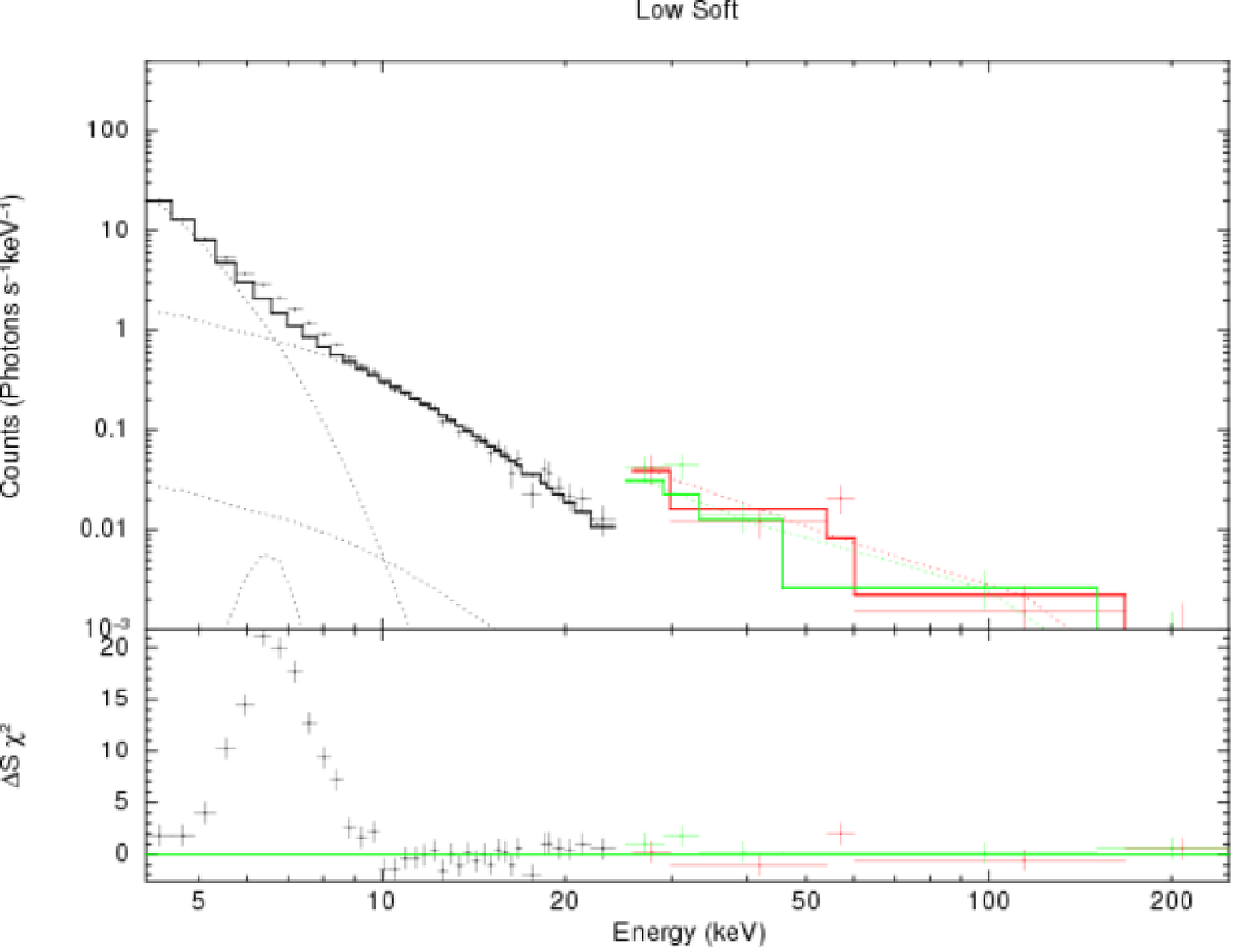}
\caption{\label{fig:nolinespectra}{\scshape top:} The {\scshape disc +
    powerlaw} models fitted to the data with no line component
  present.  {\scshape bottom:}The {\scshape disc +
    powerlaw + gaussian} models fitted to the data, and then the
  normalisation of the line component set to zero before producing the
plot.}
\end{figure*}

We show in Table \ref{tab:fits} the results of fitting the models
adopted for this analysis to the summed spectra.  We only fit the
models used in this analysis rather than more complex ones which would
be appropriate for achieving a good fit to this high signal-to-noise
data.  We also show in Fig. \ref{fig:nolinespectra} fits to the
spectra which did not include a line component, as well as the fits
used in Fig. \ref{fig:spectra} but with the line normalisation set to
zero before producing the plot.

\pagebreak

\begin{table}
\caption{}
\label{tab:fits}
\begin{tabular}{lcl}
\\
\end{tabular}
\end{table}

\begin{figure*}
\includegraphics*[height=0.99\textheight]{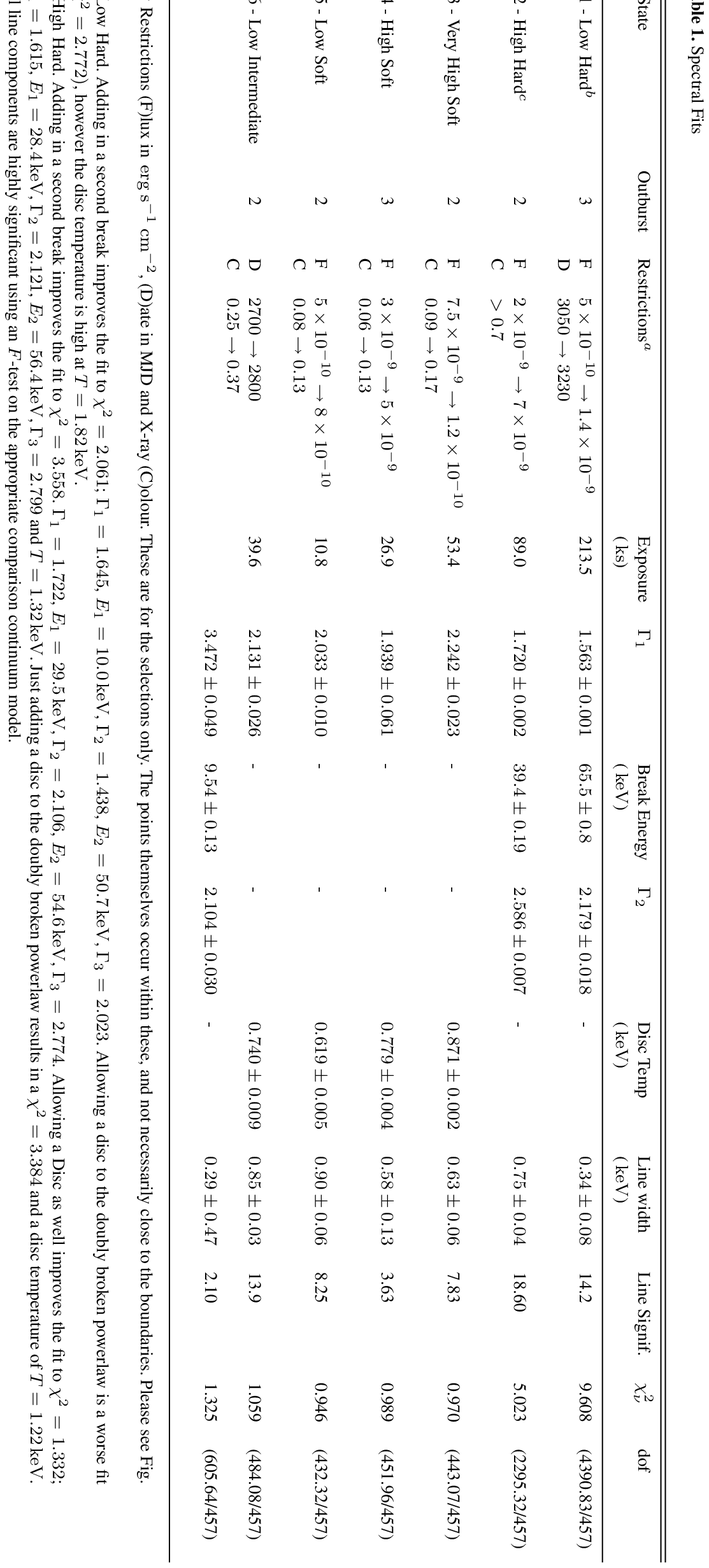}
\end{figure*}

\end{document}